\documentclass[aps,twocolumn,showpacs,superscriptaddress,nofootinbib]{revtex4-2}

\usepackage{graphicx}
\usepackage{latexsym}

\usepackage{amsmath}
\usepackage{bm}

\usepackage{color}
\newcommand{\red}[1]{\textcolor[named]{Black}{#1}}

\begin{document}
\title{
%% Differences in stochastic heat descriptions between particles and the density field
%% Stochastic heat differences between particle and density field descriptions
Stochastic heat differences between many-particle and density field descriptions
}

\author{Takuya Saito}
\email[Electric mail:]{tsaito@phys.aoyama.ac.jp}
\affiliation{Department of Physical Sciences, Aoyama Gakuin University, Chuo-ku, Sagamihara 252-5258, Japan}

\author{Yutaka Sumino}
%% \email[Electric mail:]{ysumino@rs.tus.ac.jp}
	\affiliation{%
		Department of Applied Physics, Tokyo University of Science, 6-3-1 Nijuku, Katsushika, Tokyo 125-8585, Japan}
        \affiliation{WaTUS and Division of colloid and interface science, Research Institute for Science \& Technology, Tokyo University of Science, 6-3-1 Nijuku, Katsushika, Tokyo 125-8585, Japan
	}

\def\Vec#1{\mbox{\boldmath $#1$}}
\def\degC{\kern-.2em\r{}\kern-.3em C}

\def\SimIneA{\hspace{0.3em}\raisebox{0.4ex}{$<$}\hspace{-0.75em}\raisebox{-.7ex}{$\sim$}\hspace{0.3em}} 

\def\SimIneB{\hspace{0.3em}\raisebox{0.4ex}{$>$}\hspace{-0.75em}\raisebox{-.7ex}{$\sim$}\hspace{0.3em}}

\date{\today}

\begin{abstract}
\red{This article investigates spatiotemporally discrete or continuous stochastic descriptions, where we focus on differences in heat naturally defined between the particle level and the density field. Both the descriptions are found to generally make the heat differences by the entropic term expressed just with the number density through spatial projection from the many particles' positions onto the density field. The transformation from the Langevin to Dean--Kawasaki equations is considered as the projection in the continuous descriptions, where the emergent heat differences undergo little temporal variations due to the sparse distributions of the point-particles. On the other hand, the analogous formalisms constructed in the discrete models may exhibit the explicit temporal evolutions of the entropic term. Furthermore, we develop arguments about the interpretation and applicability of the heat differences as well as the perspectives to many-polymer system.
}
\end{abstract}

%% Notably, spatiotemporal resolutions are not altered in all the above projections.


\def\degC{\kern-.2em\r{}\kern-.3em C}

\newcommand{\gsim}{\hspace{0.3em}\raisebox{0.5ex}{$>$}\hspace{-0.75em}\raisebox{-.7ex}{$\sim$}\hspace{0.3em}} 
\newcommand{\lsim}{\hspace{0.3em}\raisebox{0.5ex}{$<$}\hspace{-0.75em}\raisebox{-.7ex}{$\sim$}\hspace{0.3em}}

\maketitle

\section{Introduction}

Projection serves as an inevitable notion to address statistical issues that involve numerous degrees of freedom and might provide multiple pictures to observe identical phenomena from different perspectives.
Systems viewed from hierarchal constructions with scale-separated resolutions, coarse-graining, or the elimination of fine structures are leveraged to extract the physics of interest~\cite{Kubo_Iwanami,JCP_Zwanzig_1960,PTP_Mori_1965,JPhysA_Kawasaki_1973,PR_Robertson_1966,PRA_Kawasaki_Gunton_1973,Grabert,SekimotoBook,deGennesBook,Doi_Edwards,Khoklov_Grosberg,Doi_SoftMatter,PhysRep_Schilling_2022,JCP_Widder_2022,EPL_Glatzel_Schilling_2021,JCTC_Jin_2022,PRE_Yoshimori_2005,JStatPhys_Zwanzig_1973,arXiv_Dengler_2016}.
A celebrated example of the application is a colloidal particle with a size on the order of 1~$\mu$m~\cite{Kubo_Iwanami,PTP_Mori_1965,SekimotoBook,PhysRep_Schilling_2022,JCP_Zwanzig_1960,JStatPhys_Zwanzig_1973}. The colloid suspended in a solution exhibits Brownian motion spurred by countless and endless collisions of typically sub-nanometer- or nanometer-sized solvent molecules.
Independent of a rich variety of solution compositions, the projection creates a simple description well known as a Langevin equation~\cite{Kubo_Iwanami,SekimotoBook,PTP_Mori_1965,PhysRep_Schilling_2022,JCP_Zwanzig_1960,JStatPhys_Zwanzig_1973} that simply leaves a degree of freedom of the colloid, whereas the collisions are translated into friction and noises with a few parameters. 
Notably, a white noise appearing in the Langevin equation concerns the coarse-graining procedures of changing spatiotemporal resolutions, where specific temporal evolutions in the noise correlations are reduced to the Dirac's delta function.

As systematic mesoscopic constructions, Mori-Zwanwig's projection operators guide us into stochastic description from microscopic Hamiltonian dynamics on phase space~\cite{JCP_Zwanzig_1960,PTP_Mori_1965,JPhysA_Kawasaki_1973,arXiv_Dengler_2016,PhysRep_Schilling_2022,JCP_Widder_2022,PR_Robertson_1966,PRA_Kawasaki_Gunton_1973,Grabert,JCP_Li_2017,JCP_Widder_2022,JCP_Meyer_2019,EPL_Glatzel_Schilling_2021,JCP_Glatzel_2021,PRE_Yoshimori_2005,JStatPhys_Zwanzig_1973}.
Although followed by well-organized instructions once the projection operators are defined, a key issue is to discover the appropriate projection operators to extract variables of interests or a simple physical formulation.
This article does not restrict itself just into the standard formalisms of projection operators, but considers the family of ``projection" in \red{a broad sense} while the basic notion to cast and pull out variables of interests being maintained.

%% This article uses terminology of ''projection" in s broad sense with the basic idea kept to cast variables, but does not restrict itself just into the standard formalisms of projection operators.

%% Those projections operators share basic notion, but are not narrowed down to a unique form as organized and developed by modern framework~\cite{PhysRep_Schilling_2022,JCP_Widder_2022}.

In this context, the projections are to involve mapping quantities tied to particles onto those in field coordinates as well as eliminating degrees of freedom.
For instance, a derivation from Langevin to Fokker-Planck \red{equation} may be included.
From the wide viewpoint of field theory, the Fokker--Planck equation of the probability density ${\cal P}(t,{\bm x})$ of a particle position~\cite{Kubo_Iwanami,SekimotoBook,vanKampen,Gardiner,PhysRep_Bouchaud_Georges_1990} is considered as a (scalar) field equation.
Recall that a standard derivation from the Langevin to the Fokker--Planck equation includes a step to see an ensemble average of $\delta({\bm x}-\hat{\bm x}(t))$ with a stochastic variable $\hat{\bm x}(t)$ at each spatial point ${\bm x}$~\cite{SekimotoBook}, which results in probability density ${\cal P}(t,{\bm x})=\left< \delta ({\bm x}-\hat{\bm x}(t))\right>$; however, this procedure does not alter the resolutions, i.e., this projection procedure does not resort to changes in spatiotemporal resolutions.
%% whereas white noises appearing in the Langevin equation concern the coarse-grainings of the spatiotemporal resolutions.
In addition, a caveat is that the time evolution of the probability density is formally seen as being ``deterministic" and ``field-theoretic" on the math expression, whereas the trajectories drawn by $\hat{\bm x}(t)$ in the Langevin equation are ``stochastic" and \red{``particle-level."}
One may, then, ask for an intermediate description.
Indeed, the Dean--Kawasaki equation~\cite{JPAMG_Dean_1996,JPhysA_Archer_Rauscher_2004} essentially plays a middle role between the Langevin and the Fokker--Planck equations as a ``stochastic" and ``field-theoretic" 
equation that retains stochastic noises projected onto the field coordinate from the Langevin equation without averaging operations.

Stochastic thermodynamics describes thermodynamic structures in mesoscopic fluctuating world~\cite{PRE_Crooks_1999,PRE_Crooks_2000,RPP_Seifert_2012,PRL_Seifert_2016,PRE_Gelin_Thoss_2009,PRE_Talkner_Hanggi_2016,PRX_Jarzynski_2017,arXiv_Roldan_2023}, and has been developed along the notion of the trajectories~\cite{PRE_Crooks_1999,PRE_Crooks_2000,RPP_Seifert_2012,PRL_Seifert_2016,PRE_Talkner_Hanggi_2016,PRX_Jarzynski_2017,PRX_Ciliberto_2017,arXiv_Roldan_2023}, especially, of the particles, where
heat defined with the Langevin equation proposed by Sekimoto~\cite{SekimotoBook} has contributed to the construction of the stochastic pictures on the energy balance.
The projection also addresses noteworthy issues in stochastic thermodynamics.
The Dean--Kawasaki equation found through the projection of mapping the particles' trajectory onto the number density also retains similar forms as in the Langevin equation (e.g., it satisfies the fluctuation theorem)~\cite{JCP_Leonard_Lander_Seifert_Speck_2013,PRX_Nardini_2017}.
This similarity invites us to develop analogous arguments by defining heat in the Dean--Kawasaki equation.
Indeed, this approach is feasible.
However, the heat naturally defined in the Langevin and Dean--Kawasaki equations is not generally identical, and the difference is formally reduced to
\begin{eqnarray} 
\Delta'Q^*
&=&
\Delta'Q
-
T
\Delta S^{(\rho)},
\label{heat_heat_relation}
\end{eqnarray}
where $\Delta S^{(\rho)}$ denotes a change in the entropy of the number density (Eq.~(\ref{S_density})) and $\Delta'Q$ or $\Delta'Q^*$ is heat defined with particles' trajectories (Eq.~(\ref{heat_particle})) or the number-density field (Eq.~(\ref{heat_def_n})), respectively.
Notably, the above projection does not alter the spatiotemporal resolutions but rather creates the difference term $T\Delta S^{(\rho)}$ in the heat.

The Dean--Kawasaki \red{equation} offers a simple fascinating form based on ``instantaneous" number density expressed with a sum of the Dirac's delta functions.
However, the instantaneous number density qualitatively differs from the ``ordinary" number density averaged over the small, but finite domain. 
The heat differences quantified by Eq.~(\ref{heat_heat_relation}) and also the discrepancy between the ``instantaneous" and the ``ordinary" number density warrant close inspection, but the investigations are incomplete.

%% \cite{PRE_Crooks_1999,RPP_Seifert_2012,PRE_Gelin_Thoss_2009,PRL_Seifert_2016,PRE_Talkner_Hanggi_2016,PRX_Jarzynski_2017}

This article pays attention to a relation in heat \red{between the particle level and the density field.}
%% the time evolution at the particle level
%% of the particles' trajectories and the \red{density} field.
We begin with investigating the spatiotemporally ``discrete"  pictures in Sec.~\ref{sec_Discrete},
whereas \red{the Langevin or the Dean-Kawasaki equation} is considered as the spatiotemporally ``continuous" counterparts.
The discrete models are rather viewed as having the ``ordinary" number density and give the clear qualitative physical interpretation in Eq.~(\ref{heat_heat_relation}).
\red{The heat differences turn out to be quantified by the entropic term expressed with the number density, i.e., ``ordinary" number density, which may undergo the meaningful temporal evolution in the heat differences.
Section~\ref{sec_numerical} demonstrates numerical results of the discrete models.
Sections~\ref{Langevin},\,\ref{stochatic_field} review the Langevin and the Dean--Kawasaki \red{equations,} and then consider the analogous matters.
Section~\ref{heat_field} investigates differences in heat naturally defined by the Langevin and the Dean--Kawasaki equations, which are eventually found to take the same form of the entropic term, but the arguments of the entropy is expressed with the ``instantaneous" number density. 
Consequently, the heat differences are expected to undergo little temporal evolution due to the sparse distributions of the point particles.}
In Section~\ref{sec_FT_particle}, we remark on some points in light of the fluctuation theorem \red{and second law of thermodynamics.}
In addition,  based on the above results, the following Secs.~\ref{sec_ensemble},\,\ref{ManyPolymers} discuss relevant issues about the ensemble average linked to Fokker-Planck \red{equation} etc., and many-polymer systems.
\red{Section~\ref{Discussion_sec} also discusses its applications and perspectives.
Section~\ref{Conclusion} concludes this study.}
In addition, technical calculi are provided in the Appendix.

\begin{figure}[t]
\begin{center}
\includegraphics[scale=0.50]{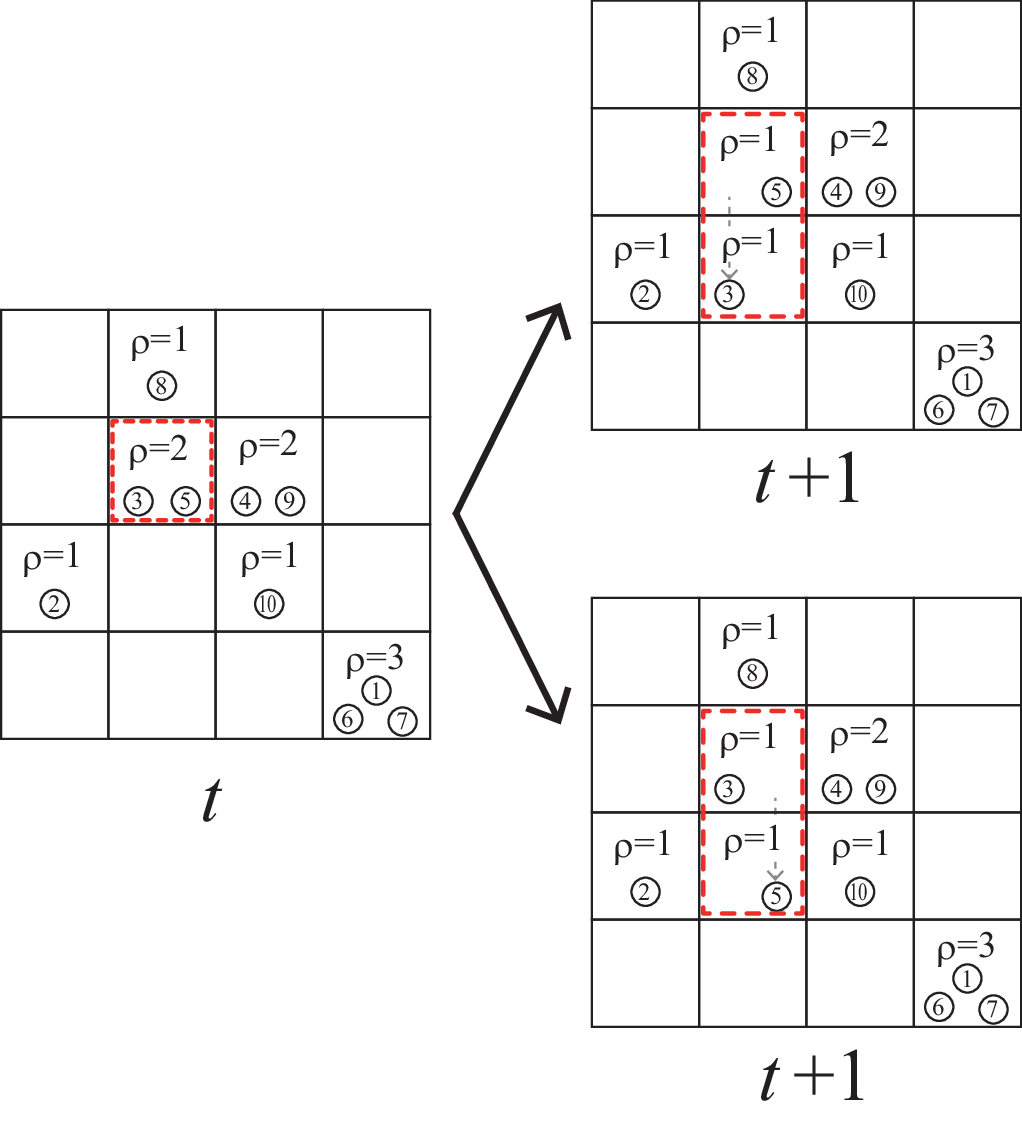}
      \caption{
	  (Color online)
	  Schematic representation of the two-dimensional discrete system.
	  The lattices, whose number density varies from $t$ to $t+1$, is indicated with red dashed lines.
	  }
\label{fig_schematic_map}
\end{center}
\end{figure}

\section{Spatiotemporally Discrete model}
\label{sec_Discrete}

%% \red{Section~\ref{sec_Discrete} shows a rigorous derivation for the discrete model, and then section~\ref{sec_numerical} provides numerical demonstrations.}

Imagine a many-particle system consisting of $M$ particles labeled with $i\in [1,M]$ in the space homogeneously divided into the square lattices (or cells) assigned with the discrete indices ``${\bm x}$".
Let the unit length be the edge size of the lattice.
Hence, for instance, on two-dimension (2D), the lattices are identified with ${\bm x}\equiv (x,y)$ for $x,\,y \in \mathrm{integers}$ (see Fig.~\ref{fig_schematic_map}).
The time is also discretized, and ``$t$" in Sec.~\ref{sec_Discrete} represents discrete time like $\cdots, t-2, t-1, t, t+1, t+2, \cdots$ by dropping the unit time for a compact notation.\footnote{
To clearly see the discreteness-continuum correspondence, the symbols ``$t,{\bm x}$" in Secs.~\ref{sec_Discrete} and \ref{sec_numerical} are employed to indicate a discrete time and spatial point, whereas many cases use ``$t,{\bm x}$" as continuous variables to be assigned to continuous time and spatial point as in Secs.~\ref{many_particles}-\ref{ManyPolymers}.
}

The number density is defined as
\begin{eqnarray}
\rho_{t,{\bm x}}=\sum_{i=1}^M \delta_{{\bm x},{\bm x}_{t,i}},
\label{rho_def_Discrete}
\end{eqnarray}
where $\delta_{{\bm x},{\bm x}'}$ denotes the Kronecker delta with $\delta_{{\bm x},{\bm x}'}=1$ for ${\bm x}={\bm x}'$, and $\delta_{{\bm x},{\bm x}'}=0$ otherwise.
Note that $\rho_{t,{\bm x}}$ may be dealt with as the integer to indicate the number of the particles being in the lattice at the spatiotemporal point $(t,{\bm x})$ because a single lattice has the unit area in 2D (or volume in 3D), but this is called the number density to more easily see the direct correspondence to \red{that (Eq.~(\ref{NumDen_Dirac_delta}))} defined in Dean-Kawasaki equation later on.
Figure~\ref{fig_schematic_map} represents the schematic \red{density field $\{ \rho_{t,{\bm x}} \}$ made by number density}, where some of the lattices are taken up by multiple particles.
The time evolution from $t$ to $t+1$ is also illustrated.
For example, either of the particles (index 3 or 5) encircled by the red dash line can move so as to maintain \red{the same density field.}
The \red{particle level} distinguishes two states at $t+1$, whereas the \red{density field} observations only with $\{ \rho_{t+1,{\bm x}} \}$ identify them as the same state.

\subsection{\red{Particle level}}

The energy balance is introduced along the discrete form compatible with the fluctuation theorem~\cite{PRE_Crooks_2000}.
The $i$-th particle's position or the external parameter at the time $t$ is denoted by ${\bm x}_{t,i}$ or $\boldsymbol{\lambda}_t$, respectively.
The total energy is constructed by summing up the particles' energies:
\begin{eqnarray}
U_T(\{{\bm x}_{t,i}\},\boldsymbol{\lambda}_{t})
=
\sum_{i=1}^M
u_T({\bm x}_{t,i},\boldsymbol{\lambda}_{t}),
\end{eqnarray}
where the energy of $i$-th particle consists of the two-body and one-body potentials:
\begin{eqnarray}
u_T({\bm x}_{t,i},\boldsymbol{\lambda}_{t})
=
\frac{1}{2}
\sum_{j=1}^Mu({\bm x}_{t,i}-{\bm x}_{t,j},\lambda_{t,2})+u_e({\bm x}_{t,i},\lambda_{t,1}).
\nonumber \\
\end{eqnarray}
The first law of thermodynamics is written as
\begin{eqnarray}
U_T(\{ {\bm x}_{t+1,i} \},\boldsymbol{\lambda}_{t+1})
-
U_T(\{ {\bm x}_{t,i} \},\boldsymbol{\lambda}_{t})
&=&
d' Q_t
+
d' W_t.
\nonumber \\
\end{eqnarray}
Note that the heat or work is defined, respectively, as 
\begin{eqnarray}
d' Q_{t}
&\equiv&
U_T(\{ {\bm x}_{t+1,i} \},\boldsymbol{\lambda}_{t})
-
U_T(\{ {\bm x}_{t,i} \},\boldsymbol{\lambda}_{t}),
\label{dicrete_Q_def}
\\
d' W_{t}
&\equiv&
U_T(\{ {\bm x}_{t+1,i} \},\boldsymbol{\lambda}_{t+1})
-
U_T(\{ {\bm x}_{t+1,i} \},\boldsymbol{\lambda}_{t})
\label{discrete_W_def}
\end{eqnarray}
with $d'(\cdot)$ employed for a difference made by a single temporal step, i.e., from $t$ to $t+1$.
As in the framework of standard stochastic thermodynamics, $\boldsymbol{\lambda}_t\equiv (\lambda_{t,1},\lambda_{t,2})$ denotes time-dependent parameters to introduce the work $W_{t}$.
In addition, the heat direction from the thermal bath to the system, or the work done on the system is assigned to be positive, respectively.

Let us here discuss relevance to the Langevin equation~(\ref{Lang_colloid}) specified in detail later.
The total heat $d'Q_t$ is divided into the local heat:
\begin{eqnarray}
d' q_{t,i}
&\equiv&
u_T({\bm x}_{t+1,i},\boldsymbol{\lambda}_{t})
-
u_T({\bm x}_{t,i},\boldsymbol{\lambda}_{t}),
\label{dicrete_local_q_def}
\end{eqnarray}
such that $d'Q_t = \sum_{i=1}^M d' q_{t,i}$, and
we find the similarity to the force balance on the Langevin equation through
\begin{eqnarray}
\frac{d' q_{t,i}}{dx}
=
\frac{
u_T({\bm x}_{t+1,i},\boldsymbol{\lambda}_{t})
-
u_T({\bm x}_{t,i},\boldsymbol{\lambda}_{t})
}{dx},
\label{discrete_force_balance}
\end{eqnarray}
where $dx$ denotes the distance traveled by $i$-th particle from $t$ to $t+1$. 
As the correspondence relation, Eq.~(\ref{discrete_force_balance}) (or Eq.~(\ref{dicrete_local_q_def})) is regarded as the fundamental equation for the system while the left hand side of Eq.~(\ref{discrete_force_balance}) includes the frictional force and the noise in the Langevin equation.
This section maintains this view and then considers $d' q_{t,i}/dx$ as the stochastic driving term to update the system in the discrete model.
$d' q_{t,i}$ is stochastically generated to meet with the detailed fluctuation theorem:
\begin{eqnarray}
\frac{{\cal P}[\{ {\bm x}_{t+1,i} \}|\{ {\bm x}_{t,i} \}]}{{\cal P}[\{ {\bm x}_{t+1,i}^\dagger \}|\{ {\bm x}_{t,i}^\dagger \}]}
&=&
\exp{\left(-\frac{
d' Q_{t}
}{k_BT}\right)}
=
\exp{\left(-\frac{
\sum_{i=1}^M d' q_{t,i}
}{k_BT}\right)},
\nonumber \\
\label{DB_Q_discrete}
\end{eqnarray}
where the superscript $\dagger$ represents the temporal reverse like ${\bm x}_{t,i}^\dagger\equiv {\bm x}_{t+1,i}$, ${\bm x}_{t+1,i}^\dagger \equiv{\bm x}_{t,i}$, $\boldsymbol{\lambda}_{t+1}^\dagger\equiv \boldsymbol{\lambda}_{t}$, and the parameters $\boldsymbol{\lambda}_{t}$, $\boldsymbol{\lambda}_{t+1}^\dagger$ are not shown in the arguments of the probabilities for notational compactness.

\subsection{\red{Density field}}

\red{Let us move on to the density-field description.}
The thermodynamic-potential-like quantity is introduced as
\begin{eqnarray}
\Phi(\{ \rho_{t,{\bm x}} \},\boldsymbol{\lambda}_{t})
&\equiv&
U_T(\{ \rho_{t,{\bm x}} \},\boldsymbol{\lambda}_{t})
+k_BT\sum_{{\bm x}} \ln{[\rho_{t,{\bm x}}!]},
\nonumber \\
\label{discrete_Phi_def}
\end{eqnarray}
where
\begin{eqnarray}
U_T(\{ \rho_{t,{\bm x}} \},\boldsymbol{\lambda}_{t})
&\equiv&
\sum_{{\bm x},{\bm x}'}\frac{1}{2}\rho_{t,{\bm x}} \rho_{t,{\bm x}'} u({\bm x}-{\bm x}',\lambda_{t,2}) 
\nonumber \\
&&
+ \sum_{{\bm x}} \rho_{t,{\bm x}} u_e({\bm x},\lambda_{t,1}).
\label{discrete_U_rho_def}
\end{eqnarray}
The system evolves according to the modified detailed fluctuation theorem:
\begin{eqnarray}
\frac{{\cal P}[\{ \rho_{t+1,{\bm x}} \}|\{ \rho_{t,{\bm x}} \}]}{{\cal P}[\{ \rho_{t+1,{\bm x}}^\dagger \}|\{ \rho_{t,{\bm x}}^\dagger \}]}
&=&
\exp{\left(-
\frac{
\Phi(\{ \rho_{t+1,{\bm x}} \},\boldsymbol{\lambda}_{t})
-
\Phi(\{ \rho_{t,{\bm x}} \},\boldsymbol{\lambda}_{t})
}{k_BT}\right)}
\nonumber \\
&=&
\exp{\left(-\frac{
d' Q^*_{t}
}{k_BT}\right)},
\label{DB_Q_rho_discrete}
\end{eqnarray}
where $\rho_{t+1,{\bm x}}^\dagger \equiv \rho_{t,{\bm x}}$ and $\rho_{t,{\bm x}}^\dagger \equiv \rho_{t+1,{\bm x}}$.
The derivation of Eq.~(\ref{DB_Q_rho_discrete}) will be discussed later on.
\red{
Our formalism assumes the first law of thermodynamics in the density field as
\begin{eqnarray}
\Phi(\{ \rho_{t+1,{\bm x}} \},\boldsymbol{\lambda}_{t+1})
-
\Phi(\{ \rho_{t,{\bm x}} \},\boldsymbol{\lambda}_{t})
&=&
d' Q_{t}^*
+
d' W_{t}^*,
\nonumber \\
\end{eqnarray}
where} the heat or the work is also defined, respectively, as 
\begin{eqnarray}
d' Q_{t}^*
&=&
\Phi(\{ \rho_{t+1,{\bm x}} \},\boldsymbol{\lambda}_{t})
-
\Phi(\{ \rho_{t,{\bm x}} \},\boldsymbol{\lambda}_{t}),
\label{dicrete_model_rho_q_def}
\\
d' W^*_{t}
&=&
\Phi(\{ \rho_{t+1,{\bm x}} \},\boldsymbol{\lambda}_{t+1})
-
\Phi(\{ \rho_{t+1,{\bm x}} \},\boldsymbol{\lambda}_{t}).
\end{eqnarray}
Note that the total work is not altered from Eq.~(\ref{discrete_W_def}), i.e., $d' W^*_t=d' W_t$ by this framework because the entropic term $k_BT\sum_{{\bm x}'} \ln{[\rho_{t,{\bm x}'}}!]$ in $\Phi(\{ \rho_{t,{\bm x}} \},\boldsymbol{\lambda}_{t})$ does not include $\boldsymbol{\lambda}_t$.
In addition, in the same way as Eq.~(\ref{discrete_force_balance}), we can view the local heat $d' q_{t,{\bm x}}^*$ as the stochastic driving term:
\begin{eqnarray}
d' q_{t,{\bm x}}^*
&\equiv&
\phi_{{\bm x}}(\rho_{t+1,{\bm x}},\boldsymbol{\lambda}_{t})
-
\phi_{{\bm x}}(\rho_{t,{\bm x}},\boldsymbol{\lambda}_{t})
\label{dicrete_local_q_rho_def}
\end{eqnarray}
with the local potential:
\begin{eqnarray}
\phi_{{\bm x}}(\{ \rho_{t,{\bm x}'} \},\boldsymbol{\lambda}_{t})
&\equiv&
\sum_{{\bm x}'}\frac{1}{2}\rho_{t,{\bm x}}\rho_{t,{\bm x}'} u({\bm x},{\bm x}',\lambda_{t,2}) 
\nonumber \\
&&
+ \rho_{t,{\bm x}} u_e({\bm x},\lambda_{t,1})
+k_BT \ln{[\rho_{t,{\bm x}}!]}
\nonumber \\
\label{dicrete_local_phi_rho_def}
\end{eqnarray}
such that the sums in both Eqs.~(\ref{dicrete_local_q_rho_def}) and (\ref{dicrete_local_phi_rho_def}) are consistent with the total quantities: $d' Q_{t}^*=\sum_{{\bm x}}d' q_{t,{\bm x}}^*$ and $\Phi=\sum_{{\bm x}}\phi_{{\bm x}}$.
Equation~(\ref{dicrete_local_q_rho_def}) is found to correspond to Eq.~(\ref{heat_def_n_local}) for the Dean-Kawasaki equation~(\ref{BasicEq_n}) introduced below.

\begin{figure}[t]
\begin{center}
\includegraphics[scale=0.20]{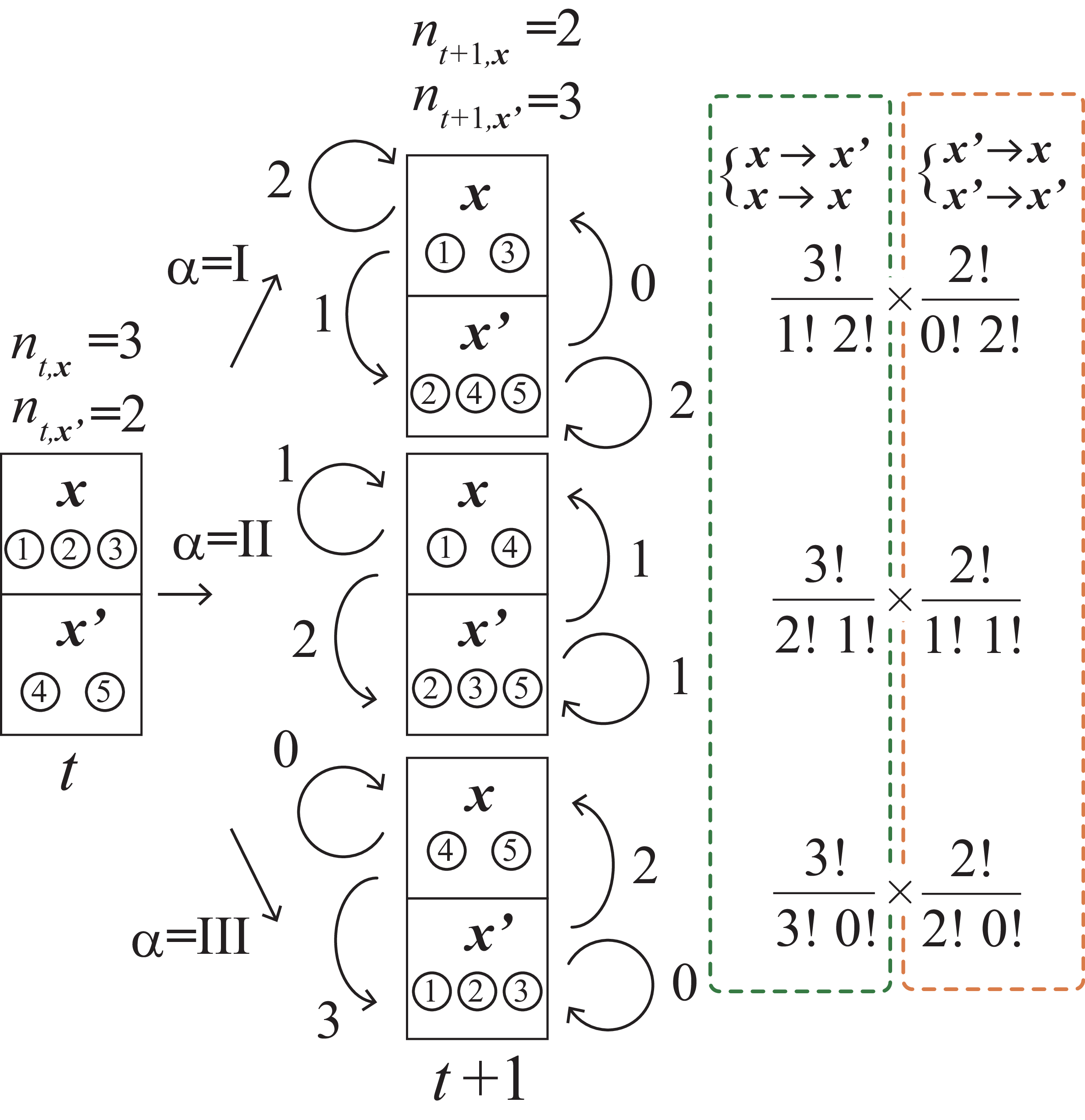}
      \caption{
	  (Color online)
	  An example to illustrate the combination $n_{t,{\bm x}}!/\prod_{{\bm x}'} n_{t,{\bm x}\rightarrow{\bm x}'}^{(\alpha)}$ with two lattices.
	  \red{The density field} starts with $(n_{t,{\bm x}},n_{t,{\bm x}'})=(3,2)$ and ends with $(n_{t+1,{\bm x}},n_{t+1,{\bm x}'})=(2,3)$ through a flow map $\{ n_{t,{\bm x}\rightarrow{\bm x}'}^{(\alpha)} \}$ identified with $\alpha=\mathrm{I}, \mathrm{II}, \mathrm{III}$.
	  For instance, $n_{t,{\bm x}\rightarrow{\bm x}'}^{(\mathrm{I})}=1$, $n_{t,{\bm x}\rightarrow{\bm x}}^{(\mathrm{I})}=2$ and $n_{t,{\bm x}'\rightarrow{\bm x}}^{(\mathrm{I})}=0$, $,n_{t,{\bm x}'\rightarrow{\bm x}'}^{(\mathrm{I})}=2$ \red{for $\alpha=\mathrm{I}$.}
	  }
\label{fig_schematic_map_2cells}
\end{center}
\end{figure}

\subsection{\red{Heat differences}}

It is time to discuss the heat difference.
Comparison of $\exp{[-d'Q_t/(k_BT)]}$ of Eq.~(\ref{DB_Q_discrete}) with $\exp{[-d'Q_t^*/(k_BT)]}$ of Eq.~(\ref{DB_Q_rho_discrete}) under Eqs.~(\ref{dicrete_Q_def}) and (\ref{discrete_Phi_def}) discovers the main consequences for the discrete model:
\begin{eqnarray}
\exp{
\left( -\frac{d'Q_t^*}{k_BT} \right)
}
&=&
\exp{
\left( -\frac{d'Q_t}{k_BT} \right)
}
\exp{
\left( \frac{dS^{(\rho)}_{t}}{k_B} \right)
},
\label{discrete_model_Q_Q_S}
\end{eqnarray}
where $S_t^{(\rho)}\equiv -k_B \sum_{\bm x} \ln{\rho_{t,{\bm x}}!}$.
This also implies
\begin{eqnarray}
\frac{{\cal P}[\{ \rho_{t+1,{\bm x}} \}|\{ \rho_{t,{\bm x}} \}]}{{\cal P}[\{ \rho_{t+1,{\bm x}}^\dagger \}| \{ \rho_{t,{\bm x}}^\dagger \}]}
&=&
\frac{{\cal P}[\{ {\bm x}_{t+1,i} \} | \{ {\bm x}_{t,i} \}]}{{\cal P}[\{ {\bm x}_{t+1,i}^\dagger \}|\{ {\bm x}_{t,i}^\dagger \}]}
\frac{
\prod_{{\bm x}}n_{t,{\bm x}}!
}
{
\prod_{{\bm x}'}n_{t,{\bm x}'}^\dagger !
}.
\nonumber \\
\label{Discrete_FT_x_rho}
\end{eqnarray}
However, Eq.~(\ref{DB_Q_rho_discrete}) to bridge the probability ratio with the heat might not have seemed trivial so far.
\red{To derive Eq.~(\ref{DB_Q_rho_discrete}),~(\ref{discrete_model_Q_Q_S}), and (\ref{Discrete_FT_x_rho})} under the assumption of Eq.~(\ref{DB_Q_discrete}), we observe a single step evolution of the system, and compare it with the reverse evolution.
Although many candidates of transition rules are conceivable, transition probabilities for the \red{particle level and the density field} are associated with
\begin{eqnarray}
{\cal P}[\{ \rho_{t+1,{\bm x}} \},\alpha|\{ \rho_{t,{\bm x}} \}]
=
{\cal P}[\{ {\bm x}_{t+1,i} \} | \{ {\bm x}_{t,i} \}]
\prod_{{\bm x}}\frac{n_{t,{\bm x}}!}{\prod_{{\bm x}'} n_{t,{\bm x}\rightarrow{\bm x}'}^{(\alpha)} !},
\nonumber \\
\label{Prho_Px_discrete}
\end{eqnarray}
where $n_{t,{\bm x}}=\rho_{t,{\bm x}}$ represents the number of the particles at the time $t$, and $n^{(\alpha)}_{t,{\bm x}\rightarrow {\bm x}'}$ denotes the number of the particles moving from ${\bm x}$ to ${\bm x}'$ identified by a flow map index $\alpha$ (see Fig.~\ref{fig_schematic_map_2cells}).
The factor $n_{t,{\bm x}}!/\prod_{{\bm x}'} n_{t,{\bm x}\rightarrow{\bm x}'}^{(\alpha)}$ in the right hand side of Eq.~(\ref{Prho_Px_discrete}) means the combination.\footnote{
If $M$ elements are distributed to $K$ places with distinguishing the elements, the number of ways to distribute the elements or the combinations is conventionally expressed as
\begin{eqnarray}
\left(
\begin{array}{ccccc}
 & & M & &
\\
n_1 & n_2 & n_3 & \cdots n_K
\end{array}
 \right)
\equiv 
\frac{M!}{\prod_{k=1}^Kn_k! }
\label{gen_comb_discrete}
\end{eqnarray}
with $\sum_{k=1}^K n_k=M$.
}
Note that the set of ${\bm x} \rightarrow {\bm x}'$ includes the quiescent process ${\bm x} \rightarrow {\bm x}$, where the particle keeps staying at the same position from $t$ to $t+1$.
Specifically, let us here see an example in Fig.~\ref{fig_schematic_map_2cells} for understanding the combination $n_{t,{\bm x}}!/\prod_{{\bm x}'} n_{t,{\bm x}\rightarrow{\bm x}'}^{(\alpha)}$ with two-lattice model.
The initial or the final \red{density field} is $(n_{t,{\bm x}},n_{t,{\bm x}'})=(3,2)$ or $(n_{t+1,{\bm x}},n_{t+1,{\bm x}'})=(2,3)$, respectively.
There are three possible flow maps labeled with $\alpha=\mathrm{I}, \mathrm{II}, \mathrm{III}$ to realize the final \red{density field} $(n_{t+1,{\bm x}},n_{t+1,{\bm x}'})=(2,3)$.
Look at $\alpha=\mathrm{I}$. 
Three particles are initially at ${\bm x}$.
One particle moves from ${\bm x}$ to ${\bm x}'$ and two particles stay at ${\bm x}$.
Thus, $\left(\begin{array}{c}
3
\\
1
\end{array}
\right)
=3!/(1!\,2!)$ ways to make a number-density change if the labeled particles are distinguished.
In the same way, none of the two particles at ${\bm x}'$ moves to ${\bm x}$, and two of them stay at ${\bm x}'$.
Consequently, we find the combination $\left(\begin{array}{c}
2
\\
0
\end{array}
\right)
=2!/(0!\,2!)$.
It is noticeable that the transition probability ${\cal P}[\{ {\bm x}_{t+1,i} \} | \{ {\bm x}_{t,i} \}]$ \red{at the particle level} is common within $\alpha=\mathrm{I}$, but is generally different from those for $\alpha=\mathrm{II}$ and $\mathrm{III}$.
Keeping this point in mind, we find a general consequence of Eq.~(\ref{Prho_Px_discrete}).

We similarly find that for the reverse processes, and a ratio of the probability of the forward processes to that of the reversible process is written as
\begin{eqnarray}
&&
\frac{{\cal P}[\{ \rho_{t+1,{\bm x}} \},\alpha|\{ \rho_{t,{\bm x}} \}]}{{\cal P}[\{ \rho_{t+1,{\bm x}}^\dagger \},\alpha^\dagger | \{ \rho_{t,{\bm x}}^\dagger \}]}
\nonumber \\
&=&
\frac{{\cal P}[\{ {\bm x}_{t+1,i} \}| \{ {\bm x}_{t,i} \}]}{{\cal P}[\{ {\bm x}_{t+1,i}^\dagger \}|\{ {\bm x}_{t,i}^\dagger \}]}
\frac{
\prod_{{\bm x}}\frac{n_{t,{\bm x}}!}{\prod_{{\bm x}'} n_{t,{\bm x}\rightarrow{\bm x}'}^{(\alpha)}!}
}
{
\prod_{{\bm x}'}\frac{n_{t,{\bm x}'}^\dagger!}{\prod_{{\bm x}} n_{t,{\bm x}'\rightarrow{\bm x}}^{(\alpha \dagger)}!}
}
\nonumber \\
&=&
\frac{{\cal P}[\{ {\bm x}_{t+1,i} \} | \{ {\bm x}_{t,i} \}]}{{\cal P}[\{ {\bm x}_{t+1,i}^\dagger \}|\{ {\bm x}_{t,i}^\dagger \}]}
\frac{
\prod_{{\bm x}}n_{t,{\bm x}}!
}
{
\prod_{{\bm x}'}n_{t,{\bm x}'}^\dagger !
},
\label{Discrete_FT_x_rho_current}
\end{eqnarray}
where $\alpha^\dagger$ denotes the reverse flow map to $\alpha$. 
The last line is followed by the one-to-one correspondence between the forward and the backward steps that satisfies $\prod_{{\bm x}}\prod_{{\bm x}'} n_{t,{\bm x}\rightarrow{\bm x}'}^{(\alpha)}=\prod_{{\bm x}}\prod_{{\bm x}'} n_{t,{\bm x}'\rightarrow {\bm x}}^{(\alpha \dagger)}$.
Recall here the fluctuation theorem (Eq.~(\ref{DB_Q_discrete})) for the particle level with Eq.~(\ref{dicrete_Q_def}), and also the definitions of $\Phi(\{ \rho_{t,{\bm x}} \},\boldsymbol{\lambda}_{t})$, $U_T(\{ \rho_{t,{\bm x}} \},\boldsymbol{\lambda}_{t})$ (Eqs.~(\ref{discrete_Phi_def}) and (\ref{discrete_U_rho_def}))).
By noting $U_T(\{ \rho_{t,{\bm x}} \},\boldsymbol{\lambda}_{t})=U_T(\{{\bm x}_{t,i}\},\boldsymbol{\lambda}_{t})$,
Eq.~(\ref{Discrete_FT_x_rho_current}) is rephrased into
\begin{eqnarray}
&&
{\cal P}[\{ \rho_{t+1,{\bm x}} \}, \alpha |\{ \rho_{t,{\bm x}} \}]
\nonumber \\
&=&
{\cal P}[\{ \rho_{t+1,{\bm x}}^{\dagger} \}, \alpha^\dagger | \{ \rho_{t,{\bm x}}^{\dagger} \}]
\nonumber \\
&&
\times 
\exp{\left(-
\frac{
\Phi(\{ \rho_{t+1,{\bm x}} \},\boldsymbol{\lambda}_{t})
-
\Phi(\{ \rho_{t,{\bm x}} \},\boldsymbol{\lambda}_{t})
}{k_BT}\right)}.
\label{Discrete_FT_x_alpha}
\end{eqnarray}
Even if the transition probabilities ${\cal P}[\{ \rho_{t+1,{\bm x}} \}, \alpha |\{ \rho_{t,{\bm x}} \}]$ for the flow maps, e.g., $\alpha=\mathrm{I}, \mathrm{II}, \mathrm{III}$ in Fig.~\ref{fig_schematic_map_2cells}, are different, the common exponential factor in the right hand side is shared, and thus, a summation of Eq.~(\ref{Discrete_FT_x_alpha}) over $\alpha$ yields
\begin{eqnarray}
&&
{\cal P}[\{ \rho_{t+1,{\bm x}} \} |\{ \rho_{t,{\bm x}} \}]
\nonumber \\
&=&
{\cal P}[\{ \rho_{t+1,{\bm x}}^{\dagger} \} | \{ \rho_{t,{\bm x}}^{\dagger} \}]
\nonumber \\
&&
\times
\exp{\left(-
\frac{
\Phi(\{ \rho_{t+1,{\bm x}} \},\boldsymbol{\lambda}_{t})
-
\Phi(\{ \rho_{t,{\bm x}} \},\boldsymbol{\lambda}_{t})
}{k_BT}\right)}.
\label{P_ration_Phi}
\end{eqnarray}
Then, applying the definition of the heat with Eq.~(\ref{dicrete_model_rho_q_def}) into Eq.~(\ref{P_ration_Phi}) immediately  leads to Eq.~(\ref{DB_Q_rho_discrete}).
Again, recalling the fluctuation theorem (Eq.~(\ref{DB_Q_discrete})) for the \red{particle level,} the definitions of $\Phi(\{ \rho_{t,{\bm x}} \},\boldsymbol{\lambda}_{t})$, $U_T(\{ \rho_{t,{\bm x}} \},\boldsymbol{\lambda}_{t})$ (Eqs.~(\ref{discrete_Phi_def}) and (\ref{discrete_U_rho_def}))), and $U_T(\{ \rho_{t,{\bm x}} \},\boldsymbol{\lambda}_{t})=U_T(\{{\bm x}_{t,i}\},\boldsymbol{\lambda}_{t})$ together with Eq.~(\ref{dicrete_Q_def}), we finally arrive at Eqs.~(\ref{discrete_model_Q_Q_S}) and (\ref{Discrete_FT_x_rho}).\footnote{\red{
This type of projection without the alternations of the resolutions from the particle-level to the density-field descriptions does not modify the entropy production $dS^{(tot)}$ on the second law of thermodynamics, i.e., 
\begin{eqnarray}
dS^{(tot)}=dS-\frac{d'Q}{T}=dS^*-\frac{d'Q^*}{T}
\end{eqnarray}
with the Shannon entropy being $S\equiv -k_B\ln{{\cal P}[\{ {\bm x}_{t,i} \}]}$ at the particle level, or $S^*\equiv -k_B \ln{{\cal P}[\{ \rho_{t,{\bm x}} \}]}$ in the density density field, respectively.
The invariant form of the entropy production is ensured by a fact that the heat difference is compensated by that of the Shannon entropy:
\begin{eqnarray}
dS^*=dS-dS^{(\rho)}
\label{Shannon_entropy_diff_discrete}
\end{eqnarray}
with $S^{(\rho)}\equiv -k_B\ln{\prod_{{\bm x}} \rho_{t,{\bm x}}!}=-k_B\sum_{{\bm x}}\ln{ \rho_{t,{\bm x}}!}$ at the time $t$.
Bear in mind that $\rho_{t,{\bm x}}=n_{t,{\bm x}}$.
The derivation of Eq.~(\ref{Shannon_entropy_diff_discrete}) based on the combination arguments is given in appendix A.
(see also pertinent arguments around Eq.~(\ref{Shannon_entropy_diff}) in Sec.~\ref{sec_FT_particle}).
}}
Furthermore, Eq.~(\ref{discrete_model_Q_Q_S}) or Eq.~(\ref{Discrete_FT_x_rho}) for multiple steps $t\rightarrow t+n$ $(n\geq 2)$ can be constructed in the similar ways as in ref.~\cite{PRE_Crooks_2000}.

It is also noted that the time evolutions are dictated not only by Eq.~(\ref{DB_Q_discrete}) (or Eq.~(\ref{DB_Q_rho_discrete})), but also by the specific dynamical dissipation mechanisms. 
Nonetheless, the discussion about the heat differences (Eq.~(\ref{heat_heat_relation})) hold regardless of the dynamical mechanisms.
Thus, the issues about persuading the dynamical mechanisms that meet with the real time evolution should be separated from the argument about the heat differences (Eq.~(\ref{heat_heat_relation})).

\if0
\begin{eqnarray}
dS^{(\rho)}
=
-k_B\ln{\frac{\prod_{{\bm x}} n_{t+1,{\bm x}}!}{\prod_{{\bm x}} n_{t,{\bm x}}!}}
\end{eqnarray}
\fi

\section{Numerical Demonstrations}
\label{sec_numerical}

This section numerically demonstrates results that facilitate understanding Eq.~(\ref{heat_heat_relation}) in the discrete model of Sec.~\ref{sec_Discrete}.

A number of algorithms have been known to numerically update the system~\cite{FrenkelBook,JCP_Fichthorn_1991,Voter_KMC}. 
Some of them are ideal while the others reproduces the more realistic dynamics.
Nonetheless, Eq.~(\ref{discrete_model_Q_Q_S}) holds in broad update rules.
Section~\ref{sec_numerical} demonstrates the time evolution of the relevant thermodynamic quantities around Eq.~(\ref{discrete_model_Q_Q_S}), but we do not put importance on what models undergo the more realistic dynamics.

In addition, in order to observe the same dynamical phenomena between the \red{particle level and the density field,} the stochastic origin or the noises are assumed to be from Eq.~(\ref{DB_Q_discrete}) in both \red{the particle level and the density field.}
This construction is harmless even to Eq.~(\ref{DB_Q_rho_discrete}), 
since Eqs.~(\ref{dicrete_Q_def}) and (\ref{DB_Q_discrete}) with $U_T(\{ \rho_{t,{\bm x}} \},\boldsymbol{\lambda}_{t})=U_T(\{{\bm x}_{t,i}\},\boldsymbol{\lambda}_{t})$ rephrase the right hand side of Eq.~(\ref{Discrete_FT_x_rho}) into expression with $\{ \rho_{t,{\bm x}} \}$ without $\{ {\bm x}_{t,i} \}$, so that we find
\begin{eqnarray}
&&
\frac{{\cal P}[\{ \rho_{t+1,{\bm x}} \}|\{ \rho_{t,{\bm x}} \}]}{{\cal P}[\{ \rho_{t+1,{\bm x}}^\dagger \} | \{ \rho_{t,{\bm x}}^\dagger \}]}
\nonumber \\
&=&
\exp{}\Bigg(
-\frac{
U_T(\{ \rho_{t+1,{\bm x}} \},\boldsymbol{\lambda}_{t})
-
U_T(\{ \rho_{t,{\bm x}} \},\boldsymbol{\lambda}_{t})
}{k_BT}
\Biggr)
\nonumber \\
&&
\times 
\exp{}\Bigg(
\frac{dS^{(\rho)}(\{ \rho_{t,{\bm x}} \})}{k_B}
\Biggr).
\label{discrete_rho_DB}
\end{eqnarray}
Equation~(\ref{discrete_rho_DB}) is a closed form only with $\{ \rho_{t,{\bm x}} \}$, and expresses the detailed balance (Eq.~(\ref{DB_Q_rho_discrete})) with $\{ \rho_{t,{\bm x}} \}$.
Thus, this implies, even if the systems are stochastically updated by Eq.~(\ref{DB_Q_discrete}), Eq.~(\ref{DB_Q_rho_discrete}) is satisfied irrespective of the stochastic dynamical mechanisms.

\begin{figure}[t]
\begin{center}
\includegraphics[scale=0.35]{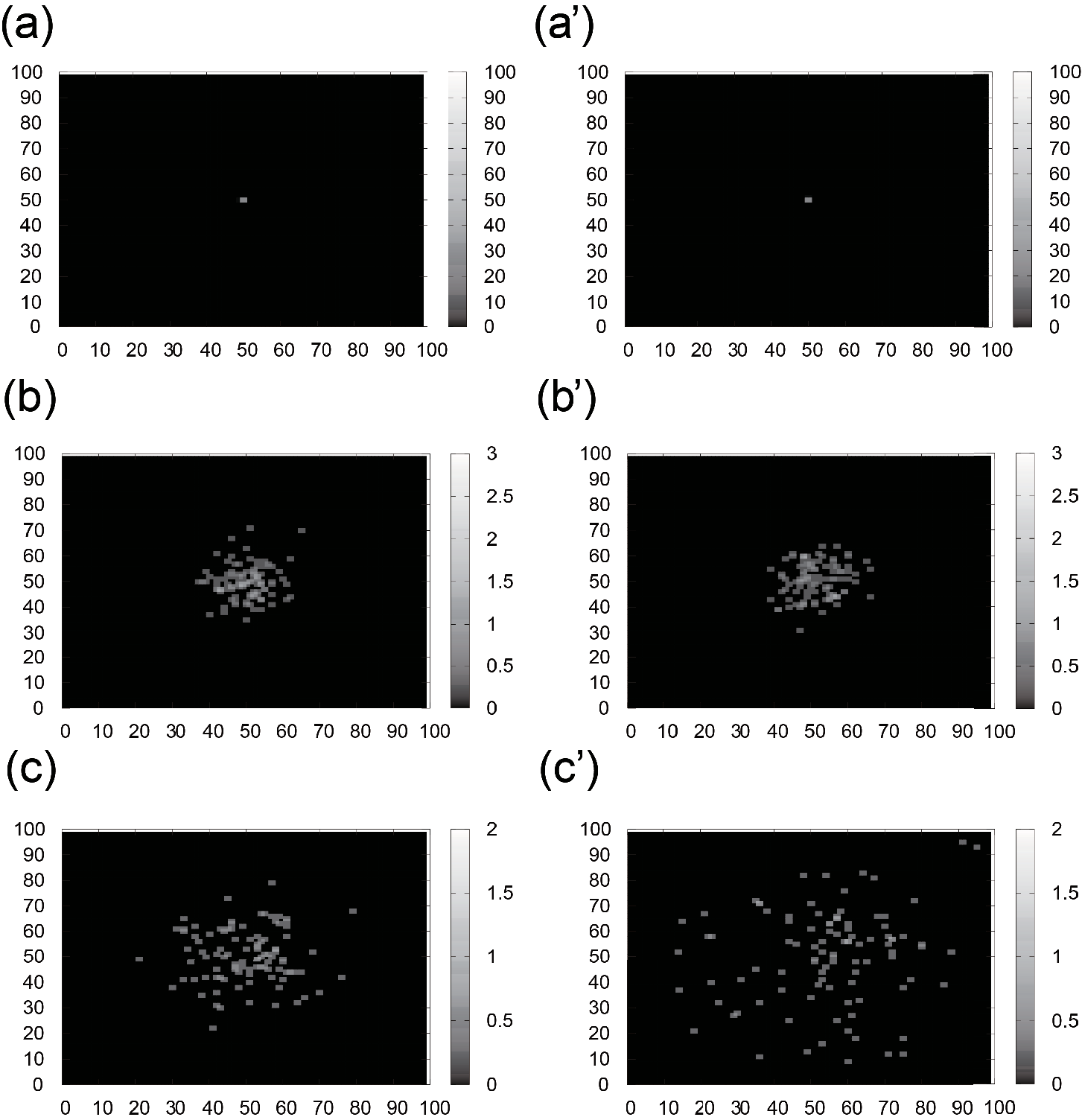}
      \caption{
	  Snapshots of the \red{density field} at $t=0$, $t=1.0\times 10^4$, and $1.0\times 10^5$ in numerical simulations.
	  The left (a)-(c) or the right (a')-(c') column shows the results for the one-body potential with Eq.~(\ref{harmonic_1body_discrete}) for the two-body potential with Eq.~(\ref{harmonic_2body_discrete}), respectively.
	  The time evolves downward.
	  }
\label{fig_density_snapshots}
\end{center}
\end{figure}

\begin{figure}[t]
\begin{center}
\includegraphics[scale=0.32]{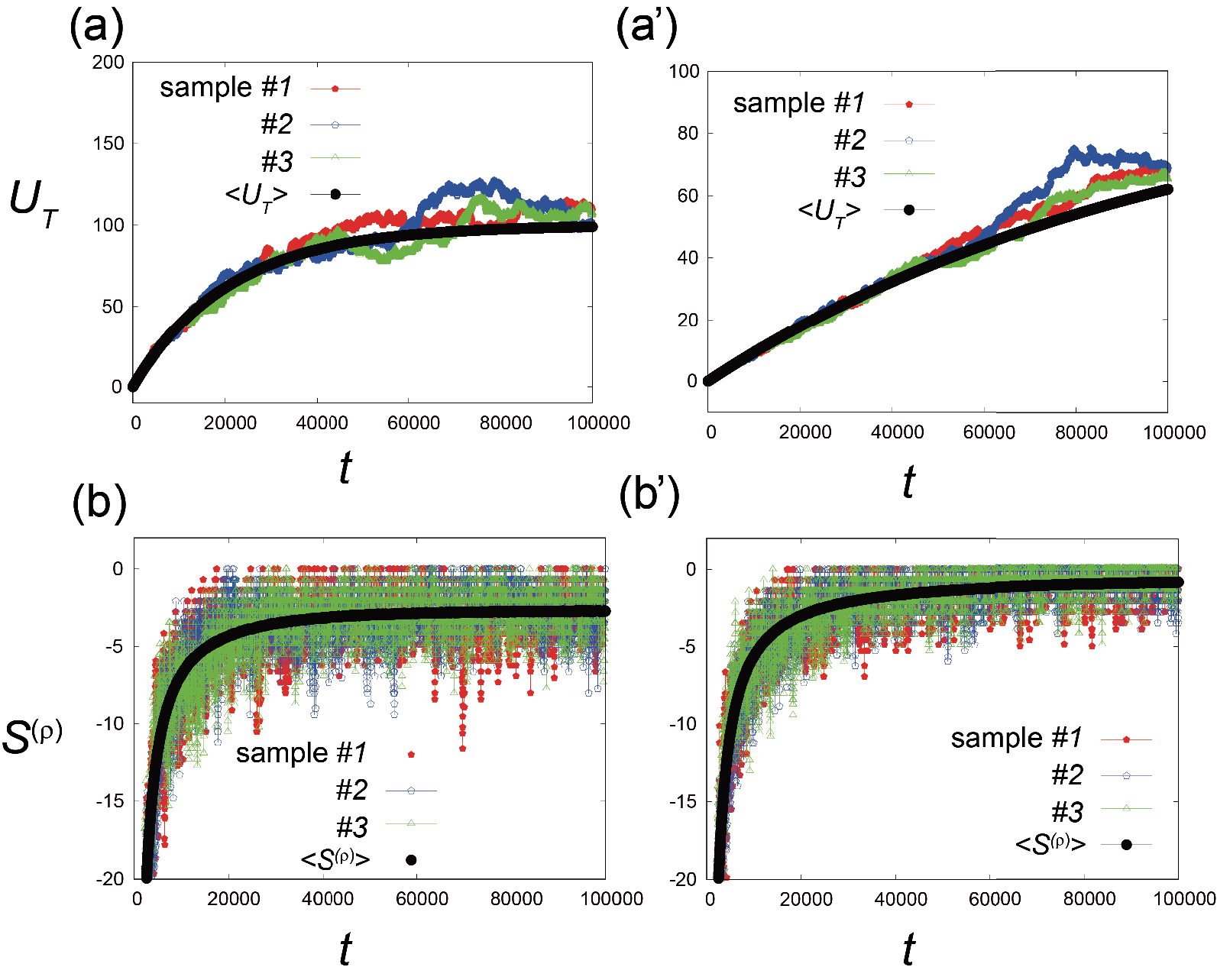}
      \caption{
	  (Color online) Numerical results of \red{the temporal evolutions} of the potential energy $U_T$ and the entropy of the number-density $S^{(\rho)}$ together with those ensemble-averaged quantities $\left< U_T\right>$ and $\langle S^{(\rho)} \rangle$.
	  (a),\,(b) are for the one-body potential with Eq.~(\ref{harmonic_1body_discrete}) , and (a'),\,(b') are for the two-body potential with Eq.~(\ref{harmonic_2body_discrete}).
	  }
\label{fig_E_S}
\end{center}
\end{figure}

\begin{figure}[t]
\begin{center}
\includegraphics[scale=0.32]{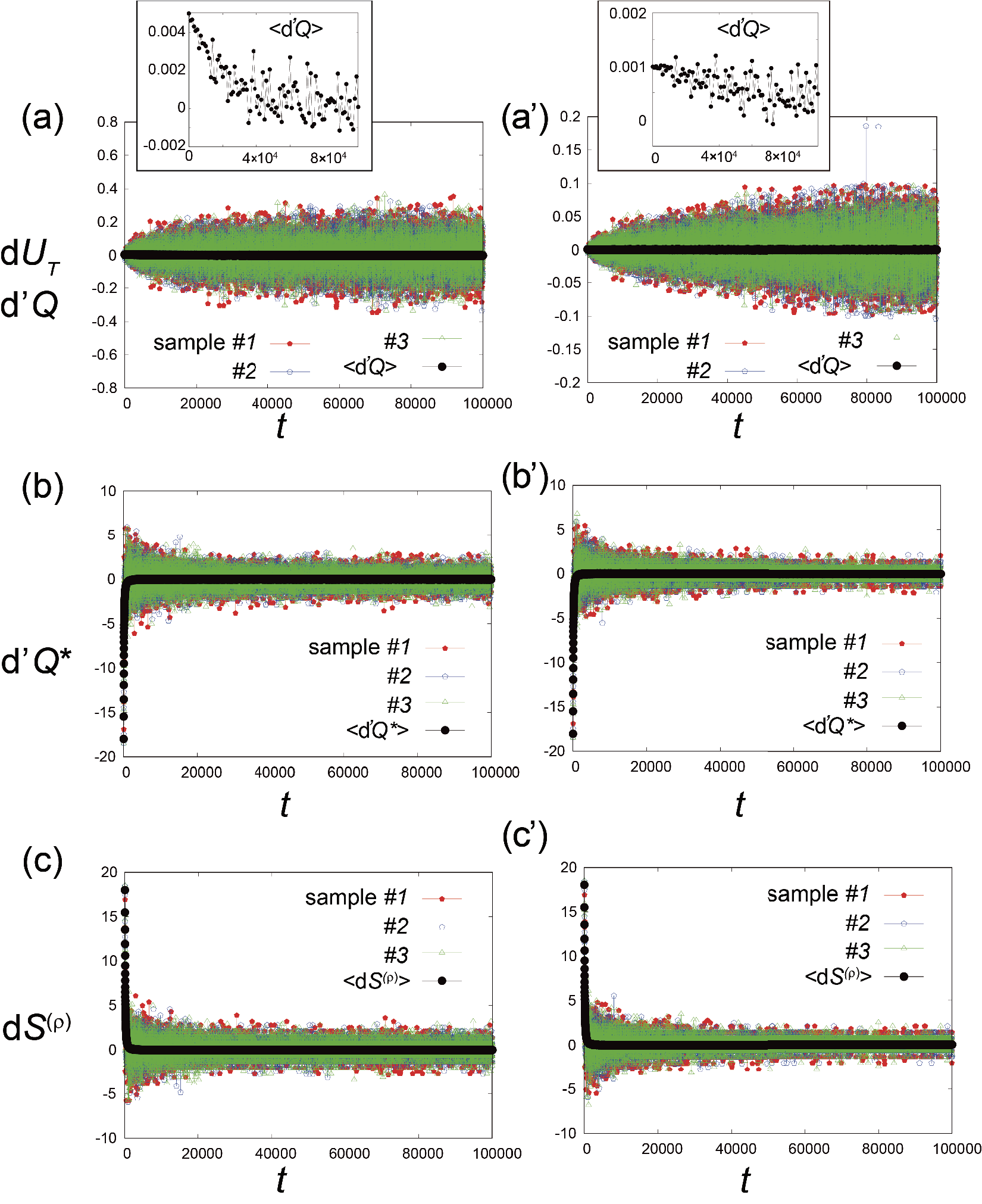}
      \caption{
	  (Color online) Numerical results of temporal evolutions of thermodynamic quantities per a single time step together with those ensemble-averaged quantities.
	  (a)-(c) are for the one-body potential with Eq.~(\ref{harmonic_1body_discrete}) , and (a')-(c') are for the two-body potential with Eq.~(\ref{harmonic_2body_discrete}).
	  (a),(a') changes in energy $dU_T$ and heat $d'Q$ for the \red{particle level,} (b),(b') $d'Q^*$ for the \red{density field,} and (c),(c') changes in \red{the entropy of the number density} $dS^{(\rho)}$.
	  Insets in (a),\,(a') are vertically magnified figures of $\left< d'Q \right>$ with the data points reduced.
	  }
\label{fig_dE_Q_Q_dS_evolution}
\end{center}
\end{figure}

Let us move on to specific numerical results  for the many-particle system shown in Figs.~\ref{fig_density_snapshots}-\ref{fig_dE_Q_Q_dS_evolution}.
The numerical study adopts Metropolis algorithm in Monte Carlo methods, where the acceptance and the rejection are determined with Eq.~(\ref{DB_Q_discrete}) as mentioned above.
This section~\ref{sec_numerical} normalizes the time, the length, and the energy by the single step, the edge length of the single lattice, or the thermal energy, respectively.
In addition, the Boltzmann constant is set as $k_B=1$.
The $M=100$ particles are in the two-dimensional system with $100 \times 100$ square lattices, where ${\bm x}=(l,m)$ ($l,m \in [0,99]$) are assigned to the lattices' addresses.
All the particles are initially at the ${\bm x}_c=(50,50)$, which is around the position of the center of the system.
At each step, a candidate particle to move is picked in the uniform random way, and the nearest neighbor lattices that the particle would be transferred into is also uniform-randomly chosen.
If the particles go out of the system, these processes are rejected. 
For a simple demonstration, work is not considered, i.e., always $d'W=0$, where the energy balance $dU_T=d'Q$ or $d\Phi=d'Q^*$ are always observed.

Two cases are present.
The first one uses the one-body potential:
\begin{eqnarray}
U_T
&=&
\sum_{i=1}^M
\frac{k_1}{2}
({\bm x}_i(t)-{\bm x}_c)^2
=
\sum_{{\bm x}}
\rho(t,{\bm x})
\frac{k_1}{2}
({\bm x}-{\bm x}_c)^2.
\nonumber \\
\label{harmonic_1body_discrete}
\end{eqnarray}
The other one employs two-body potential:
\begin{eqnarray}
U_T
&=&
\sum_{(i,j)}
\frac{k_2}{2}
({\bm x}_i(t)-{\bm x}_j(t))^2
\nonumber \\
&=&
\sum_{({\bm x},{\bm x}')}
\rho(t,{\bm x})
\rho(t,{\bm x}')
\frac{k_2}{2}
({\bm x}-{\bm x}')^2,
\label{harmonic_2body_discrete}
\end{eqnarray}
with the spring constants being $k_1=1.0\times 10^{-2}$ and $k_2=2.0\times 10^{-5}$.
The summation is taken over all the combination of $(i,j)$ or $({\bm x},{\bm x}')$, respectively.

Figure~\ref{fig_density_snapshots} demonstrates the temporal sequence of the snapshots of the \red{density field} in gray-scale, where the dense cells are drawn more lightly.
The left or the right column displays the temporally-sequential \red{density field} from one-body potential with Eq.~(\ref{harmonic_1body_discrete}) or two-body potential with Eq.~(\ref{harmonic_2body_discrete}), respectively.
We can see the transient processes with time from $t=0$ to $t=1.0\times 10^5$.
Note that the time $t$ (the number of the steps) includes the rejection in the Metropolis algorithm.

Figure~\ref{fig_E_S} shows the the stochastic time evolutions of $U_T$ and $S^{(\rho)}$ drawn by the three samples.
In addition, the ensemble-averaged quantities $\left< U_T \right>$, $\left< S^{(\rho)} \right>$ over $10^4$ realizations are also plotted.
In either potential (Eq.~(\ref{harmonic_1body_discrete}) or Eq.~(\ref{harmonic_2body_discrete})), the particles are initially localized at the center of the harmonic potential(s), so that the potential energy sets from $U_T=0$.
Then, the mean potential energy $\langle U_T \rangle$ relaxes to go up as see in Fig.~\ref{fig_E_S}\,(a),\,(a'). Especially, the later stage in (a) looks like that the system enters into the equilibrium state, where $\langle U_T \rangle$ is supposed to plateau.
In addition, as shown in Fig.~\ref{fig_density_snapshots}, the particles get dispersed in the systems \red{with time, which is} quantified by the increases of the mean entropy term $\langle S^{(\rho)} \rangle$ in Fig.~\ref{fig_E_S}\,(b),\,(b'). 
In any plots, we can observe $U_T$ and $S^{(\rho)}$ for each stochastic process fluctuate around the ensemble-averaged quantities.

Figure~\ref{fig_dE_Q_Q_dS_evolution} shows the change in the thermodynamic quantities $dU_T$, $d'Q$, $d'Q^*$ and $dS^{(\rho)}$ from the particles' trajectories and from the number density per a single step together with the ensemble-averaged values $\left< dU_T \right>$, $\left<d'Q\right>$, $\left<d'Q^*\right>$ and $\left<dS^{(\rho)}\right>$.
Note that we use $d'Q^*=d'Q-TdS^{(\rho)}$ to numerically calculate $d'Q^*$.
In each run, $d'Q\,(=dU_T)$, $d'Q^*$ and $dS^{(\rho)}$ fluctuate around the respective averaged quantities $\langle d'Q \rangle\,(=\langle dU_T \rangle)$, $\langle d'Q^*\rangle$ and $\langle dS^{(\rho)}\rangle$, which tend to go towards zero from the nonzero values at the beginning as expected in the transient processes.
These numerical results serve as a guide to see the difference between $d'Q$ and $d'Q^*$ appearing in our statement with Eq.~(\ref{heat_heat_relation}).

Figures~\ref{fig_density_snapshots}-\ref{fig_dE_Q_Q_dS_evolution} have employed the Monte Carlo steps as $t$-axes while there are schemes to convert from the Monte Carlo steps to the real elapsed time~\cite{FrenkelBook,JCP_Fichthorn_1991,Voter_KMC}.
As mentioned, this section does not focus on a specific time evolution of the system.
However, Eq.~(\ref{heat_heat_relation}) is not modified even if the time conversion schemes are included.

\section{Langevin \& Dean--Kawsasaki equations}
\label{many_particles}

Let us move on to the Langevin and the Dean--Kawasaki equations formulated on the continuous time and space.
In the same way as the discrete framework, the heat difference naturally defined by the Langevin and the Dean-Kawasaki equations satisfies Eq.~(\ref{heat_heat_relation}).

However, the qualitative gap from the discrete to the continuous formulations demand a great caution.
The Dean--Kawasaki equation employs the instantaneous number density with the sum of the Dirac's delta functions making its formal structure very simple and transparent,
whereas the particles on the instantaneous number density are ``sparsely" distributed in the real space so that $\Delta S^{(\rho)}$ obtained from the integral over the real space for a set of the almost isolated point-like particles may exhibit little temporal change under plausible physical conditions.\footnote{\red{
This picture of almost isolated particles leads us to expect that the particles' trajectories are traced even in the Dean-Kawasaki equation.
This article develops rather intuitive analytical arguments, while rigorous studies about the mathematical structures of Dean-Kawsasaki equation~\cite{ECP_Konarovski_2019,JStatPhys_Konarovski_2020} have been reported.
}}

Section~\ref{many_particles} observes Eq.~(\ref{heat_heat_relation}) in the continuous spatiotemporal structures.

\subsection{\red{Particle level}}
\label{Langevin}

Employing a corresponding setup to that of the discrete model in Sec.~\ref{sec_Discrete}, but discarding the unit time and length in the discrete models, we consider a many-particle system in the continuous space indexed with a continuous variable ${\bm x}$ along the continuous time $t$.

Let $M$ colloids suspended in solution.
The particles are indexed with $i \in [1,M]$, and a set of the variables is succinctly expressed by $\{ {\bm x}_i \}\equiv \{ {\bm x}_1,{\bm x}_2,\cdots,{\bm x}_{M}\}$.
The motion of the $i$-th particle in the solution is governed by the overdamped Langevin equation:
\begin{eqnarray}
\gamma \frac{d\hat{\bm x}_i(t)}{dt}
=
-{\bm \nabla}_i U_T(\{ \hat{\bm x}_{i'} \},\boldsymbol{\lambda})
+
{\boldsymbol \zeta}_i(t),
\label{Lang_colloid}
\end{eqnarray}
where $i$-th particle's position at time $t$ is denoted by, e.g., $\hat{\bm x}_i(t)=(\hat{x}_i(t),\hat{y}_i(t),\hat{z}_i(t))$ in three-dimensional space.
Let $\gamma$ denote the frictional coefficient per particle, where Eq.~(\ref{Lang_colloid}) does not assume hydrodynamic interactions (HIs) to make arguments simple.
If symbols can function as both stochastic variables and fixed values, adding $\hat{(\,)}$ onto the symbol,  such as $\hat{\bm x}_i$, indicates stochastic variables to distinguish the fixed value ${\bm x}_i$.
The noise ${\boldsymbol \zeta}_i(t)$ is Gaussian-distributed white noise with a variance amenable to the fluctuation-dissipation relation (FDR) of the second kind:
\begin{eqnarray}
\left< {\boldsymbol \zeta}_i(t)\otimes {\boldsymbol \zeta}_j(t')\right>
=
2\gamma k_BT \delta (t-t') \delta_{ij} {\sf I}
\label{FDR_element_Lang}
\end{eqnarray}
with ${\sf I}$ being the $3\times 3$ unit tensor in three-dimensional space.
Analogously to the discrete models, the total potential $U_T$ is divided into pairwise or additive parts:
\begin{eqnarray}
U_T(\{ {\bm x}_{i'} \},\boldsymbol{\lambda})=U(\{ {\bm x}_{i'}-{\bm x}_{j'} \},\lambda_2) +U_e(\{ {\bm x}_{i'} \},\lambda_1),
\label{tot_potentlal_Lang}
\end{eqnarray}
where
\begin{eqnarray}
U(\{ {\bm x}_{i'}-{\bm x}_{j'} \},\lambda_2)
&=&
\frac{1}{2}
\sum_{i=1}^M
\sum_{j=1}^M
u({\bm x}_i-{\bm x}_j,\lambda_2)
\label{U_pair}
\\
U_e(\{ {\bm x}_{i'} \},\lambda_1)
&=&
\sum_{i=1}^M
u_e({\bm x}_i,\lambda_1)
\label{U_ex}
\end{eqnarray}
Throughout the article, the spatial differential operators are performed with respect to subscript variables, e.g., ${\bm \nabla}_{{\bm x}_i}\equiv (\partial/\partial x_i,\partial/\partial y_i,\partial/\partial z_i)$, specified in a Cartesian coordinate system.
Notably, we often simplify them like ${\bm \nabla}_i\equiv {\bm \nabla}_{{\bm x}_i}$ unless confusion is a concern.

Similarly to the discrete model, the work is defined as
\begin{eqnarray}
\Delta' W(t)
=
\int_{t_0}^{t_f} dt'\,
\frac{\partial U_T(\{ \hat{\bm x}_{i'} \},\boldsymbol{\lambda})}{ \red{\partial} \boldsymbol{\lambda}} \cdot \frac{d\boldsymbol{\lambda}}{dt'}.
\label{work_particle}
\end{eqnarray}

The particles undergo the Markov processes obeyed by the Langevin equation~(\ref{Lang_colloid}) in the presence of the Gaussian-distributed white noise.
%% with mean zero and the variance of Eq.~(\ref{FDR_element_Lang}).
According to Sekimoto~\cite{SekimotoBook}, heat during the finite time interval from $t_0$ to $t_f$ is defined as
\begin{eqnarray}
\Delta' Q(t)
=
\int_{t_0}^{t_f} dt'\,
\sum_{i=1}^{M}
\left( -\gamma \frac{d\hat{\bm x}_i}{dt'} +{\boldsymbol \zeta}_i(t') \right) \odot \frac{d\hat{\bm x}_i}{dt'},
\label{heat_particle}
\end{eqnarray}
which quantifies the heat along a trajectory integrated by a sequence of the infinitesimal segments $\{ \hat{\bm x}_i(t_0) \} \rightarrow \{ \hat{\bm x}_i(t_1) \} \rightarrow \{ \hat{\bm x}_i(t_2) \} \rightarrow \cdots \rightarrow \{ \hat{\bm x}_i(t_f) \}$.
With a conventional notation $\circ$ denoting the Stratonovich multiplication, the symbol $\odot$ sandwiched by two vectors is defined to include both the dot product $\cdot$ and the Stratonovich multiplication $\circ$ as ${\bm A}\odot {\bm B}=A_x\circ B_x+A_y\circ B_y+A_z\circ B_z$ with ${\bm A}=(A_x,A_y,A_z)$ and ${\bm B}=(B_x,B_y,B_z)$ in the Cartesian coordinate system.
Instead, the Ito multiplication is expressed without symbols.

\subsection{\red{Density field}}
\label{stochatic_field}

We now move on to a field description by projecting the set of the configurations $\{ \hat{\bm x}_i(t) \}$ to the stochastic number density defined as
\begin{eqnarray}
\rho(t,{\bm x})=\sum_{i=1}^{M} \delta ({\bm x}-\hat{\bm x}_{i}(t)).
\label{NumDen_Dirac_delta}
\end{eqnarray}
Bear in mind that the Kronecker delta in Eq.~(\ref{rho_def_Discrete}) is replaced with the Dirac delta function in Eq.~(\ref{NumDen_Dirac_delta}) from the discrete to the continuous space.
Applying the Ito lemma to Eq.~(\ref{NumDen_Dirac_delta}), we arrive at the Dean--Kawasaki or the stochastic field equation with \red{the density field} ${\rho}(t,{\bm x})$~\cite{JPAMG_Dean_1996} as
\begin{eqnarray}
\frac{\partial {\rho}(t,{\bm x})}{\partial t}
&=&
{\bm \nabla}_{{\bm x}}\cdot \frac{1}{\gamma} {\rho}(t,{\bm x}) {\bm \nabla}_{{\bm x}}{\mu}(t,{\bm x},\boldsymbol{\lambda})
+
\xi^{(\rho)}(t,{\bm x}),
\label{BasicEq_n}
\end{eqnarray}
where \red{the chemical potential is defined as
\begin{eqnarray}
{\mu}(t,{\bm x},\boldsymbol{\lambda})
&\equiv&
\int_\Omega d{\bm x}'\, {\rho}(t,{\bm x}') u({\bm x}-{\bm x}',\lambda_2)
+
u_e({\bm x},\lambda_1)
\nonumber \\
&&
+
k_BT\ln{{\rho}(t,{\bm x})}
\label{chemical_potential_def}
\end{eqnarray}
with $\Omega$ denoting the entire spatial domain of the system.}

The number of particles is locally conserved. 
Hence, Eq.~(\ref{BasicEq_n}) is interpreted as the equation of continuity $\partial {\rho}(t,{\bm x})/\partial t=-{\bm \nabla} \cdot [{\bm j}(t,{\bm x})+\delta{\bm j}(t,{\bm x})]$, where the argument for the divergence in the right-hand side arises from the particle's number current and is classified into the organized ${\bm j}(t,{\bm x})$ and the fluctuating $\delta{\bm j}(t,{\bm x})$.

The first component ${\bm j}(t,{\bm x})$ is the current driven by a gradient of chemical potential as
\begin{eqnarray}
{\bm j}(t,{\bm x})
&=&
-\frac{1}{\gamma}
{\rho}(t,{\bm x}) {\bm \nabla}_{{\bm x}}{\mu}(t,{\bm x},\boldsymbol{\lambda}).
\end{eqnarray}

The other component, $\delta{\bm j}(t,{\bm x})$, comes directly from the thermal noises $\{ {\boldsymbol \zeta}_i(t) \}$ in the Langevin equation~(\ref{Lang_colloid}).
In the fundamental equation~(\ref{BasicEq_n}) for the scalar field of $\rho(t,{\bm x})$, the noise $\xi^{(\rho)}(t,{\bm x})=-{\bm \nabla} \cdot \delta{\bm j}(t,{\bm x})$ takes the mean zero $\left< \xi^{(\rho)}(t,{\bm x})\right>= -\left< {\bm \nabla}\cdot \delta{\bm j}(t,{\bm x})\right>=0$ and satisfies the FDR:\footnote{
\red{
The density-field noises are explicitly written by using the particle-level noises $\{ {\bm \zeta}_i(t) \}$:
\begin{eqnarray}
\xi^{(\rho)}(t,{\bm x})
=
-\frac{1}{\gamma}
\sum_i {\bm \nabla}_i \cdot (\rho_i(t,{\bm x}){\bm \zeta}_i(t)).
\end{eqnarray}
This does not look like a closed form with the density field ${\rho}(t,{\bm x})$.
However, the statistics of the field noises $\{ \xi^{(\rho)}(t,{\bm x}) \}$ obeys the Gaussian distributions dictated only with mean values and variances, so that the closed form is represented by $\left< \xi^{(\rho)}(t,{\bm x}) \right>=0$ and Eq.~(\ref{FDR2nd_field}) only with ${\rho}(t,{\bm x})$. 
}}
\begin{eqnarray}
\left< \xi^{(\rho)}(t,{\bm x}) \xi^{(\rho)}(t',{\bm x}') \right>
&=&
\frac{2k_BT}{\gamma}\delta(t-t')
\nonumber \\
&& 
\times {\bm \nabla}_{{\bm x}} \cdot {\bm \nabla}_{{\bm x}'}
{\rho}({\bm x},t)
\delta({\bm x}-{\bm x}').
\nonumber \\
\label{FDR2nd_field}
\end{eqnarray}
This equation is rather the FDR of the second kind to fit with the form of Eq.~(\ref{BasicEq_n}) in the sense of the explicit expression with the noise correlation \red{(see appendix B).}

Our next attempt is to find the FDR associated with a response function with respect to the chemical potential difference $\Delta {\mu} (t,{\bm x'},\boldsymbol{\lambda}) \equiv {\mu} (t,{\bm x'},\boldsymbol{\lambda})-\mu_0$ with $\mu_0=const.$ (see \red{appendix C} for details related to the response function).
Although thermodynamic force is produced by a gradient of a chemical potential ${\bm \nabla}_{{\bm x}}{\mu}(t,{\bm x},\boldsymbol{\lambda})$, a chemical potential difference $\Delta {\mu}(t,{\bm x},\boldsymbol{\lambda})$ is rather appropriate as a conjugate variable to ${\rho}(t,{\bm x})$ with respect to the energy density (i.e., [number density]$\times$[chemical potential]=[energy density]).
This approach leads us to a natural definition of a response function $R(t-s,{\bm x},{\bm x}')\Theta(t-s)$ as
\begin{eqnarray}
\frac{\delta \left< \partial_t {\rho}(t,{\bm x}) \right>_\xi}{\delta (-\Delta {\mu}(s,{\bm x}',\boldsymbol{\lambda}))}
=
R(t-s,{\bm x},{\bm x}')\Theta(t-s),
\label{M_def}
\end{eqnarray}
where $\left< (\cdot) \right>_\xi$ represents the ensemble average over the thermal noises at a fixed \red{density field} $\{ {\rho}(t,\cdot) \}$; also, $\Theta(t)$ denotes the Heaviside step function defined as $\Theta (t)=1$ for $t>0$, $\Theta (t)=0$ for $t<0$, and $\Theta (t)=1/2$ for $t=0$.
For later notational convenience, the factor $R(t-s,{\bm x},{\bm x}')$ in the response function is defined such that a temporal reversal symmetry $R(t-s,{\bm x},{\bm x}')=R(s-t,{\bm x},{\bm x}')$ holds.
Explicitly, that for the system is specified by
\begin{eqnarray}
R(t-s,{\bm x},{\bm x}')=\frac{2}{\gamma}\delta(t-s){\bm \nabla}_{{\bm x}}\cdot {\bm \nabla}_{{\bm x}'} {\rho}(t,{\bm x}) \delta({\bm x}-{\bm x}').
\nonumber \\
\label{R_explicit_form}
\end{eqnarray}

The response function $R(t,{\bm x},{\bm x}')$ rewrites the fundamental Eq.~(\ref{BasicEq_n}) with Eq.~(\ref{M_def}) into
\begin{eqnarray}
\frac{\partial {\rho}(t,{\bm x})}{\partial t} 
&=&
\int_{-\infty}^t ds\int_\Omega d{\bm x'}\,R(t-s,{\bm x},{\bm x'}) 
(-\Delta {\mu} (s,{\bm x'},\boldsymbol{\lambda}))
\nonumber \\
&&
+\xi^{(\rho)}(t,{\bm x}),
\label{EOM_field_rho}
\end{eqnarray}
Using $R(t-s,{\bm x},{\bm x}')$, the FDR in the form of the field noise correlation (Eq.~(\ref{FDR2nd_field})) is also replaced by
\begin{eqnarray}
\left< \xi^{(\rho)}(t,{\bm x}) \xi^{(\rho)}(t',{\bm x}') \right>
=
k_BT R(t-t',{\bm x},{\bm x}').
\label{FDR1st_xi}
\end{eqnarray}
Although the chemical potential incorporated into the integral kernel in Eq.~(\ref{EOM_field_rho}) might look unfamiliar, a local dipole-like factor ${\bm \nabla}_{{\bm x}'} {\rho}(t,{\bm x}) \delta({\bm x}-{\bm x}')$ in Eq.~(\ref{R_explicit_form}) intrinsically extracts the gradient of the chemical potential (mathematically through integration by parts) such that Eq.~(\ref{EOM_field_rho}) is found to be equivalent to Eq.~(\ref{BasicEq_n}) (see also \red{appendix C}).

\subsection{\red{Heat differences}}
\label{heat_field}

We now attempt to discover an appropriate heat definition in the fundamental field equation~(\ref{EOM_field_rho}).
Equation~(\ref{EOM_field_rho}) is solved with respect to $\Delta \mu (t,{\bm x})$ by introducing $R^{-1}(t-s,{\bm x},{\bm x'})$ so as to satisfy $\int_{-\infty}^{+\infty}dt''\int_{\Omega}d{\bm x}''\, R^{-1}(t-t'',{\bm x},{\bm x''})R(t''-t',{\bm x}'',{\bm x'})=4\delta(t-t')\delta({\bm x}-{\bm x'}) $.
To proceed further, we focus on the fact that the kernel $R(t,{\bm x},{\bm x}')$ is constructed by the product of the temporal $\delta(t-s)$ and the configurational \red{${\bm \nabla}_{{\bm x}}\cdot {\bm \nabla}_{{\bm x}'} {\rho}(t,{\bm x}) \delta({\bm x}-{\bm x}')$} factors so that a separation can be implemented as $R(t,{\bm x},{\bm x}')\equiv 2\delta (t)\sigma({\bm x},{\bm x}')$.
Then, defining $\int_{\Omega}d{\bm x}''\, \sigma^{-1}({\bm x},{\bm x''})\sigma({\bm x}'',{\bm x'})=\delta({\bm x}-{\bm x'})$,\footnote{
The inverse kernel $R^{-1}(t-s,{\bm x},{\bm x'})=2\delta (t-s)\sigma^{-1}({\bm x},{\bm x'})$ is expressed with $\sigma^{-1}({\bm x},{\bm x'})$.
Applying this into $R_\Theta^{-1}(t-t'',{\bm x},{\bm x''})$ defined as $\int_{-\infty}^{+\infty}dt''\int_{\Omega}d{\bm x}''\, R_\Theta^{-1}(t-t'',{\bm x},{\bm x''})R(t''-t',{\bm x}'',{\bm x'})\Theta(t''-t')=\delta(t-t')\delta({\bm x}-{\bm x'})$, we find $R_\Theta^{-1}(t-t'',{\bm x},{\bm x''})=(1/2)R^{-1}(t-t'',{\bm x},{\bm x''})=\delta (t-s)\sigma^{-1}({\bm x},{\bm x''})$.
Noting
\begin{eqnarray}
&&\int_{-\infty}^{+\infty} ds\int_\Omega d{\bm x'}\,R_\Theta^{-1}(t-s,{\bm x},{\bm x'})\psi(s,{\bm x},{\bm x'})
\nonumber \\
&=&
\int_{-\infty}^{+\infty} ds\int_\Omega d{\bm x'}\,\frac{1}{2}R^{-1}(t-s,{\bm x},{\bm x'})\psi(s,{\bm x},{\bm x'})
\nonumber \\
&=&
\int_\Omega d{\bm x'}\,\sigma^{-1}({\bm x},{\bm x'})\psi(t,{\bm x},{\bm x'})
\end{eqnarray}
with an arbitrary function $\psi(s,{\bm x},{\bm x'})$, we arrive at Eq.~(\ref{EOM_GLE_mu}).}
we eventually arrive at
\begin{eqnarray}
\int_{-\infty}^tdt' \int_\Omega d{\bm x'} && R^{-1}(t-t',{\bm x},{\bm x'})
\nonumber \\
&& \times \frac{\partial {\rho}(t',{\bm x}')}{\partial t'}
=
-\Delta {\mu} (t,{\bm x,\boldsymbol{\lambda}})
+
\xi^{(\mu)}(t,{\bm x}),
\nonumber \\
\label{EOM_GLE_mu}
\end{eqnarray}
where
\begin{eqnarray}
{\xi}^{(\mu)}(t,{\bm x})
\equiv
\frac{1}{2}\int_{-\infty}^{+\infty} ds\int_\Omega d{\bm x''}\,
R^{-1}(s-t,{\bm x},{\bm x''})
\xi^{(\rho)}(s,{\bm x}'')
\nonumber \\
\end{eqnarray}
Equations~(\ref{EOM_field_rho}),\,(\ref{EOM_GLE_mu}) are invertible together with mean zero $\left< \xi^{(\mu)}(t,{\bm x}) \right>=0$ and the FDR:
\begin{eqnarray}
\left< \xi^{(\mu)}(t,{\bm x})\xi^{(\mu)}(t',{\bm x}') \right>=k_BTR^{-1}(t-t',{\bm x},{\bm x'}).
\end{eqnarray}

Finally, we are prepared to define the heat viewed from Eq.~(\ref{EOM_field_rho}).
Comparing Langevin Eq.~(\ref{Lang_colloid}) with Eq.~(\ref{EOM_GLE_mu}) implies a conjugacy $({\bm x},{\bm f})\Leftrightarrow (\rho,-\Delta \mu)$ with ${\bm f}=-{\bm \nabla}U_T$.
Therefore, the analogy proposes the heat defined with Eq.~(\ref{EOM_GLE_mu}):
\begin{eqnarray} 
\Delta'Q^*
&\equiv&
\int_{t_0}^{t_f}dt
\int_\Omega d{\bm x} \,
\frac{d'q^*(t,{\bm x})}{dt'},
\label{heat_def_n}
\end{eqnarray}
\begin{eqnarray} 
\frac{d'q^*(t,{\bm x})}{dt}
&=&
\Bigg[
-\int_\Omega d{\bm x'}\,\sigma^{-1}({\bm x},{\bm x'}) \frac{\partial {\rho}(t,{\bm x}' )}{\partial t}
\nonumber \\
&&
+
\xi^{(\mu)}(t,{\bm x})
\Biggr]
\circ
\frac{\partial {\rho}(t,{\bm x})}{\partial t},
\label{heat_def_n_local}
\end{eqnarray}
where the temporal integral in the square bracket of Eq.~(\ref{heat_def_n_local}) is eliminated with $-\int_{-\infty}^t dt'\int_\Omega d{\bm x'}\,R^{-1}(t-t',{\bm x},{\bm x'}) \partial {\rho}(t',{\bm x}')/\partial t=-\int_\Omega d{\bm x'}\,\sigma^{-1}({\bm x},{\bm x'}) \partial {\rho}(t,{\bm x}')/\partial t$.
Equation~(\ref{heat_def_n}) or (\ref{heat_def_n_local}) represents the whole heat or the rate of the local heat, respectively.

The heat defined by Eqs.~(\ref{heat_def_n}) and (\ref{heat_def_n_local}) is reasonable on the basis of the energy balance. For example, let us consider an equilibrium state.
Substituting Eq.~(\ref{EOM_GLE_mu}) into Eqs.~(\ref{heat_def_n}) and (\ref{heat_def_n_local}), we obtain
$d'Q^*/dt=\int_{\Omega}d{\bm x}\,\Delta {\mu} (t,{\bm x})\circ(\partial {\rho}(t,{\bm x})/\partial t)$.
Algebraic manipulation reveals that the rate of heat is equal to that of a thermodynamic-potential-like function, whose mean should be zero in equilibrium. 
Hence, we are naturally led to $\left<d'Q^*/dt\right>=0$ in this condition (see a footnote\footnote{
Using the explicit expression (Eq.~(\ref{chemical_potential_def})) with conservation of the number of particles $(\partial/\partial t)\int_\Omega d{\bm x}\, {\rho}(t,{\bm x})=0$, we transform $d'Q^*/dt=\int_{\Omega}d{\bm x}\,\Delta {\mu} (t,{\bm x})(\partial {\rho}(t,{\bm x})/\partial t)$ into
\begin{eqnarray}
\frac{d'Q^*}{dt}
&=&
\frac{d \Phi(\{ \rho \},\boldsymbol{\lambda})}{dt} 
-
\frac{\partial \Phi(\{ \rho \},\boldsymbol{\lambda})}{\partial \boldsymbol{\lambda}}  
\cdot
\frac{d \boldsymbol{\lambda}}{dt}
,
\label{heat_number_density_chem_diff}
\end{eqnarray}
where $\Phi(\{ \rho \},\boldsymbol{\lambda})$ is defined by Eq.~(\ref{thermodynamic_potential_rho}).
$\Phi(\{ \rho \},\boldsymbol{\lambda})$ evoke a thermodynamic potential similar to the Helmholtz free energy. A caveat is that the thermodynamic-potential-like quantity is not deterministic but fluctuates.
However, its average $\left< \Phi(\{ \rho \},\boldsymbol{\lambda}) \right>$ in the equilibrium under $d\boldsymbol{\lambda}/dt={\bm 0}$ should become constant.
}).

One of the key points in this article is the difference in heat between the \red{particle level} and the \red{density field.}
Interestingly, as shown in Eq.~(\ref{heat_heat_relation}), \red{the heat difference is quantified with the entropy being as a function of the number density:}
\begin{eqnarray} 
S^{(\rho)}(t) \equiv -k_B \int_\Omega d{\bm x}\, {\rho}(t,{\bm x})\ln{{\rho}(t,{\bm x})},
\label{S_density}
\end{eqnarray}
which fluctuates in conjunction with the stochastic evolution of \red{${\rho}(t,{\bm x})$} (refer to \red{appendix D} for the derivation).
Notably, Eq.~(\ref{heat_heat_relation}) corresponds to the expression several lines after Eq.~(12) in the paper of Leonard et al.~\cite{JCP_Leonard_Lander_Seifert_Speck_2013}; however, this equation should be distinct from Eq.~(\ref{S_density}) \red{represented with the ``instantaneous" number density} because the coarse-graining is not carried out in $S^{(\rho)}$.

We also note that, although $S^{(\rho)}$ evokes the Shannon entropy  with the probability distribution like e.g., $S=-k_B\ln{{\cal P}[\{ {\bm x}_{t,i} \}]}$ or $S^*=-k_B\ln{{\cal P}[\{ \rho_{t,{\bm x}} \}]}$, 
\red{Eq.~(\ref{S_density}) does not have appear to have a direct form to that of the Shannon entropy because Eq.~(\ref{S_density}) has the number density as the argument but not the probability.
However, as clear discussions are given in the discrete descriptions, $S^{(\rho)}(t)$ in Eq.~(\ref{S_density}) may be interpreted as the Shannon entropy.}\footnote{\red{
The direct form of $S^{(\rho)}$ to the Shannon entropy is clearer in the discrete descriptions. 
If we only know the density field without the particles' labelings, the uniform distribution of the labelings in the probability are plausible.
We then have ${\cal P}[\{ i \}]=\prod_{{\bm x}}\rho_{t,{\bm x}!}/M!$ with the probability of the particles' labelings denoted by ${\cal P}[\{ i \}]$.
The entropy of the number density is expressed as $S^{(\rho)}=-k_B\ln{{\cal P}[\{ i \}]}-k_B\ln{M!}$.
%% See also the footnote of Sec.~\ref{Distance_SD}.
}
},\footnote{\red{
For the discrete descriptions, if $\rho_{t,{\bm x}}=n_{t,{\bm x}} \gg 1$, the Stirling's formula is applicable to 
\begin{eqnarray}
S^{(\rho)}
=
-k_B \sum_{{\bm x}}\ln{\rho_{t,{\bm x}}!}
\simeq 
-k_B \sum_{{\bm x}} \rho_{t,{\bm x}} \ln{\rho_{t,{\bm x}}} +k_B M,
\label{S_rho_discrte_aymptotic}
\end{eqnarray}
where $\ln{\rho_{t,{\bm x}}!}\simeq \rho_{t,{\bm x}} \ln{\rho_{t,{\bm x}}} -\rho_{t,{\bm x}}$ and $\sum_{{\bm x}}\rho_{t,{\bm x}}=M$ are used.
The asymptotic expression (Eq.~(\ref{S_rho_discrte_aymptotic})) corresponds to Eq.~(\ref{S_density}) for the continuous descriptions.
In addition, it is remarkable that the math form of $\rho_{t,{\bm x}} \ln{\rho_{t,{\bm x}}}$ appears at a single realization before the ensemble average.
As a consequence, $\left< S^{(\rho)}\right>\simeq -k_B \left< \sum_{{\bm x}} \rho_{t,{\bm x}} \ln{\rho_{t,{\bm x}}} \right> +k_B M$ gives contrast to $\left< S\right>=-k_B\left< \ln{{\cal P}[\{ {\bm x}_{t,i} \}]} \right>$ or $\left< S^*\right>=-k_B\left< \ln{{\cal P}[\{ \rho_{t,{\bm x}} \}]} \right>$.
}
}

In addition, together with the heat $\Delta' Q^*$, the work is analogously defined in the number-density field as
\begin{eqnarray}
\Delta' W^*(t)
=
\int_{t_0}^{t_f} dt'\,
\frac{\partial \Phi(\{ \rho \},\boldsymbol{\lambda})}{ \partial \boldsymbol{\lambda}} \cdot \frac{d\boldsymbol{\lambda}}{dt'},
\label{work_rho}
\end{eqnarray}
where 
\begin{eqnarray}
\Phi(\{ \rho \},\boldsymbol{\lambda})
&\equiv&
U_T(\{ \rho \},\boldsymbol{\lambda})
\nonumber \\
&&
+
k_BT \int_\Omega d{\bm x}\, \rho(t,{\bm x}) \ln{\rho(t,{\bm x})}.
\label{thermodynamic_potential_rho}
\end{eqnarray}
It is noticeable that this definition does not alter the work from the \red{particle level to the density field descriptions}: \red{$d' W^*=(\partial \Phi(\{ \rho \},\boldsymbol{\lambda})/\partial \boldsymbol{\lambda}) \cdot (d\boldsymbol{\lambda}/dt)=(\partial U_T(t,{\bm x},\boldsymbol{\lambda})/\partial \boldsymbol{\lambda}) \cdot (d\boldsymbol{\lambda}/dt)=d'W$} because $k_BT\ln{\rho(t,{\bm x})}$
 does not explicitly include $\boldsymbol{\lambda}$.

\section{View of Eq.~(\ref{heat_heat_relation}) from fluctuation theorem and second law of thermodynamics}
\label{sec_FT_particle}

%% \red{Let us here see the fluctuation theorem and second law of thermodynamics for the ``spatiotemporally continuous" models represented by the Langevin and the Dean--Kawasaki equations.}

As investigated in the literature~\cite{JCP_Leonard_Lander_Seifert_Speck_2013,PRX_Nardini_2017}, one can find the fluctuation theorem for the \red{density field,} where the ratio between the path probability of forward evolution of the \red{density field} to that of the reverse evolution is related to $\Delta' Q-T\Delta S^{(\rho)}$.
In addition, we have just defined the heat for the \red{density field} $\Delta' Q^*$ on the basis of Eqs.~(\ref{heat_def_n}) and (\ref{heat_def_n_local}).
Recalling the consequence appearing as Eq.~(\ref{heat_heat_relation}), we discover that an analogous form of the fluctuation theorem (Eq.~(\ref{FT_field})) holds even for the \red{density field} with $\Delta' Q^*\,(=\Delta' Q-T\Delta S^{(\rho)})$.
Furthermore, the variant (Eq.~(\ref{FT_x_to_n_exact})) of the consequential relation between the Langevin and Dean--Kawasaki equations will be found to correspond to Eq.~(\ref{Discrete_FT_x_rho}) in the discrete model.
\red{This section~\ref{sec_FT_particle} observes Eq.~(\ref{FT_x_to_n_exact})  for the ``spatiotemporally continuous" models described with the Langevin and the Dean--Kawasaki equations.
In addition, we remark its resultant compatibility with second law of thermodynamics.}

The conventional fluctuation theorem~\cite{PRE_Crooks_1999,RPP_Seifert_2012} identifies the particles and traces the particles' trajectories $\{ \hat{\bm x}_i \}_{t_0}^{t_f}\equiv\{ \{ \hat{\bm x}_i(t_0) \} \rightarrow \{ \hat{\bm x}_i(t_1) \} \rightarrow \{ \hat{\bm x}_i(t_2) \} \rightarrow \cdots \rightarrow \{ \hat{\bm x}_i(t_f) \} \}$ constructed by jointing successive infinitesimal evolutions, to consider the path probability ${\cal P}[\{ \hat{\bm x}_i \}_{t_0}^{t_f}|\{ \hat{\bm x}_{i} \}_{t_0}]$ given an initial condition $\{ \hat{\bm x}_i \}_{t_0} \equiv \{ \hat{\bm x}_i(t_0) \}$.
On the other hand, in the field description that does not distinguish the particles,
we observe a temporal sequence of the \red{density field,} i.e., $\{ \{ {\rho} \}_{t_0}^{t_f} \}\equiv\{ \{ {\rho}(t_0,{\bm x}) \} \rightarrow \{ {\rho}(t_1,{\bm x}) \} \rightarrow \{ {\rho}(t_2,{\bm x}) \} \rightarrow \cdots \rightarrow \{ {\rho}(t_f,{\bm x}) \} \}$.
The path probability of the temporal sequence given an initial \red{density field} $\{ \rho \}_{t_0}\equiv\{ {\rho}(t_0,{\bm x}) \}$ is denoted by ${\cal P}[ \{ {\rho} \}_{t_0}^{t_f} | \{ \rho \}_{t_0} ]$.
Similarly, let the reverse process denote $\{ \{ {\rho}^\dagger \}_{t_0}^{t_f} \}\equiv\{ {\rho}^\dagger(t_0,{\bm x}) \rightarrow {\rho}^\dagger(t_1,{\bm x}) \rightarrow {\rho}^\dagger(t_2,{\bm x}) \rightarrow \cdots \rightarrow {\rho}^\dagger(t_f,{\bm x}) \}=\{ {\rho}(t_f,{\bm x}) \rightarrow {\rho}(t_{f-1},{\bm x}) \rightarrow {\rho}(t_{f-2},{\bm x}) \rightarrow \cdots \rightarrow {\rho}(t_0,{\bm x}) \}$, and we consider the reverse path probability ${\cal P}[ \{ {\rho}^\dagger \}_{t_0}^{t_f} | \{\rho^\dagger\}_{t_0} ]$ given a reverse initial \red{density field} $\{ {\rho}^\dagger\}_{t_0} \equiv \{ {\rho}^\dagger(t_0,{\bm x}) \}$.
Note that the superscript $\dagger$ denotes the number density of the temporal reverse ${\rho}^\dagger(t,{\bm x})\equiv{\rho}(t_f-(t-t_0),{\bm x})$ and also $t^\dagger\equiv t_f-(t-t_0)$.
The parameters $\boldsymbol{\lambda}$ also evolve backward as $\boldsymbol{\lambda}^\dagger$ in the reverse processes, but not shown in the arguments for the compact notation.

Recalling the fluctuation theorem for the number-density field~\cite{JCP_Leonard_Lander_Seifert_Speck_2013,PRX_Nardini_2017} and Eq.~(\ref{heat_heat_relation}), we arrive at an analogous formalism in the fluctuation theorem as
\begin{eqnarray}
\frac{{\cal P}[ { \{ \rho} \}_{t_0}^{t_f} | \{ {\rho} \}_{t_0} ]}{{\cal P}[ \{ \rho^\dagger \}_{t_0}^{t_f} | \{ \rho^\dagger \}_{t_0} ]}
=
\exp{\left( -\frac{\Delta'Q^*}{k_BT} \right)}.
\label{FT_field}
\end{eqnarray}
Also, the conventional fluctuation theorem \red{at the particle level} is explicitly written as
\begin{eqnarray}
\frac{{\cal P}[\{ \hat{\bm x}_i \}_{t_0}^{t_f} |\{ \hat{\bm x}_i \}_{t_0}]}{{\cal P}[\{ \hat{\bm x}_i^\dagger\}_{t_0}^{t_f}|\{ \hat{\bm x}_i^\dagger \}_{t_0}]}
=
\exp{\left( -\frac{\Delta'Q}{k_BT} \right)},
\label{FT_particle}
\end{eqnarray}
where notations similar to those in Eq.~(\ref{FT_field}) are used (see \red{appendix E}).

We now compare Eq.~(\ref{FT_field}) with Eq.~(\ref{FT_particle}) from Eq.~(\ref{heat_heat_relation}).
A little algebra proposes an exact expression in a different form:
\begin{eqnarray} 
\frac{{\cal P}[ { \{ \rho} \}_{t_0}^{t_f} | \{ {\rho} \}_{t_0} ]}{{\cal P}[ \{ \rho^\dagger \}_{t_0}^{t_f} | \{ \rho^\dagger \}_{t_0} ]}
&=&
\frac{
{\cal P}[\{ \hat{\bm x}_i \}_{t_0}^{t_f} |\{ \hat{\bm x}_i \}_{t_0}]
e^{S^{(\rho)}(t_f)/k_B}
}{
{\cal P}[\{ \hat{\bm x}_i^\dagger\}_{t_0}^{t_f}|\{ \hat{\bm x}_i^\dagger \}_{t_0}]
e^{S^{(\rho)}(t_f^\dagger)/k_B} 
}.
\label{FT_x_to_n_exact}
\end{eqnarray}
Recall Eq.~(\ref{Discrete_FT_x_rho}).
Viewing a correspondence by $1/\prod_{{\bm x}} n_{t,{\bm x}}! \rightarrow e^{S^{(\rho)}(t_f)/k_B}$ and $1/\prod_{{\bm x}'}n_{t,{\bm x}'}^\dagger ! \rightarrow e^{S^{(\rho)}(t_f^\dagger)/k_B} $, 
we find that Eq.~(\ref{FT_x_to_n_exact}) in the continuous descriptions is a counterpart to Eq.~(\ref{Discrete_FT_x_rho}) for the discrete models.
Note that Eq.~(\ref{Discrete_FT_x_rho}) needs to be appropriately expanded from a single step form ($t\rightarrow t+1$) to that of multiple steps ($t\rightarrow t+n$, ($n\geq 2$)).

\red{
With the heat centered, we have so far discussed first law of thermodynamics  on the many-particle or the density field descriptions, which are summarized as, respectively,
\begin{eqnarray}
\Delta U_T =\Delta' Q +\Delta' W,
\label{1stlaw_particle}
\\
\Delta \Phi =\Delta' Q^* +\Delta' W^*.
\label{1stlaw_DF}
\end{eqnarray}
%% Note that the arguments drop for compact representations.
The stochastic thermodynamics conventionally also provides the framework of second law of thermodynamics with the entropy production along a single trajectory for the particle level~\cite{RPP_Seifert_2012,arXiv_Roldan_2023}:
\begin{eqnarray}
\Delta S^{(tot)}
=
\Delta S
-
\frac{\Delta'Q}{T},
\label{2ndlaw_particle}
\end{eqnarray}
where $\Delta S^{(tot)}$ denotes the entropy production for the whole consisting of the system and the thermal bath, $-\Delta' Q/T$ is interpreted as the entropy change in the thermal baths, and the entropy for the system at the particle level is quantified by the Shannon entropy $S$ defined as
\begin{eqnarray}
S(t) = -k_B\ln{{\cal P}[\{ \hat{\bm x}_{t,i} \}]}.
\label{ShannonEntropy_particle}
\end{eqnarray}
Note that $\Delta S=S(t_f)-S(t_0)$ in Eq.~(\ref{2ndlaw_particle}), and the ensemble average exhibits the nonnegativity led by the Gibbs inequality, i.e., $\left<\Delta S^{(tot)}\right> \geq 0$, which corresponds to temporal asymmetricity appearing in the second law of thermodynamics or Clausius inequality.
Besides that at the particle level,
Eq.~(\ref{2ndlaw_particle}) on the density field is replaced with
\begin{eqnarray}
\Delta S^{(tot)}
=
\Delta S^*
-
\frac{\Delta'Q^*}{T}
\label{2ndlaw_DF}
\end{eqnarray}
with the Shannon entropy in the density field defined as
\begin{eqnarray}
S^*(t) = -k_B\ln{{\cal P}[\{ \rho_{t,{\bm x}} \}]}
\label{ShnnonEntropy_DF}
\end{eqnarray}
and the difference $\Delta S^*=S^*(t_f)-S^*(t_0)$.
As seen in Sec.~\ref{sec_ensemble} later on, the probability density ${\cal P}(\{ \hat{\bm x}_{t,i} \})={\cal P}(t,\{ {\bm x}_{i}=\hat{\bm x}_{i} \})$ or ${\cal P}[\{ \rho \}_{t}]={\cal P}[t,\{ \varrho \}]$ is a solution to the Fokker--Planck equation~(\ref{ensemble_FE_all_particle}) at the particle level or the Fokker--Planck equation~(\ref{FP_density}) in the density filed, respectively.\footnote{
\red{Section~\ref{sec_ensemble} expresses the probability density for the density filed with ${\cal P}[t,\{ \varrho \}]\,(={\cal P}[\{ \rho \}_{t}])$ for notational correspondence.}}
As in the discrete models, the continuous models should formally show the differences in the Shannon entropy:
\begin{eqnarray}
\Delta S^*=\Delta S -\Delta S^{(\rho)}.
\label{Shannon_entropy_diff}
\end{eqnarray}
The conversion from the particle level to the density field generally transforms the Shannon entropy and the heat.
It is, however, remarkable that the entropy production should be invariant between Eqs.~(\ref{2ndlaw_particle}) and (\ref{2ndlaw_DF}) by the projection with Eq.~(\ref{NumDen_Dirac_delta}) as easily verified in the discrete descriptions by that with Eq.~(\ref{rho_def_Discrete}).
The formalisms with Eqs.~\,(\ref{1stlaw_particle}) and (\ref{1stlaw_DF}), Eqs.~(\ref{2ndlaw_particle}) and (\ref{2ndlaw_DF}), and Eqs.~(\ref{heat_heat_relation}) and (\ref{Shannon_entropy_diff}) are demanded in both the discrete and the continuous descriptions, but, as claimed repeatedly, the discrete descriptions may exhibit the explicit temporal changes in the entropy of the number density $\Delta S^{(\rho)}$, whereas the continuous descriptions represented by the Langevin and the Dean--Kawasaki equations show little change due to almost no particles' overlap.
%% On the other hand, the discrete models like those in Sec.~\ref{sec_Discrete} offer clear meaning in terms of the indistinguishability of the particles sitting at the same places.
}

\begin{table*}[t]
\begin{tabular}{|c|c|c|c|c|} \hline
many-particle  & & \multicolumn{3}{c|}{many-particle }
\\ 
Langevin equation~(\ref{Lang_colloid}) & $\rightarrow$ & \multicolumn{3}{c|}{Fokker-Planck equation~(\ref{ensemble_FE_all_particle})}
\\ 
$\Delta' Q$  & (B)~~$\{ \hat{\bm x}_i(t) \}$ to ${\cal P}(t,\{ {\bm x}_i \})$ & \multicolumn{3}{c|}{$\left< \Delta' Q \right>$}
 \\
\hline
 &  &  & & 
\\
$\downarrow$ (A)~~$\{ \hat{\bm x}_i(t) \}$ to $\rho(t,{\bm x})$ &  & $\downarrow$ (C)~~${\cal P}(t,\{ {\bm x}_i \})$ to ${\cal P}[t,\{ \varrho \}]$ & & 
\\ 
 & & & & $\downarrow$ (E)~~${\cal P}(t,\{ {\bm x}_i \})$ to $\left<\rho(t,{\bm x})\right>$
\\
\cline{1-1}\cline{2-2} \cline{3-3} 
 & & & &
\\
 & $\rightarrow$ & Fokker-Planck equation~(\ref{FP_density}) for $\{\rho(t,{\bm x}) \}$ & & 
\\ 
Dean-Kawasaki equation~(\ref{BasicEq_n}) & (D)~~$\rho(t,{\bm x})$ to ${\cal P}[t,\{ \varrho \}]$ & $\left<\Delta' Q^*\right>$ & &
\\ 
$\Delta' Q^*$ & & & &
\\ \cline{2-2} \cline{3-3} \cline{4-4} \cline{5-5}
 & \multicolumn{3}{c|}{} & 
\\ 
 & \multicolumn{3}{c|}{$\rightarrow$} & DDFT equation~(\ref{ensemble_FE_density})
\\
 & \multicolumn{3}{c|}{(F)~~$\rho(t,{\bm x})$ to $\left<\rho(t,{\bm x})\right>$} & $\Delta' \overline{Q}$
\\
\hline
\end{tabular}
\caption{
Flowchart of the fundamental equations and heat \red{on continuous descriptions}.}
\label{flowchart_derivations}
\end{table*}

\section{Ensemble Average}
\label{sec_ensemble}

Sections~\ref{sec_ensemble},\,\ref{ManyPolymers} provide relevant issues around the main claim (Eq.~(\ref{heat_heat_relation})) made in Secs.~\ref{many_particles}-\ref{sec_FT_particle} \red{for the continuous descriptions.}

The Dean--Kawasaki \red{equation} has a simple fascinating form apart from difficulties due to the instantaneous number density, and has been inspiring a lot of pertinent theoretical and numerical studies~\cite{JCP_Archer_Evans_2004,JCP_Uneyama_2007,JSRJ_Uneyama_2020,JCP_Reguera_Reiss_2004,JPhysA_Frusawa_2000} (although just a few literatures are here cited).
%% pathological
One of the key debates was to construct the whole map of the derivation routes to relate the Dean--Kawasaki \red{equation} with the Fokker--Planck \red{equation} or the dynamical density functional theory (DDFT), etc.
The whole derivation map was established by combining the ensemble average according to the literature~\cite{JPhysA_Archer_Rauscher_2004}, whose flowchart motivated us to consider analogous guide on the energy balance, especially, of heat as shown in the flowchart~\ref{flowchart_derivations}.

Throughout this section, a key observation to identify heat is a total derivative of potential:
\begin{eqnarray}
d(\mathrm{potential})
=
d'(\mathrm{heat})
+
d'(\mathrm{work}),
\label{Energy_balance_outline}
\end{eqnarray}\red{
which is a basis to find first law of thermodynamics for each description.}
For respective picture with, e.g., Eq.~(\ref{Lang_colloid}) or Eq.~(\ref{BasicEq_n}), the potential $U_T$ or $\Phi$ is well defined and plugged into the (potential) in Eq.~(\ref{Energy_balance_outline}) to specify respective heat.
Keep in mind this basic idea with Eq.~(\ref{Energy_balance_outline}) (see Eqs.~(\ref{FP_x_dU}),\,(\ref{dif_mean_chem}), and (\ref{DDFT_dPhi}) in footnotes).

\subsection*{\red{Route (A): $\{ \hat{\bm x}_i(t) \}$ to $\rho(t,{\bm x})$}}

The succeeding section~\ref{many_particles} has gone along the route A that connects many-particle Langevin equation~(\ref{Lang_colloid}) and Dean-Kawasaki equation~(\ref{BasicEq_n}), where the heat difference extracts the stochastic $\Delta S^{\rho}$ \red{expressed by Eq.~(\ref{heat_heat_relation}).}

Its ensemble average of Eq.~(\ref{heat_heat_relation}) via the route A turns out to be\red{
\begin{eqnarray}
\left<\Delta'Q^*\right>
&=&
\left<\Delta'Q\right>
-
T
\left<\Delta S^{(\rho)}\right>.
\label{heat_heat_relation_ensemble}
\end{eqnarray}
Equation~(\ref{heat_heat_relation_ensemble})} is also extracted by the Fokker-Planck equations between the the particles' positions~(Eq.(\ref{ensemble_FE_all_particle})) and the number density (Eq.~(\ref{FP_density})) that connect the route C.

\subsection*{\red{Route (C): ${\cal P}(t,\{ {\bm x}_i \})$ to ${\cal P}[t,\{ \varrho \}]$}}

Let us here see the mean heat difference through the route C.
We begin with the many-particle Fokker-Planck equation corresponding to Eqs.~(\ref{Lang_colloid}) and (\ref{BasicEq_n}).
By applying the Ito formula into ${\cal P}(t,\{ {\bm x}_i \})=\left<\prod_{i=1}^M \delta ({\bm x}-\hat{\bm x}_i)\right>$ with many-particle Langevin equation~(\ref{Lang_colloid}), we find the many-particle Fokker-Planck equation~(\ref{ensemble_FE_all_particle}):
\begin{eqnarray}
\frac{\partial {\cal P}(t,\{ {\bm x}_i \})}{\partial t}
&=&
-
\sum_{i=1}^M
{\bm \nabla}_i 
\cdot
{\bm J}_i(t,\{ {\bm x}_{i'} \}),
\label{ensemble_FE_all_particle}
\end{eqnarray}
where $i$-th particle's probability flow is denoted by
\begin{eqnarray} 
{\bm J}_i(t,\{ {\bm x}_{i'} \})
&=&
-
\frac{{\cal P}(t,\{ {\bm x}_i \})}{\gamma}
{\bm \nabla}_i
\biggl(
U_T(\{ {\bm x}_{i'} \},\boldsymbol{\lambda})
\nonumber \\
&&
+
k_BT
\ln{{\cal P}(t,\{ {\bm x}_{i'} \})}
\biggr).
\end{eqnarray}
According to the stochastic energetics~\cite{SekimotoBook}, the mean infinitesimal heat is also given just by quantities appearing on the Fokker-Planck equations:\footnote{
An infinitesimal change in the averaged energy reads
\begin{eqnarray}
\left< dU_T(\{ {\bm x}_{i'} \},\boldsymbol{\lambda}) \right>
&=&
d \left(
\int d{\bm x}_{1} \cdots d{\bm x}_{M}\, U_T(\{ {\bm x}_{i'} \},\boldsymbol{\lambda}) 
{\cal P}(t,\{ {\bm x}_i \})
\right)
\nonumber \\
&=&
\int d{\bm x}_{1} \cdots d{\bm x}_{M}\,
\frac{\partial U_T(\{ {\bm x}_{i'} \},\boldsymbol{\lambda}) }{\partial t} 
{\cal P}(t,\{ {\bm x}_i \})
dt
\nonumber \\
&&
+
\int d{\bm x}_{1} \cdots d{\bm x}_{M}\, U_T(\{ {\bm x}_{i'} \},\boldsymbol{\lambda}) 
\frac{\partial {\cal P}(t,\{ {\bm x}_i \})}{\partial t}
dt.
\nonumber \\
\label{FP_x_dU}
\end{eqnarray}
In the right hand side of the last equation above, the first term is interpreted as mean infinitesimal work $\left< d'W \right>=\int d{\bm x}_{1} \cdots d{\bm x}_{M}\,(\partial U_T(\{ {\bm x}_{i'} \},\boldsymbol{\lambda})/\partial \boldsymbol{\lambda})\cdot(d\boldsymbol{\lambda}/dt){\cal P}(t,\{ {\bm x}_i \})dt$, and the other term corresponds to the mean heat $\left< d'Q \right>=\int d{\bm x}_{1} \cdots d{\bm x}_{M}\, U_T(\{ {\bm x}_{i'} \},\boldsymbol{\lambda})(\partial {\cal P}(t,\{ {\bm x}_i \})/\partial t)dt$.
The heat component is transformed with the equation~(\ref{ensemble_FE_all_particle}) of continuity as
\begin{eqnarray}
\left< d'Q \right>
&=&
-\int d{\bm x}_{1} \cdots d{\bm x}_{M}\, U_T(\{ {\bm x}_{i'} \},\boldsymbol{\lambda}) 
\sum_{i'=1}^M {\bm \nabla}_{i'} \cdot {\bm J}_{i'}(t,\{ {\bm x}_i \})
dt
\nonumber \\
&=&
\int d{\bm x}_{1} \cdots d{\bm x}_{M}\, 
\sum_{i'=1}^M {\bm \nabla}_{i'}U_T(\{ {\bm x}_{i'} \},\boldsymbol{\lambda})  \cdot {\bm J}_{i'}(t,\{ {\bm x}_i \})
dt.
\nonumber \\
\end{eqnarray}
Note that the last line assumes that boundary terms are negligible after a use of an integral by parts.
}
\begin{eqnarray} 
\left<d'Q\right>
&=&
\sum_{i=1}^M
\int  d{\bm x}_{1} \cdots d{\bm x}_{M}\, 
{\bm \nabla}_i U_T(\{ {\bm x}_{i'} \},\boldsymbol{\lambda})
\cdot
{\bm J}_i(t,\{ {\bm x}_{i'} \})dt
\nonumber \\
\label{mean_particle_heat}
\end{eqnarray}
Note that only the conservative force is assumed to act on the system for simplicity.
The mean heat $\left<\Delta'Q\right>=\int \left< d'Q \right>$ is identified by the ensemble average of $\Delta'Q =\int d'Q$ in Langevin equation~(\ref{Lang_colloid}) or directly by equation~(\ref{mean_particle_heat}) in the Fokker-Planck equation~(\ref{ensemble_FE_all_particle}).

In addition, we consider an analogous Fokker-Planck formalism for the density map $\{ \varrho(t,{\bm x}) \}$~\cite{JPhysA_Frusawa_2000,JCP_Archer_Evans_2004} with the probability distribution ${\cal P}[t,\{\varrho\}]$ defined as
\begin{eqnarray}
{\cal P}[t,\{\varrho\}]
&\equiv&
\int  d{\bm x}_{1} \cdots d{\bm x}_{M}\, 
\nonumber \\
&&
\times \delta({\varrho}(t,{\bm x})-\rho(t,\{ {\bm x}_i \}))
{\cal P}(t,\{ {\bm x}_{i'} \}),
\end{eqnarray}
where $\int D\varrho\, {\cal P}[t,\{\varrho\}]=1$. 
While Eq.~(\ref{BasicEq_n}) is stochastic equation, the Fokker-Planck equation paired with Eq.~(\ref{BasicEq_n})~\cite{JPhysA_Frusawa_2000,JCP_Archer_Evans_2004} is given as the deterministic form:
\begin{eqnarray}
\frac{\partial {\cal P}[t,\{\varrho\}]}{\partial t}
&=&
-
\int_\Omega d{\bm x}\, 
\frac{\delta}{\delta \varrho(t,{\bm x})}
{\bm \nabla}\cdot 
{\bm J}^{(\rho)}[t,\{ \varrho \}],
\label{FP_density}
\end{eqnarray}
where the probability flow is defined as
\begin{eqnarray}
{\bm J}^{(\rho)}[t,\{ \varrho \}]
&=&
\frac{1}{\gamma}
\varrho(t,{\bm x})
\biggl[
{\cal P}[t,\{\varrho\}]
{\bm \nabla}
\biggl[
\frac{\delta U_T(\{ \varrho \},\boldsymbol{\lambda})}{\delta \varrho(t,{\bm x})}
{\cal P}[t,\{\varrho\}
\biggr]
\nonumber \\
&&
+
k_BT{\bm \nabla}\frac{\delta {\cal P}[t,\{\varrho\}]}{\delta \varrho(t,{\bm x})}
\biggr].
\end{eqnarray}
This description also allows us to naturally introduce infinitesimal mean heat on the energy balance:\footnote{
The potential $\Phi(\{ \varrho \},\boldsymbol{\lambda})$ behaves like the internal energy in the Dean-Kawasaki equation~(\ref{BasicEq_n}).
Accordingly, the mean potential $\left< \Phi(\{ \varrho \},\boldsymbol{\lambda})\right>$ plays the analogous role in the Fokker-Planck equation~(\ref{FP_density}) for $\{\rho(t,{\bm x}) \}$.
Its infinitesimal change turns to be
\begin{eqnarray}
\left< d\Phi(\{ \varrho \},\boldsymbol{\lambda})\right>
&=&
d\left(
\int D\varrho\, \Phi(\{ \varrho \},\boldsymbol{\lambda}){\cal P}[t,\{\varrho\}]
\right)
\nonumber \\
&=&
\int D\varrho\, \Phi(\{ \varrho \},\boldsymbol{\lambda})\frac{\partial {\cal P}[t,\{\varrho\}]}{\partial t} dt
\nonumber \\
&&
+
\int D\varrho\, \frac{\partial \Phi(\{ \varrho \},\boldsymbol{\lambda})}{\partial t} {\cal P}[t,\{\varrho\}] dt
\label{dif_mean_chem}
\end{eqnarray}
As in a standard setup in stochastic thermodynamics, the work is defined through the time-dependent parameters $\boldsymbol{\lambda}(t)$ in $U_T({\bm x},\boldsymbol{\lambda})$ as
\begin{eqnarray}
\left< d'W^* \right>
&\equiv&
\int D\varrho\, \frac{\partial \Phi(\{ \varrho \},\boldsymbol{\lambda})}{\partial t} {\cal P}[t,\{\varrho\}]dt
\nonumber \\
&=&
\int D\varrho\, \frac{\partial (U_T(\{\varrho\},\boldsymbol{\lambda}))}{\partial \boldsymbol{\lambda}} \cdot \frac{d\boldsymbol{\lambda}}{dt} {\cal P}[t,\{\varrho\}]dt.
\end{eqnarray}
The last term in the right hand side of the last equation~(\ref{dif_mean_chem}) corresponds to the work.
Furthermore, we find $\left< d'W^* \right>=\left< d'W \right>$ because $\boldsymbol{\lambda}$ does not explicitly enter into the entropic term $k_BT\ln{\rho(t,{\bm x})}$ in $\Phi(\{ \varrho \},\boldsymbol{\lambda})$.
The other term in Eq.~(\ref{dif_mean_chem}) amounts to mean infinitesimal heat, which is transformed with the equation~(\ref{FP_density}) of the continuity into
\begin{eqnarray}
\left< d'Q^*\right>
&=&
\int D\varrho\, \Phi(\{ \varrho \},\boldsymbol{\lambda})\frac{\partial {\cal P}[t,\{\varrho\}]}{\partial t}dt
\nonumber \\
&=&
-
\int D\varrho\,
\Phi(\{ \varrho \},\boldsymbol{\lambda})
\int d{\bm x}'\,
\frac{\delta}{\delta \varrho(t,{\bm x}')}
{\bm \nabla}_{{\bm x}'}
\cdot
{\bm J}^{(\rho)}[t,\{ \varrho \}]dt
\nonumber \\
&=&
-
\int D\varrho\,
\left(
\int d{\bm x}'\,
{\bm \nabla}_{{\bm x}'}
\frac{\delta}{\delta \varrho(t,{\bm x}')}
\Phi(\{ \varrho \},\boldsymbol{\lambda})
\right)
\cdot
{\bm J}^{(\rho)}[t,\{ \varrho \}]dt
\nonumber \\
\end{eqnarray}
Note that the boundary terms appearing in the integral by parts are assumed to be negligible.
}
\begin{eqnarray}
\left<d'Q^*\right>
&=&
-\int D\varrho\, 
\int d{\bm x}'\,
\nonumber \\
&&
\times
{\bm \nabla}_{{\bm x}'}\frac{\delta}{\delta \varrho(t,{\bm x}')}
\Phi(\{ \varrho \},\boldsymbol{\lambda})
\cdot
{\bm J}^{(\rho)}[t,\{ \varrho \}]dt.
\nonumber \\
\label{mean_Dean_heat}
\end{eqnarray}

The mean heat \red{represented with Eq.~(\ref{mean_particle_heat}) or Eq.~(\ref{mean_Dean_heat})} corresponds to the mean heat in the Langevin equation~(\ref{Lang_colloid}) or the Dean-Kawasaki equation~(\ref{BasicEq_n}), respectively.
Hence, the mean change in the entropic term $\left<dS^{(\rho)}\right>$ is also extracted by comparing the mean heat in the Fokker-Planck equations~(\ref{ensemble_FE_all_particle}),\,(\ref{FP_density}).

\subsection*{\red{Route (E) or (F): To DDFT heat with $\left< \rho(t,{\bm x}) \right>$}}

On the other hand, the number density in dynamical density functional theory (DDFT) needs a distinct story.
According to the interpretation scenario organized in literature~\cite{JPhysA_Archer_Rauscher_2004}, the number density in the DDFT is the number density averaged over realizations.
The following fundamental equation~(\ref{ensemble_FE_density}) for the DDFT is obtained from Dean-Kawasaki equation~(\ref{BasicEq_n}) or the many-particle Fokker-Planck equation~(\ref{ensemble_FE_all_particle})~\cite{JPhysA_Frusawa_2000,JCP_Archer_Evans_2004} along the route F or E in the flowchart~\ref{flowchart_derivations}.
%% \red{In addition, although the flowchart~\ref{flowchart_derivations} does not include the derivations from microscopic Hamiltonian, its construction with the projection operators were discussed in the literature~\cite{PRE_Yoshimori_2005}.}
The time evolution equation in the DDFT obeys a deterministic form as
\begin{eqnarray}
\frac{\partial \left< \rho(t,{\bm x}) \right>}{\partial t}
&=&
-{\bm \nabla} \cdot \overline{\bm J}(t,{\bm x}).
\label{ensemble_FE_density}
\end{eqnarray}
Note that the conventional form of the DDFT is recast by equation of continuity with the mean-number-density flow defined as
\begin{eqnarray} 
\overline{\bm J}(t,{\bm x})
&\equiv&
-\frac{1}{\gamma}
\left<\rho(t,{\bm x})\right>
{\bm \nabla}
\frac{\delta
\overline{\Phi}(\{ \left<\rho\right> \},\boldsymbol{\lambda})
}{\delta \left<\rho(t,{\bm x})\right>}
\label{flow_DDFT}
\end{eqnarray}
and with the analogous thermodynamic potential:
\begin{eqnarray}
\overline{\Phi}(\{ \left<\rho\right> \},\boldsymbol{\lambda})
&\equiv&
\overline{\Phi}_{2}(\{ \left<\rho\right> \},\lambda_2)
+
\int d{\bm x}\,
\left<\rho(t,{\bm x})\right> u_e ({\bm x},\lambda_1)
\nonumber \\
&&
+
k_BT \int d{\bm x}\,
\left<\rho(t,{\bm x})\right> \ln{\left<\rho(t,{\bm x})\right>}.
\label{DDFT_Phi}
\end{eqnarray}The standard DDFT program further employs the relation:\footnote{The expression with the direct correlation $c({\bm x})\equiv (k_BT)^{-1}\delta \overline{\Phi}_{2}(\{ \left<\rho\right> \},\lambda_2)/\delta \left<\rho(t,{\bm x})\right>$ is written as
\begin{eqnarray}
k_BT \left< \rho(t,{\bm x})\right> {\bm \nabla}c({\bm x})
&=&
\int d{\bm x}'\, \left< \rho(t,{\bm x})\rho(t,{\bm x}') \right> {\bm \nabla}u({\bm x}-{\bm x}',\lambda_2).
\nonumber \\
\end{eqnarray}
}
\begin{eqnarray}
\left<\rho(t,{\bm x})\right>
\frac{\delta \overline{\Phi}_{2}(\{ \left<\rho\right> \},\lambda_2)}{\delta \left<\rho(t,{\bm x})\right>}
&=&
-\int d{\bm x}'\, \left< \rho(t,{\bm x})\rho(t,{\bm x}') \right> 
\nonumber \\
&&
\times {\bm \nabla}u({\bm x}-{\bm x}',\lambda_2),
\label{DDFT_ansatz}
\end{eqnarray}
whose $\left< \rho(t,{\bm x})\rho(t,{\bm x}') \right> $ is exact in the equilibrium and assumed in the nonequilibrium.
To solve Eq.~(\ref{ensemble_FE_density}), we need more equation, but identifying the extra equation is a different problem from defining the heat on Eq.~(\ref{ensemble_FE_density}).

By following the similar arguments around Eqs.~(\ref{mean_particle_heat}) and (\ref{mean_Dean_heat}),\footnote{
The total derivative of the potential $\overline{\Phi}$ leads us to the natural identification of the heat and the work on the energy balance:
\begin{eqnarray}
d\overline{\Phi}
&=&
\int d{\bm x}\,
\frac{\partial \left<\rho(t,{\bm x})\right>}{\partial t}
\frac{\delta}{\delta \left<\rho(t,{\bm x})\right>}
\overline{\Phi}(\{ \left<\rho\right> \},\boldsymbol{\lambda}) dt
+
\frac{\partial \overline{\Phi}}{\partial {\bm \lambda}}\cdot d{\bm \lambda}
\label{DDFT_dPhi}
\end{eqnarray}
The last term in the right hand side is defined as work:
\begin{eqnarray}
d'\overline{W} \equiv \frac{\partial \overline{\Phi}}{\partial {\bm \lambda}}\cdot d{\bm \lambda}
\end{eqnarray}
The other component is defined as the heat, and a little calculation yields
\begin{eqnarray}
\int d{\bm x}\,
\frac{\partial \left<\rho(t,{\bm x})\right>}{\partial t}
\frac{\delta \overline{\Phi}(\{ \left<\rho\right> \},\boldsymbol{\lambda})}{\delta \left<\rho(t,{\bm x})\right>}dt
&=&
-
\int d{\bm x}\,
{\bm \nabla} \cdot
\overline{\bm J}(t,{\bm x})
\frac{\delta \overline{\Phi}(\{ \left<\rho\right> \},\boldsymbol{\lambda})}{\delta \left<\rho(t,{\bm x})\right>}dt
\nonumber \\
&=&
\int d{\bm x}\,
{\bm \nabla} \frac{\delta \overline{\Phi}(\{ \left<\rho\right> \},\boldsymbol{\lambda})}{\delta \left<\rho(t,{\bm x})\right>}
\cdot
\overline{\bm J}(t,{\bm x})dt.
\nonumber\\
\end{eqnarray}
Note that the equation of continuity (Eq.~(\ref{ensemble_FE_density})) is used in the first line, \red{and then the integral by parts is performed while assuming the boundary terms are negligible.
The result} is equal to the right hand side of Eq.~(\ref{mean_field_heat}).
}
the infinitesimal heat for the DDFT is also introduced under the formulation with Eqs.~(\ref{ensemble_FE_density})-(\ref{DDFT_ansatz}) as
\begin{eqnarray} 
d'\overline{Q}
&\equiv&
\int d{\bm x}\,
{\bm \nabla} \frac{\delta \overline{\Phi}(\{ \left<\rho\right> \},\boldsymbol{\lambda})}{\delta \left<\rho(t,{\bm x})\right>}
\cdot
\overline{\bm J}(t,{\bm x})dt.
\label{mean_field_heat}
\end{eqnarray}
%% It is noticeable that the right hand side of Eq.~(\ref{mean_field_heat}) can be transformed into a function form relying only on $\left<\rho(t,{\bm x})\right>$ and $\left< \rho(t,{\bm x})\rho(t,{\bm x}') \right>$ by eliminating $\overline{\bm J}(t,{\bm x})$, $\delta\overline{\Phi}(\{ \left<\rho\right> \},\boldsymbol{\lambda})/\delta \left<\rho(t,{\bm x})\right>$, and $\delta \overline{\Phi}_{2}(\{ \left<\rho\right> \},\lambda_2)/\delta \left<\rho(t,{\bm x})\right>$ with eqs.~(\ref{flow_DDFT}),\,(\ref{DDFT_Phi}),\,(\ref{DDFT_ansatz}).
A remarkable consequence is generally $\Delta'\overline{Q}\neq \left< \Delta'Q^{*} \right>$.
Thus, the DDFT framework does not show a direct relation to Eq.~(\ref{heat_heat_relation_ensemble}) or $\left< \Delta S^{(\rho)} \right>$ based on Eq.~(\ref{heat_heat_relation}).

\section{Many-polymer solution}
\label{ManyPolymers}

Section~\ref{ManyPolymers} provides a side topic from projection issues brought up by polymer systems.

Although a particle is idealized as not having internal structures, a polymer consists of many internal degrees of freedom.
The feature of the polymeric chain architecture enables quite different multiple projections to be performed qualitatively without changing the spatiotemporal resolutions.
This section observes two kinds of projections inherent in the polymeric systems.

\subsection{Projection of internal degrees of freedom}

One of the projections is to trace just a single monomer by eliminating the degrees of freedom of the other monomers.
Although this is neither the conventional Mori's nor Zwanzig's operator~(see a review~\cite{PhysRep_Schilling_2022}),
the projections to trace only a single monomer is considered as the variant of the projection operators.
For simplicity, let a linear polymer consisting of $N+1$ monomers be in one-dimensional space.
The projection operator to extract $n$-th monomer is defined as
\begin{eqnarray}
{\sf P}_n
{\bm x}^*
=
\frac{{\bm x}^*\cdot{\bm e}_n}{{\bm e}_n\cdot{\bm e}_n}
{\bm e}_n,
\label{Projection_tag_monomer}
\end{eqnarray}
where a configuration of the single polymer assigned by $(x_0,x_1,\cdots,x_N)$ from the one chain end is specified by
\begin{eqnarray}
{\bm x}^* &\equiv& \sum_{n'=0}^N x_{n'}{\bm e}_{n'}= x_0{\bm e}_0 +x_1{\bm e}_1 +\cdots +x_N{\bm e}_N.
\end{eqnarray}
Note that $\{ {\bm e}_i \}$ forms an orthogonal basis such that ${\bm e}_i\cdot{\bm e}_j=\delta_{ij}$, where the right hand side of Eq.~(\ref{Projection_tag_monomer}) evidently outputs ${\sf P}_n{\bm x}^*=x_n{\bm e}_n$.
This projection reduces a set of the (memoryless) Langevin equations with an instantaneous response in the presence of white noise to only a single equation formulated by a generalized Langevin equation with the finite-time correlated response in the colored noise that generates anomalous diffusion as seen in the literatures~\cite{JStatMech_Panja_2010,PRE_Sakaue_2013,PRE_Maes_Thomas_2013,PRE_Saito_2015,PRE_Saito_Sakaue_2017,PRE_Saito_2017,PRE_Saito_2022_1,PRE_Lizanna_Barkai_2010}.
This choice of the projections demonstrates a consequence led by a general fact that eliminating some out of a complete set of the variables to describe Markov processes manifests nonMarkov processes~\cite{Kubo_Iwanami}.

Although Eq.~(\ref{Projection_tag_monomer}) uses inner product,
the formulation of Eq.~(\ref{Projection_tag_monomer}) differs from the conventional Mori's or Zwanzig's operator, where, notably, the Mori-type's projection operator with an inner product derives the FDR of second kind in linear GLE from the Hamiltonian dynamics~\cite{PhysRep_Schilling_2022}.
Nonetheless, Eq.~(\ref{Projection_tag_monomer}) deserves attention because, as investigated in the preceding studies, the projection links the memoryless Langevin equation to the generalized Langevin equation (GLE) with holding the FDR of second kind~\cite{JStatMech_Panja_2010,PRE_Sakaue_2013,PRE_Maes_Thomas_2013,PRE_Saito_2015,PRE_Saito_Sakaue_2017,PRE_Saito_2017,PRE_Saito_2022_1,PRE_Lizanna_Barkai_2010}.

The derivation of the GLE has been observed mainly on the force balance.
Let us here observe the alternations not only on the force balance, but also on energy balance as formulated in stochastic energetics~\cite{SekimotoBook,JStatMech_Ohkuma_Ohta_2007,arXiv_Roldan_2023}.
From natural perspectives, the elastic-energy (entropic-elastic) potentials appearing on the Langevin equation are transferred by Eq.~(\ref{Projection_tag_monomer}) from internal energy to heat on first law of thermodynamics~\cite{PRE_Saito_2022_1}.
In addition, a view from the second law of thermodynamics is noteworthy since the modified FDR of second kind for the GLE makes a change in path probability through the Onsager-Machlup approaches, which leads to the redefined heat compatible with the fluctuation theorem.
According to second law of thermodynamics in stochastic thermodynamics, 
\red{the whole entropy production for the system of interest with the environment is interpreted as the sum of the Shannon entropy change for the system of interest and the (minus) heat divided by temperature~\cite{RPP_Seifert_2012,arXiv_Roldan_2023}.
The projection \red{defined with} Eq.~(\ref{Projection_tag_monomer}) also clearly alters the view of the Shannon entropy for the system of interest.}
In the case of the polymer formed by physically connecting monomers, the projection \red{with} Eq.~(\ref{Projection_tag_monomer}) transfers the other monomers except $n$ (i.e., $x_{n'}$ for $n'\neq n$) from the system into the environment or the exterior.
This implies that the monomers, whose degrees of freedom are eliminated, are transferred from the system into the exterior. 
Hence, the energy balance modified by the projections for polymer systems is found to be involved in the interesting interpretation issue about where the system spans.

\subsection{Projection onto the hyperdimensional field}

Although we wish to develop Eq.~(\ref{heat_heat_relation}) towards the polymer systems, relating rigorously Eq.~(\ref{Projection_tag_monomer}) to Eq.~(\ref{heat_heat_relation}) did not seem straightforward in our attempt.
Instead, we here consider this issue with another projection defined in many-polymer system from the polymer configurations onto the field:
\begin{eqnarray}
{\rho}(t,{\bm X}^*)
&=&
\sum_{i=1}^M \delta({\bm X}^{*}_{i}-\hat{\bm X}^{*}_{i}),
\label{temp_rho_def_polymer}
\end{eqnarray}
which is compatible with Eq.~(\ref{heat_heat_relation}).
Note that Eq.~(\ref{temp_rho_def_polymer}) is copied from the part of Eq.~(\ref{NumDen_polymer_2}) as mentioned in detail later on.
While the relation between the projections formulated by Eqs.~(\ref{Projection_tag_monomer}) and (\ref{temp_rho_def_polymer}) is left for the future work, the rest of this section is mainly devoted into the compatibility of the projection \red{defined by} Eq.~(\ref{temp_rho_def_polymer}) (or Eq.~(\ref{NumDen_polymer_2})) with Eq.~(\ref{heat_heat_relation}).

Imagine that $M$ identical polymers are suspended in a solution.
Each polymer forms a linear chain of length $Na$.
The polymers are indexed with integers $i\in [1,M]$, whereas monomers (segments) in each polymer are assigned as a continuous variable with $n \in [0,N]$ from one chain end to the other.
The notational combination $(i,n)$ specifies a monomer, whose configurations are identified by ${\bm x}_{i,n}$.

A main analytical tool leveraged in this section is mode analysis~\cite{Doi_Edwards,JStatMech_Panja_2010,PRE_Sakaue_2013,PRE_Saito_2015,PRE_Lizanna_Barkai_2010,PRE_Saito_Sakaue_2017,PRE_Saito_2017,PRE_Saito_2022_1,PRE_Saito_2022_2,PRE_Saito_2023}, which has the advantage of enabling simultaneously diagonalization of frictions, noises, and restoring forces (due to the chain structure) on the equation of motion.
A normal mode, such as a $q$-th mode, is conventionally denoted by uppercase ${\bm X}_{i,q}$. 
The equation of motion in the mode space is rigorously or approximately governed by
\begin{eqnarray}
\gamma_q\frac{d\hat{\bm X}_{i,q}}{dt}
=
-k_q \hat{\bm X}_{i,q}
+{\bm F}_{i,q}
+
{\bm Z}_{i,q},
\label{EOM_polymer_1}
\end{eqnarray}
where a transformation between real and mode space is defined as
\begin{eqnarray}
\hat{\bm X}_{i,q} \equiv \int_0^N dn\, \hat{\bm x}_{i,n}h_{q,n},\quad \hat{\bm x}_{i,n} \equiv \sum_{q=0}^{\infty} \hat{\bm X}_{i,q}h_{q,n}^\dagger.
\label{xn_Xq_def}
\end{eqnarray}
(Refer to the literature~\cite{Doi_Edwards,PRE_Saito_Sakaue_2017}, a footnote\footnote{
The Rouse polymer is quite often employed as an instructive polymer model~\cite{deGennesBook,Doi_Edwards,Khoklov_Grosberg}.
We here illustrate a single Rouse polymer \red{while disregarding} the subscript $i$.
The fundamental equation in real space under a continuous limit along $n$ is rigorously given by
\begin{eqnarray}
\gamma \frac{\partial \hat{\bm x}_n}{\partial t}
=
k\frac{\partial^2 \hat{\bm x}_n}{\partial n^2}
+{\bm f}_n
+{\bm \zeta}_n(t),
\label{Rouse_BE}
\end{eqnarray}
where ${\bm f}_n$ or ${\bm \zeta}_n(t)$ denotes external force and noise acting on the $n$-th monomer, respectively.
The ${\bm \zeta}_n(t)$ is Gaussian-distributed noises with mean zero and the FDR of the second kind:
 \begin{eqnarray}
\left< {\bm \zeta}_n(t)\otimes {\bm \zeta}_{n'}(t')\right>=2\gamma k_BT \delta(n-n')\delta (t-t'){\sf I}.
\label{Rouse_FDR}
\end{eqnarray}
Equation~(\ref{xn_Xq_def}) transforms Eq.~(\ref{Rouse_BE}), Eq.~(\ref{Rouse_FDR}), and $\left< {\bm \zeta}_n(t) \right>={\bm 0}$ into Eq.~(\ref{EOM_polymer_1}), Eq.~(\ref{FDR_polymer_mode_1}), and $\left< {\bm Z}_q(t) \right>={\bm 0}$, respectively, with $k_q= k\pi^2(q/N)^2$ and $\gamma_q =\gamma$.
}, or \red{appendix F} for readers unfamiliar with the mode analyses of the polymer.)
The linear polymer chain satisfies the free boundary condition $\partial \hat{x}_n/\partial n|_{n=0}=\partial \hat{x}_n/\partial n|_{n=N}=0$ ensured by $h_{q,n}\equiv (1/N)\cos{(\pi qn/N)}$, $h_{q,n}^\dagger \equiv (1/c_q)\cos{(\pi qn/N)}$ with $c_q\equiv (1+\delta_{q0})/2$.
The $q$-dependent coefficients for $q\geq 1$ display the power-law function~\cite{Doi_Edwards}: 
\begin{eqnarray}
k_q \simeq k\left(\frac{q}{N}\right)^{1+2\nu},
\qquad
\gamma_q \simeq \gamma \left( \frac{q}{N} \right)^{1-(z-2)\nu},
\end{eqnarray}
where $\nu$ or $z$ stands for the Flory exponent or the dynamical exponent, respectively.
For $q=0$, the frictional coefficient is assumed to be $\gamma_0=\gamma_1$ and the center-of-mass motion has zero spring constant $k_{q=0}=0$, exhibiting the translational center-of-mass motion.

On the right-hand side of Eq.~(\ref{EOM_polymer_1}), the restoring force $-k_q \hat{\bm X}_{i,q}$ is produced by the harmonic potentials $u_{q}(\hat{\bm X}_{i,q})=(1/2)k_q \hat{\bm X}_{i,q}^2$ to retain a chain architecture for $q\geq 1$.
The Rouse polymer takes $\nu=1/2$, whereas the self-avoiding (SA) polymer rigorously displays $\nu=3/4$ in two dimensions or approximately displays $\nu\simeq 0.588$ in three dimensions~\cite{deGennesBook}.

For simplicity, this section does not include interactions between the polymers (i.e., no polymer--polymer interaction is considered).
However, the present formalism can be extended to account for the intrachain interactions (i.e., the monomer--monomer hydrodynamic interaction on the same polymer) by introducing the dynamical exponent $z$.
Parameter $z$ characterizes the scaling in the relaxation time of each polymer $\tau_1\sim N^{z\nu}\sim R^z$~\cite{deGennesBook}, with \red{$\tau_q\equiv \gamma_q/k_q\sim(N/q)^{z\nu}$} or $R\sim N^\nu$ being the characteristic relaxation time of the $q$-normal mode or the spatial size in the solution, respectively.
The $q$-dependent coefficients $\gamma_q$ with $z=3$ for non-draining approximately represent the intra-chain hydrodynamic interactions (HIs), whereas those with $z=2+(1/\nu)$ for free-draining exclude any HIs, as shown by the local friction $\gamma_q \simeq \gamma$.
A specific instructive example is the Rouse polymer that has the free-draining (no HIs) frictional coefficient $\gamma_q=\gamma$ kept constant with $z=4$ at $\nu=1/2$.

The noises have a Gaussian distribution with mean zero $\left< {\bm Z}_{i,q}(t) \right>=0$, and the FDR in the mode space is diagonalized into
\begin{eqnarray}
\left<  {\bm Z}_{i,q}(t)\otimes {\bm Z}_{i',q'}(t') \right>
=
2\gamma_q
\frac{c_q k_BT}{N} \delta_{ii'}\delta_{qq'}\delta (t-t') {\sf I},
\nonumber \\
\label{FDR_polymer_mode_1}
\end{eqnarray}
where ${\sf I}$ denotes a $3\times 3$ identity tensor in three dimensions. Note that this convention modifies the temperature in the thermal bath for the normal modes with the factor $1/N$.

Let ${\bm F}_{i,q}$ be external force.
In the present work, we assume that ${\bm F}_{i,q}$ is a conservative force arising from the energy potential.

We now introduce the \red{configuration density} for the $i$-th polymer identified with the product of the Dirac delta functions of the monomer positions or of the normal modes:
\begin{eqnarray}
{\rho}_i(t,{\bm x}^*)d{\bm x}^*
&=&
\prod_{n} \delta ({\bm x}_n-\hat{\bm x}_{i,n})d{\bm x}^*
=
\prod_{q=0}^\infty \delta ({\bm X}_q-\hat{\bm X}_{i,q})d{\bm X}^*,
\nonumber \\
\label{number_density_ith_polymer}
\end{eqnarray}
where ${\bm x}^*\equiv \oplus_{n\in [0,N]} {\bm x}_n$.
In the notational rule, the replacement of the argument from ${\bm x}^*$ to ${\bm X}^*$ represents the corresponding density such that ${\rho}_i(t,{\bm X}^*)d{\bm X}^*={\rho}_i(t,{\bm x}^*)d{\bm x}^*$, where we define ${\bm X}^*\equiv \oplus_{q\in [0,\infty)} {\bm X}_q \equiv{\bm X}_{0}\oplus{\bm X}_{1}\oplus{\bm X}_{2}\oplus{\bm X}_{3}\oplus\cdots$ in hyperdimension space.
The vector ${\bm X}^*$ has multiple components built up from the center-of-mass and the internal degrees of freedom of the polymer.
To clearly observe the configuration number density in hyperspace, we compare a polymer with a particle in three dimensions.
A single particle indexed with $i$ has three Cartesian components of $x_i$, $y_i$, $z_i$, where the number density is explicitly expressed as ${\rho}_i(t,{\bm x})=\delta (x_i-\hat{x}_i)\delta (y_i-\hat{y}_i)\delta (z_i-\hat{z}_i)$.
However, a single polymer indexed with $i$ has configurational components $\{X_{i,q}^{(x)}\}$, $\{X_{i,q}^{(y)}\}$, $\{X_{i,q}^{(z)}\}$ over $q\in [0,\infty)$, where the configuration number density becomes ${\rho}_i(t,{\bm X^*})=\prod_{q=0}^\infty\delta (X_{i,q}^{(x)}-\hat{X}_{i,q}^{(x)})\delta (X_{i,q}^{(y)}-\hat{X}_{i,q}^{(y)})\delta (X_{i,q}^{(z)}-\hat{X}_{i,q}^{(z)})$.
Hence, including an extension in the configurational dimensions enables the many-polymer system to be considered an analogue of the many-particle system.

The analogy between the particle system and the polymer system proceeds further to enable us to apply a replacement in the equation of motion, and the FDR as
\begin{eqnarray}
{\bm \gamma}^* \frac{d\hat{\bm X}^*_i(t)}{dt}
=
-{\sf c}{\bm \nabla}_{\hat{\bm X}_i^*} U_T^*(\{ \hat{\bm X}_{i'}^* \},\boldsymbol{\lambda})
+
{\bm Z}_{i}^*(t),
\label{EOM_polymer_2}
\end{eqnarray}
\begin{eqnarray}
\left<  {\bm Z}_{i}^*(t)\otimes{\bm Z}_{i'}^*(t') \right>
&=&
2{\sf c}{\bm \gamma}^*
\frac{k_BT}{N} \delta_{ii'}\delta (t-t'),
\label{FDR_polymer_mode_2}
\end{eqnarray}
where ${\sf c}\equiv \otimes_{q=0}^{\infty} c_q {\sf I}$ denotes a diagonalized tensor, $\hat{\bm X}_{i}^*(t)\equiv \oplus_{q} \hat{\bm X}_{i,q}(t) =\hat{\bm X}_{i,0}\oplus\hat{\bm X}_{i,1}\oplus\hat{\bm X}_{i,2}\oplus\hat{\bm X}_{i,3}\oplus\cdots$, ${\bm Z}_{i}^*(t)\equiv \oplus_{q} {\bm Z}_{i,q}(t)$, and ${\bm F}_{i}^*(t)\equiv \oplus_{q} {\bm F}_{i,q}(t)$. 
A notation for the potentials is introduced to maintain correspondence with Eq.~(\ref{Lang_colloid}) as
\begin{eqnarray}
U_T^*(\{ \hat{\bm X}_{i'}^* \},\boldsymbol{\lambda}) &\equiv& \sum_{i=1}^M u_T^*({\bm X}_i^*,\boldsymbol{\lambda}),
\label{total_U_polymer}
\\
u_T^*({\bm X}_i^*,\boldsymbol{\lambda}) &\equiv& u^*({\bm X}_i^*,\lambda_2) +u_e^*({\bm X}_i^*,\lambda_1) 
\label{u_polymer}
\\
u^*({\bm X}^*,\lambda_2) 
&\equiv& \sum_{q=0}^\infty c_q^{-1}u_q({\bm X}_{q},\lambda_2)=\sum_{q=0}^\infty \frac{1}{2} c_q^{-1}k_q{\bm X}_{q}^2,
\nonumber \\
\label{harmonic_chain_q}
\\
u_e^*({\bm X}^*,\lambda_1) 
&\equiv& \sum_{q=0}^\infty c_q^{-1}u_{e,q}({\bm X}_{q},\lambda_1).
\label{external_q}
\end{eqnarray}
Equations~(\ref{total_U_polymer})--(\ref{external_q}) yield conservative force $-{\sf c}{\bm \nabla}_{\hat{\bm X}_i^*} U_T^*(\{\hat{\bm X}_{i'}^*\},\boldsymbol{\lambda})=-{\sf c}{\bm \nabla}_{\hat{\bm X}_i^*} u_T^*(\hat{\bm X}_i^*,\boldsymbol{\lambda})=-{\bm k}^*\hat{\bm X}_i^*+{\bm F}_i^*$ acting on the $i$-th polymer.
Note that ${\bm \gamma}^*$ or ${\bm k}^*$ serves as a tensor.
The components of the coefficient tensors are denoted by $({\bm \gamma}^*)_{qq'}\equiv \gamma_{q}\delta_{qq'}{\sf I}$ and $({\bm k}^*)_{qq'}\equiv k_q\delta_{qq'}{\sf I}$, where the subscripts $q$, $q'$ dictate the mode components of the coefficient tensor.

\subsubsection{Mode space density field}

Along the replacements from many particles to many polymers, the analogy defines the \red{density field} as
\begin{eqnarray}
{\rho}(t,{\bm X}^*)
&=&
\sum_{i=1}^M{\rho}_i(t,{\bm X}^{*}_{i})
=
\sum_{i=1}^M \delta({\bm X}^{*}_{i}-\hat{\bm X}^{*}_{i}).
\label{NumDen_polymer_2}
\end{eqnarray}
Here, we observe that Eqs.~(\ref{EOM_polymer_2}),\,(\ref{FDR_polymer_mode_2}), and (\ref{NumDen_polymer_2}) have the extended form of Eqs.~(\ref{Lang_colloid}),\,(\ref{FDR_element_Lang}), and (\ref{NumDen_Dirac_delta}).\footnote{
The \red{configuration density} is converted between the real and the mode space as follows:
\begin{eqnarray}
\rho(t,{\bm x}^*)d{\bm x}^*
&=&
\sum_{i=1}^M
\prod_{n} \delta({\bm x}_{n}-\hat{\bm x}_{i,n}) \prod_{n}d{\bm x}_{n}
\nonumber \\
&=&
\sum_{i=1}^M
\prod_{q=0}^\infty |J^{-1}| \delta({\bm X}_{q}-\hat{\bm X}_{i,q}) |J| \prod_{q'=0}^\infty d{\bm X}_{q'}
\nonumber \\
&=&
\rho(t,{\bm X}^*)d{\bm X}^*,
%% \sum_{i=1}^M \prod_{q=0}^\infty \rho_{i,q}(t,{\bm X}_{q})d{\bm X}^*,
\label{rho_xn_Xq}
\end{eqnarray}
where the Jacobian is $J={\rm det}(\partial x_n/\partial X_q)$.
The Jacobian is canceled from the second line to the last line.} 
Hence, the procedure to find the field equation~(\ref{BasicEq_n}) is available by changing the dimensions.
%% As in the particles, 
The density field $\rho(t,{\bm X}^*)=\sum_{i=1}^M \delta ({\bm X}^*-\hat{\bm X}_{i}^*)$ for the polymers obeys the Dean--Kawasaki equation for \red{the density field
$\{ {\rho}(t,{\bm X}^*) \}$:}
\begin{eqnarray}
\frac{\partial {\rho}(t,{\bm X}^*)}{\partial t}
&=&
{\bm \nabla}_{{\bm X}^*} \cdot
\left(
{\sf c}{\bm \gamma}^{*\,-1}
{\rho}(t,{\bm X}^*)
{\bm \nabla}_{{\bm X}^*}
{\mu}(t,{\bm X}^*,\boldsymbol{\lambda})
\right)
\nonumber \\
&&
+
{\Xi}^{*(\rho)}(t,{\bm X}^*),
\label{BE_rho_q}
\end{eqnarray}
where ${\bm \gamma}^{*\,-1}$ denotes the inverse of ${\bm \gamma}^{*}$ such that ${\bm \gamma}^{*\,-1}{\bm \gamma}^{*}=\otimes_{q=0}^{\infty} {\sf I}$, and a chemical potential or a noise is respectively defined as
\begin{eqnarray}
{\mu}(t,{\bm X}^*,\boldsymbol{\lambda})
&\equiv&
u^*({\bm X}^*,\lambda_1)
+
u^*_e({\bm X}^*,\lambda_2)
\nonumber \\
&&
+
\frac{k_BT}{N}
\ln{{\rho}(t,{\bm X}^*)},
\label{BE_mu_q}
\end{eqnarray}
\begin{eqnarray}
{\Xi}^{*(\rho)}(t,{\bm X}^*)
&=&
-{\bm \nabla}_{{\bm X}^*} \cdot [{\rho}(t,{\bm X}^*) ({\bm \gamma}^*)^{-1} {\bm Z}^*(t)].
\label{BE_Xi_q}
\end{eqnarray}
The noise correlations satisfy the FDR in the \red{density field:}
\begin{eqnarray}
\left<
{\Xi}^{*(\rho)}(t,{\bm X}^*)
{\Xi}'^{*(\rho)}(t',{\bm X}'^*)
\right>
&=&
2\frac{k_BT}{N}\delta (t-t')
\nonumber \\
&&
\times 
({\bm \nabla}_{{\bm X}^*} \cdot)
({\bm \nabla}_{{\bm X}'^*} \cdot)
{\sf c}{\bm \gamma}^{*\,-1}
\nonumber \\
&&
\times 
{\rho}(t,{\bm X}^*)
\delta ({\bm X}-{\bm X}'^*).
\nonumber \\
\label{FDR_field_mode}
\end{eqnarray}
We note that the chain length $N$ and also ${\sf c}$ enter into Eq.~(\ref{FDR_field_mode}) arising from the FDR Eq.~(\ref{FDR_polymer_mode_2}) (see \red{appendix G}).

\subsubsection{Heat differences}

A heat rate along the polymeric configuration trajectory is written in the sense of Sekimoto's definition as
\begin{eqnarray}
\frac{d'Q}{dt}
&=&
\sum_{i=1}^M
\left[ 
N{\sf c}^{-1}
\left( 
-{\bm \gamma}^* \frac{d{\bm X}_{i}^*(t)}{dt}
+
{\bm Z}_{i}^*(t)
\right)
\right]
\odot
\frac{d{\bm X}_i^*(t)}{dt},
\nonumber \\
\label{heat_rate_polymer}
\end{eqnarray}
which are given in the mode space.
Recall that the dot $(\cdot)$ stands for the combination of dot product and Stratonovich multiplication, such as ${\bm X}^*\odot {\bm Y}^*\equiv \sum_{q=0}^{\infty}( X_{q}^{(x)}\circ Y_{q}^{(x)} +X_{q}^{(y)}\circ Y_{q}^{(y)}+X_{q}^{(z)}\circ Y_{q}^{(z)} )$.
Once the heat definition is applied to the mode space, we arrive at the heat form in the real space through variable transformation~\cite{PRE_Saito_2022_1,PRE_Saito_2022_2} (see \red{appendix F} for the real-space expression (Eq.~(\ref{heat_rate_polymer_real_space}))).

The next step is to find the heat rate in the field equation.
To define the heat, Eq.~(\ref{BE_rho_q}) with respect to $\partial {\rho} (t,{\bm X}^*)/\partial t$ is solved for $-\Delta \mu(t,{\bm X}^*,\boldsymbol{\lambda})$ and organized as
\begin{eqnarray}
\int d{\bm X}'^*\,\sigma^{-1}({\bm X}^*,{\bm X}'^*)
\frac{\partial {\rho}(t,{\bm X}'^*)}{\partial t}
&=&
-\Delta \mu(t,{\bm X}^*,\boldsymbol{\lambda})
\nonumber \\
&&
+
{\Xi}^{*(\mu)}(t,{\bm X}^*),
\nonumber \\
\label{BE_rho_poly_3}
\end{eqnarray}
which corresponds to Eq.~(\ref{EOM_GLE_mu}).
Note that $\sigma^{-1}({\bm X}^*,{\bm X}'^*)$ is defined through an inverse function of $\sigma({\bm X}'^*,{\bm X}''^*)$ appearing in the response function $\delta \left< {\rho}(t,{\bm X}^*)\right>_\Xi/\delta (-\Delta {\mu}(s,{\bm X}'^*,\boldsymbol{\lambda}))=R(t-s,{\bm X}^*,{\bm X}'^*)\Theta(t-s)=2\delta (t-s)\sigma ({\bm X}^*,{\bm X}'^*)\Theta(t-s)$ in the same manner as the arguments around Eq.~(\ref{EOM_GLE_mu}).
%% \footnote{The response functions in the mode space are separated as a product of temporal and configurational components, where we have $R(t-s,{\bm X}^*,{\bm X}'^*)\equiv 2\delta (t-s)\sigma ({\bm X}^*,{\bm X}'^*)$ as in Section~\ref{heat_field}. Then, $\sigma^{-1}({\bm X}^*,{\bm X}''^*)$ is defined such that $\int_\Omega d{\bm X}''^*\,\sigma^{-1}({\bm X}^*,{\bm X}''^*)\sigma({\bm X}''^*,{\bm X}'^*)=\delta({\bm X}^*-{\bm X}'^*)$.}

Let us proceed further along the formalism in Sec.~\ref{heat_field}.
Although the variables are modified as ${\bm x}\rightarrow {\bm X}^*$, the analogous form is retained.
Hence, the definition of the heat rate in the field is employed as
\begin{eqnarray}
\frac{d'Q^*}{dt}
&=&
\int d(N{\sf c}^{-1}{\bm X}^*)\,
\Biggl[
\nonumber \\
&&
-\int d{\bm X}'^*\,\sigma^{-1}({\bm X}^*,{\bm X}'^*) \frac{\partial {\rho}(t,{\bm X}'^*)}{\partial t}
\nonumber \\
&&
+
\Xi^{*(\mu)}(t,{\bm X}^*)
\Biggr]
\circ
\frac{\partial {\rho}(t,{\bm X}^*)}{\partial t}.
\label{heat_polymer_field_def}
\end{eqnarray}

Analogously, the differences in the heat (temporal integral of Eqs.~(\ref{heat_rate_polymer}) and (\ref{heat_polymer_field_def})) are organized into Eq.~(\ref{heat_heat_relation}) with the entropy of the \red{number density} defined as 
\begin{eqnarray} 
S^{(\rho)}
&=&
-k_B \int d{\bm X}^*\, {\rho}(t,{\bm X}^*)\ln{{\rho}(t,{\bm X}^*)}.
\label{S_polymer}
\end{eqnarray}
Note that \red{the entropy of the number density $S^{(\rho)}$} is specified in hyperdimension space (see \red{appendix H}).

We can also find the many-polymer version of the fluctuation theorem with Eqs.~(\ref{FT_field}) and (\ref{FT_particle}) by simply replacing ${\bm x}$ with ${\bm X}^*$ likewise.
Equation~(\ref{heat_heat_relation}) is then transformed into the polymeric equation~(\ref{FT_x_to_n_exact}).

We have found that the projections \red{defined by Eq.~(\ref{NumDen_polymer_2}) are} compatible with Eq.~(\ref{heat_heat_relation}).
Equation~(\ref{S_polymer}) does not, however, have a practical form because of \red{the instantaneous (configuration) number density and also the hyperdimensions.
Even if the instantaneous number density is circumvented by employing the discrete models, the hyperdimensions hamper the practical usage (see appendix I).
}
Further projections are required to make it more tractable, and one of the \red{directions} is to combine two kinds of projections (Eq.~(\ref{NumDen_polymer_2}) and Eq.~(\ref{Projection_tag_monomer})) in a rigorous or an approximate way.
This is the future issue.
It would be interesting if the projection studies for the polymers are given further insights by focusing on the energy balance or on a change in the thermodynamic quantities like entropy.

%% Although the above results are exact, they might not be suitable for practical purposes.
%% In Section~\ref{Discussion_sec}, we present arguments about comparisons with analyses developed in the polymer theory.

\begin{table}[t]
\begin{tabular}{|lll|}
\hline
%% \multicolumn{3}{|l|}{(i) Hamiltonian dynamics}
(i) Hamiltonian dynamics & $\Rightarrow$ & 
\\ 
\qquad $\Downarrow$\, Projection operators & & \rotatebox{45}{$\Downarrow$}
\\
(ii) GLE & & \quad $\Downarrow$ Projection operators
\\
\qquad $\Downarrow$\, Coarse graining & & \quad $\Downarrow$ \qquad \qquad $+$
\\ 
\qquad ~~~ (Markovian limit) & & \quad $\Downarrow$ Coarse graining
\\
(iii) memoryless  & & ~~~$\Downarrow$ (Markovian limit)
\\
\qquad Langevin equation & &  \quad $\Downarrow$
\\
\qquad (Eqs.~(\ref{Lang_colloid}) and (\ref{FDR_element_Lang})) & &  \quad $\Downarrow$
\\
\qquad $\Downarrow$\, Ito formula & & \rotatebox{315}{$\Downarrow$}
\\
(iv) Dean-Kawasaki equation & $\Leftarrow$ &
\\
\hline
\end{tabular}
\caption{\red{Roadmap to construct the conventional Dean-Kawasaki equation from Hamiltonian dynamics}}
\label{chart_GLE}
\end{table}

\section{Discussion}
\label{Discussion_sec}

In Sec.~\ref{Discussion_sec}, we make arguments or give perspectives about the consequences obtained in the previous sections. The remarks are listed below.

%% \subsection{\red{Time-dependent FDR of second kind}}

\subsection{Colored noises and memories}

Projection operators have been highly developed to derive the Langevin equation, or the generalized Langevin equations (GLE) from microscopic Hamiltonian picture~\cite{Kubo_Iwanami,JCP_Zwanzig_1960,PTP_Mori_1965,JPhysA_Kawasaki_1973,PhysRep_Schilling_2022,PR_Robertson_1966,PRA_Kawasaki_Gunton_1973,Grabert,arXiv_Dengler_2016,PhysRep_Schilling_2022,PRE_Yoshimori_2005,SekimotoBook}.
Flowchart~\ref{chart_GLE} sketches a roadmap to arrive at the conventional Dean--Kawasaki equation.

The multiple-layered route (i) Hamiltonian dynamics $\Rightarrow$ (ii) GLE $\Rightarrow$ (iii) the memoryless Langevin equation $\Rightarrow$ (iv) the Dean--Kawasaki equation passes through the stochastic particle picture formulated by the Langevin equation.
The step from (i) to (ii) is implemented on rigorous mathematical instructions with the projection operators to derive the linear and the nonlinear GLE~\cite{JCP_Zwanzig_1960,PTP_Mori_1965,JPhysA_Kawasaki_1973,Grabert,PhysRep_Schilling_2022,arXiv_Dengler_2016}.
The next step from (ii) to (iii) inserts the Markovian limit considered as the coarse-grainings in this article.
Note that potential of mean force often implicitly employed in modeling soft matter like polymers needs particular attention because the standard FDR of second kind expressed with a linear memory kernel is not necessarily ensured~\cite{PhysRep_Schilling_2022,EPL_Glatzel_Schilling_2021}.
%%  but a crucial notion for physics in a sense to extract the universal features from a rich variety of the systems.
Although those controversial points warrant careful inspections, if the system of interest is phenomenologically well described by memoryless Langevin equation with the FDR of second kind, the idealization from the microscopic picture should require the temporal resolution change (Markovian limit) in any coarse-graining strategies.
Incidentally, a projection scheme with the harmonic bath survives the bare external force potential intact and does not get involved in this issue~\cite{JStatPhys_Zwanzig_1973,PhysRep_Schilling_2022}.
After the Markovian limit being taken, the temporally local formulation eventually allows us to apply the Ito formula into stochastic differential equations on the way from (iii) to (iv), although the colored noises and the memory integral kernels in the GLE are now formidable to deal with.
%% The step from (ii) to (iii) demands at least the Markovian limit considered as a change in the temporal resolution.

Besides the multiple-layered one, the direct route from (i) to (iv) discussed with the projection operators and the Markovian limit in the literature~\cite{PRE_Yoshimori_2005} finds the Dean--Kawasaki equation while not going through the stochastic particle picture expressed by the Langevin equation.

Both the multiple-layered and the direct routes to get the conventional Dean-Kawasaki equation employ the Markovian limit viewed as the temporal resolution change.
While the conventional Dean--Kawasaki equation is rather temporally local fundamental equation, discovering the extended Dean--Kawasaki equation with the memory or the colored noises is an interesting challenge.

\subsection{Non-stationary GLE}

Recent progresses created projection operators that can obey the explicitly time-dependent Liouvillian or Hamiltonian, 
resulting in non-stationary GLE~\cite{PhysRep_Schilling_2022,JCP_Meyer_2019,JCP_Glatzel_2021,JCP_Widder_2022}.
As inspired by the preceding studies, 
this section gives perspectives about the stochastic descriptions constructed from the explicitly time-dependent Hamiltonian.
%% Liouvillian.
%% not have time-translation symmetry.

If we wish to get Eq.~(\ref{heat_heat_relation}) given in the main text from the explicitly time-dependent Hamiltonian, 
\red{the projection scheme experiences several steps in a hierarchy of the fundamental equations like flowchart~\ref{chart_GLE}.}
The exact analyses from (i) the Hamiltonian dynamics to (ii) non-stationary GLE are discussed~\cite{PhysRep_Schilling_2022,JCP_Meyer_2019,JCP_Glatzel_2021,JCP_Widder_2022}.
However, if we are restricted to closed formalisms at the respective hierarchial levels, the further step from (ii) non-stationary GLE (or even the non-stationary linear GLE) to the closed fundamental equation \red{on the density field} like the Dean-Kawasaki equation seems hard at present because of the problems, one of which is that the simple application of the Ito formula in stochastic differential equations does not go well in the colored noises appearing in the GLE.

%% The focus in this article focuses on the step (iii) $\rightarrow$ (iv) by .
Assuming the validity of the Markovian limit as a tentative prescription, we separate the issues about the step (ii) $\rightarrow$ (iii) from (iii) $\rightarrow$ (iv).
Once the memoryless Langevin equation with the white noises is admitted after the change in the temporal resolutions from (ii), this coarse-graining allows us to apply the Ito formula, so that we can go along with the analogous line towards the step (iv).
For simplicity, say, in the absence of nonconservative force,
the plausible phenomenological description would correspond to the one with the time-dependent parameters, where the friction coefficient or the noise is replaced as $\gamma \rightarrow \gamma (t,t_0)$ or ${\boldsymbol \zeta}_i(t) \rightarrow {\boldsymbol \zeta}_i(t,t_0)$ in the conventional stationary Langevin equation~(\ref{Lang_colloid}), respectively, as well as the the time-dependent potentials with $t_0$ denoting the origin of time:
\begin{eqnarray}
\gamma(t,t_0)
\frac{d\hat{\bm x}_i(t)}{dt}
=
-{\bm \nabla}_i U_T(\{ \hat{\bm x}_{i'} \},\boldsymbol{\lambda})
+
{\boldsymbol \zeta}_i(t,t_0).
\label{nsLangevin_FE}
\end{eqnarray}
Whereas the presence of the potential of mean force~\cite{PhysRep_Schilling_2022,EPL_Glatzel_Schilling_2021} or the non-equilibrium processes do not generally satisfy the standard FDR of the second kind, if it holds, or if the different coarse-graining schemes to satisfy it such as harmonic bath proposed by Zwanwig~\cite{JStatPhys_Zwanzig_1973} are effective, we analogously assume that the noise is identified by the gaussian distribution with $\left< {\boldsymbol \zeta}_i(t,t_0) \right>={\bm 0}$ and the FDR of the second kind:
\begin{eqnarray}
\left< {\boldsymbol \zeta}_i(t,t_0)\otimes {\boldsymbol \zeta}_j(t',t_0)\right>
=
2\gamma (t,t_0) k_BT \delta (t-t') \delta_{ij} {\sf I}
\nonumber \\
\label{TD_FDR_element_Lang}
\end{eqnarray}
These replacements lead us to the analogous time-dependent Dean-Kawasaki equation \red{(see appendix J),} and do not modify the consequence expressed by Eq.~(\ref{heat_heat_relation}).
The arguments from (ii) to (iii) are, however, intuitive here.
In nonequilibrium conditions, how to rigorously \red{formulate the particle-level pictures} needs further developments with the projection operators from the microscopic Hamiltonian.

\subsection{Hydrodynamic interactions}

We have not as yet considered the HIs between the particles or the polymers (i.e., the disjointed objects) because the position-dependent friction that represents the HIs poses technical difficulties.
An optimistic observation suggests that the formalism of the heat difference (Eq.~(\ref{heat_heat_relation})) would not be modified.
%% , although this article does not attempt to find explicit derivations.
These close inspections are left as a topic for future work.

\subsection{Multinomial diffusion model}

Section~\ref{sec_Discrete} has investigated the spatiotemporally discrete model that may produce the explicit time evolution of \red{the entropy of the number density $S^{(\rho)}$} unlike the spatiotemporally continuous models represented by the Langevin and the Dean--Kawasaki equations.
In addition, a remarkable point in Sec.~\ref{sec_Discrete} is that dynamical mechanisms are not specified, so that the main statement of Eq.~(\ref{heat_heat_relation}) is applicable to the other stochastic discrete models.
%% A caveat is, however, that the discrete time do not have to be assumed although the spacious discreteness is inevitable to observe the explicit time evolution of the number-density entropy in plausible conditions.
Keeping these points in mind, we can find that one of the interesting perspectives is multinomial diffusion model and its family~\cite{JCP_Lampoudi_2009,JCP_Gillespie_2013,JCP_Tian_Burrage_2004,PRE_Balter_Tartakovsky_2011} to deal with reaction and diffusion fields, where the space is discretely divided into subspaces with finite volume.
%% while the temporal axis continuously extends.
%% The schemes to manipulate the stochastic diffusion-reaction in spatially inhomogeneous systems have been discussed with multinomial diffusion model~\cite{JCP_Lampoudi_2009,JCP_Gillespie_2013,JCP_Tian_Burrage_2004,PRE_Balter_Tartakovsky_2011}.
The developments of Eq.~(\ref{heat_heat_relation}) towards the multinomial  diffusion model would be one of the promising directions.

\subsection{Coarse-grainings}

This article has dealt with (a) the discrete or (b) the continuous picture,  within each of which, the spatiotemporal resolutions have not been altered in the processes of derivation of Eq.~(\ref{heat_heat_relation}).
However, changing the spatiotemporal resolutions before formulating (a) the \red{discrete particle-level} picture with Eqs.~(\ref{dicrete_local_q_def}) and (\ref{DB_Q_discrete}), or (b) the \red{continuous one} with the Langevin equation~(\ref{Lang_colloid}), does not affect Eq.~(\ref{heat_heat_relation}).

We here focus on the former coarse-graining (a), which may include a transformation of the Langevin equation~(\ref{Lang_colloid}) into the discrete models governed by Eqs.~(\ref{dicrete_local_q_def}) and (\ref{DB_Q_discrete}).
If the coarse-graining (a) is systematically built up, it would provide solider ground for the dynamical mechanisms although this article does not go deeply into this construction issue.

\subsection{\red{Distance on stochastic descriptions}}
\label{Distance_SD}

\red{
Suitable characterizations to figure out distance between two probabilistic distributions are crucial subjects in statistical physics~\cite{JCP_ScottShell_2008,JCP_Rudzinski_2011,JCP_Foley_2015,PRR_Nakazato_Ito_2021}.
One of the important issues is to systematically quantify coarse-grainings, which have been proposed with relative entropy~\cite{JCP_ScottShell_2008,JCP_Rudzinski_2011,JCP_Foley_2015}.}

\red{
The differences in the Shannon entropy from the particle-level to the density field descriptions may share the analogous expressions in light of relative entropy~\cite{JCP_ScottShell_2008,JCP_Rudzinski_2011,JCP_Foley_2015}.
This section introduces a conceivable scenarios to characterize the distance between stochastic formalisms.
%%  from which we can characterize the distance between stochastic formalisms.
To make math formalisms easier, we pay attention to the discrete descriptions while the basic ideas are applicable to the continuous descriptions.
}

\red{
Let the particles' labelings stochastically chosen given the density field.
Therein, all the labelings are uniformly distributed in the same density field, where the probability of the labelings is denoted by an inverse of the multinomial distribution ${\cal P}[\{ i \}]\equiv \prod_{\bm x} \rho_{t,{\bm x}}! /M!$.
The relative entropy is defined in the context of the present article as
\begin{eqnarray}
\left< S_{rel}[\{ {\bm x}_{t,i} \},\{ \rho_{t,{\bm x}} \}] \right>
&\equiv&
k_B
\sum_{ \{ {\bm x}_{t,i} \} \in \Omega^M } \,
{\cal P}[\{ {\bm x}_{t,i} \}]
\nonumber \\
&&
\times
\ln{
\frac{{\cal P}[\{ {\bm x}_{t,i} \}]}{{\cal P}[\{ \rho_{t,{\bm x}} \}]{\cal P}[\{ i \}]}
},
\label{relative_entropy}
\end{eqnarray}
where we attach $\left< (\cdot) \right>$ in $\left< S_{rel} \right>$ to represent the averaged quantity.  
Note that $\{ {\bm x}_{t,i} \}$ identifies the density field $\{ \rho_{t,{\bm x}} \}$ so that ${\cal P}[\{ \rho_{t,{\bm x}} \}]$ corresponding to ${\cal P}[\{ {\bm x}_{t,i} \}]$ can be uniquely determined.
Since the many-particle systems for the discrete descriptions specify the number of the particles' overlappings also with multinomial distributions $M!/\prod_{{\bm x}} \rho_{t,{\bm x}}!$ (cf. Eq.~(\ref{Pro_ini_x_rho})), we have 
\begin{eqnarray}
S^*(t)=S(t)-S^{(\rho)} -k_B \ln{M!}.
\label{SPrho_SPx_Srho_M}
\end{eqnarray}
Thus, Eq.~(\ref{relative_entropy}) turns out to be $\left< S_{rel} \right>=0$.\footnote{\red{
If Eq.~(\ref{relative_entropy}) is organized into the mutual-information-like form between the density field and the particles' labelings, we find $\left< S_{rel} \right>=\left< I^* \right>$ such that
\begin{eqnarray}
\left< I^* \right>
&\equiv&
k_B
\sum_{ (\{ \rho_{t,{\bm x}} \}, \{ i \} ) } \,
{\cal P}[\{ \rho_{t,{\bm x}} \}, \{ i \} ]
\ln{
\frac{{\cal P}[\{ \rho_{t,{\bm x}} \}, \{ i \} ]}{{\cal P}[\{ \rho_{t,{\bm x}} \}] {\cal P}[\{ i \}] }
},
\label{mutual_info_like}
\end{eqnarray}
where the joint probability ${\cal P}[\{ \rho_{t,{\bm x}} \}, \{ i \} ]$ denotes the probability of the positional distributions of the labeled particles, i.e., ${\cal P}[\{ \rho_{t,{\bm x}} \}, \{ i \} ] =  {\cal P}[\{ {\bm x}_{t,i} \}]$.
A caution is, however, that $\left< I^* \right>$ is not the conventional mutual information since ${\cal P}[\{ i \}]={\cal P}[\{ {\bm x}_{t,i} \}\,|\,\{ \rho_{t,{\bm x}} \}]$ is the conditional probability of $\{ {\bm x}_{t,i} \}$ given $\{ \rho_{t,{\bm x}} \}$, and also $\sum_{ \{ \rho_{t,{\bm x}} \} } {\cal P}[\{ \rho_{t,{\bm x}} \}, \{ i \} ]$ is not generally reduced to ${\cal P}[\{ i \}]$ as the marginal distribution relation.
The identity ${\cal P}[\{ {\bm x}_{t,i} \}]={\cal P}[\{ {\bm x}_{t,i} \}\,|\,\{ \rho_{t,{\bm x}} \}]{\cal P}[\{ \rho_{t,{\bm x}} \} ]$ leads to
\begin{eqnarray}
-\ln{{\cal P}[\{ \rho_{t,{\bm x}} \} ]}=-\ln{{\cal P}[\{ {\bm x}_{t,i} \}]} +\ln{{\cal P}[\{ {\bm x}_{t,i} \}\,|\,\{ \rho_{t,{\bm x}} \}]},
\end{eqnarray}
which immediately finds a direct correspondence to Eq.~(\ref{SPrho_SPx_Srho_M}) and $\left< S_{rel} \right>=\left< I^* \right>=0$.
}}
}

\red{
Subsequently, we consider more general cases involving spatial-resolution changes.
The $K$ lattices at the fine resolutions form a block lattice assigned with the coarse-grained coordinate $\overline{\bm x}$, where the notation ${\bm x} \in \overline{\bm x}$ is adopted if the coordinates ${\bm x}$ at the fine resolution belong to $\overline{\bm x}$.
Let $\overline{\rho}_{t,\overline{\bm x}}=(1/K)\sum_{{\bm x}\in \overline{\bm x}} {\rho}_{t,{\bm x}}$ be the averaged number density for the block lattice at $\overline{\bm x}$.
}

\red{
By putting aside the particles' labelings for simplicity, our first try is devoted to the spatial coarse-graining of the density field, whose definition of the relative entropy is naturally given by
\begin{eqnarray}
\left< S_{rel}[\{ \rho_{t,{\bm x}} \},\{ \overline{\rho}_{t,\overline{\bm x}} \}] \right>
&\equiv&
k_B
\sum_{ \{ {\rho}_{t,{\bm x}} | {\bm x} \in \Omega  \} } \,
\,
{\cal P}[\{ \rho_{t,{\bm x}} \}]
\nonumber \\
&&
\times 
\ln{
\frac{{\cal P}[\{ \rho_{t,{\bm x}} \}]}{{\cal P}[\{ \overline{\rho}_{t,\overline{\bm x}} \}]{\cal Z}_{CG}[\{ \overline{\rho}_{t,\overline{\bm x}} \}]}
}.
\nonumber \\
\label{relative_entropy_DFs}
\end{eqnarray}
The probability of the coarse-grained density field is obtained from
\begin{eqnarray}
{\cal P}[\{ \overline{\rho}_{t,\overline{\bm x}} \}]
=
\sum_{ \{ {\rho}_{t,{\bm x}} | {\bm x} \in \Omega  \} } \,
{\cal P}[\{ \rho_{t,{\bm x}} \}]
\prod_{\overline{\bm x}}
\delta \left( \overline{\rho}_{t,\overline{\bm x}},\frac{1}{K}\sum_{{\bm x}\in \overline{\bm x}} {\rho}_{t,{\bm x}} \right),
\nonumber \\
\end{eqnarray}
where $\delta(\alpha,\beta)\equiv \delta_{\alpha\beta}$ is employed to enlarge arguments.
Note that the particles inside the block lattices are assumed to be uniformly distributed, and the number of the configurations are denoted and specified by $1/{\cal Z}_{CG}[\{ \overline{\rho}_{t,\overline{\bm x}} \}]$ with ${\cal Z}_{CG}[\{ \overline{\rho}_{t,\overline{\bm x}} \}]=\prod_{\overline{\bm x}} \left[ \prod_{{\bm x}\in\overline{\bm x}} \rho_{t,{\bm x}}!/(K\overline{\rho}_{t,\overline{\bm x}})! \right]$.
The normalization conditions are summarized as $\sum_{ \{ {\rho}_{t,{\bm x}} | {\bm x} \in \Omega  \} } {\cal P}[\{ \rho_{t,{\bm x}} \}]=1$, and $\sum_{ \{ \overline{\rho}_{t,\overline{\bm x}} | \overline{\bm x} \in \Omega  \} } {\cal P}[\{ \overline{\rho}_{t,\overline{\bm x}} \}]=\sum_{ \{ {\rho}_{t,{\bm x}} | {\bm x} \in \Omega  \} } {\cal P}[\{ \overline{\rho}_{t,\overline{\bm x}} \}]{\cal Z}_{CG}[\{ \overline{\rho}_{t,\overline{\bm x}} \}]=1$.
Equation~(\ref{relative_entropy_DFs}) is found to take nonnegative values from the Gibbs inequality, and presents one of the measures to quantify degrees of the spatial coarse-graining in the density field.
While no coarse-graining exhibits $\left< S_{rel} \right>=0$, the higher $\left< S_{rel} \right>$ alludes to the coarser resolutions.
}

\red{
A last step in this section is to incorporate particles' labelings.
Mediated by the relative entropy between $\{ {\bm x}_{t,i} \}$ and $\{ \overline{\bm x}_{t,i} \}$, 
a candidate of the relative entropy between $\{ {\bm x}_{t,i} \}$ and $\{ \overline{\rho}_{t,\overline{\bm x}} \}$ is introduced as
\begin{eqnarray}
\left< S_{rel}[\{ {\bm x}_{t,i} \},\{ \overline{\bm x}_{t,i} \}] \right>
&\equiv&
k_B
\sum_{ \{ {\bm x}_{t,i} \} \in \Omega^N }
{\cal P}[\{ {\bm x}_{t,i} \}]
\nonumber \\
&&
\times
\ln{
\frac{{\cal P}[\{ {\bm x}_{t,i} \}]}{{\cal P}[\{ \overline{\bm x}_{t,i} \}]{\cal Z}_{CG}[\{ \overline{\bm x}_{t,i} \}]}
}.
\nonumber \\
&=&
k_B
\sum_{ \{ {\bm x}_{t,i} \} \in \Omega^N }
{\cal P}[\{ {\bm x}_{t,i} \}]
\nonumber \\
&&
\times
\ln{
\frac{{\cal P}[\{ {\bm x}_{t,i} \}]}{{\cal P}[\{ \overline{\rho}_{t,\overline{\bm x}} \}]{\cal P}[\{ \overline{i} \}]{\cal Z}_{CG}[\{ \overline{\bm x}_{t,i} \}]}
},
\nonumber \\
\label{relative_entropy_labeling_CGDF}
\end{eqnarray}
where $\sum_{ \{ \overline{\bm x}_{t,i} \} \in \Omega^N }{\cal P}[\{ \overline{\bm x}_{t,i} \}]=\sum_{ \{ {\bm x}_{t,i} \}\in \Omega^N  } {\cal P}[\{ \overline{\bm x}_{t,i} \}]{\cal Z}_{CG}[\{ \overline{\bm x}_{t,i} \}]=1$ with ${\cal Z}_{CG}[\{ \overline{\bm x}_{t,i} \}]=1/K^M$.
Note that the number of the fine-resolution configurations denoted by ${\bm x}_{t,i}$ for ${\bm x}_{t,i} \in \overline{\bm x}_{t,i}$ in a block lattice at $\overline{\bm x}_{t,i}$ is $1/{\cal Z}_{CG}[\{ \overline{\bm x}_{t,i} \}]=K^M$.
In Eq.~(\ref{relative_entropy_labeling_CGDF}), by using ${\cal P}[\{ \overline{\bm x}_{t,i} \}]={\cal P}[\{ \overline{\rho}_{t,\overline{\bm x}} \}]{\cal P}[\{ \overline{i} \}]$ with ${\cal P}[\{ \overline{i} \}]\equiv \prod_{\overline{\bm x}} \overline{\rho}_{t,\overline{\bm x}}! /M!$, 
through the first equality to show a comparison of $\{ {\bm x}_{t,i} \}$ with $\{ \overline{\bm x}_{t,i} \}$,
we arrive at the last equality, where the relative entropy quantify the distance in the probabilities between $\{ {\bm x}_{t,i} \}$ and $\{ \overline{\rho}_{t,\overline{\bm x}} \}$.
Equation~(\ref{relative_entropy_labeling_CGDF}) is similarly found to exhibit the nonnegativity from Gibbs inequality, and no spatial resolution change with $K=1$ reduces Eq.~(\ref{relative_entropy_labeling_CGDF}) to Eq.~(\ref{relative_entropy}).
Although Eq.~(\ref{relative_entropy_labeling_CGDF}) is a conceivable form to see the degrees of the coarse-grainings, applications, availabilities and the better definitions are elusive, and the further studies are awaited.
}

\subsection{Mean-field picture}

In many polymer systems, the Flory-Huggins framework based on the mean-field pictures has been utilized~\cite{deGennesBook}.
Therein, the free energy is assumed to be an additive form of the phenomenological interaction energy and the mixing entropy. 
The latter is quantified by the volume fraction considered as locally averaged quantity made from the volume of monomers and the solvent molecules.
Note that the picture even before the average operation is already simplified because the space is exclusively allocated to the monomers or the solvent molecules without the vacant place.
Evidently, some coarse-grainings or changes in the spatiotemporal resolution need to be inserted if the construction of the Flory-Huggins framework built up from Eq.~(\ref{NumDen_polymer_2}) is possible.

Probably, Eq.~(\ref{NumDen_polymer_2}) is a natural formalism to satisfy Eq.~(\ref{heat_heat_relation}).
\red{
The density field defined by Eq.~(\ref{heat_heat_relation}) seems, however, formalistic.
Indeed, as well as the Flory-Huggins framework, the practical modern polymer models exemplified by self-consistent field theory~\cite{JPCB_Delaney_Fredrickson_2016,JCP_Panagiotou_Fredrickson_2019,Macromolecules_Vigil_2021} are compatible with more familiar form of the density field defined as\footnote{
Equation~(\ref{rho_poly_continuous_def}) is for ``continuous" monomer labelings in the ``continuous" space.
The discrete descriptions replace Eq.~(\ref{rho_poly_continuous_def}) with 
\begin{eqnarray}
\varphi(t,{\bm x})
&\equiv&
\sum_{i=1}^M
\sum_{n=1}^N\,
\delta_{{\bm x},\hat{\bm x}_{i,n}}.
\end{eqnarray}
for the ``discrete" monomer labelings in the ``discrete" space, or 
\begin{eqnarray}
\varphi(t,{\bm x})
&\equiv&
\sum_{i=1}^M
\sum_{n=1}^N\,
\delta({\bm x}-\hat{\bm x}_{i,n}).
\end{eqnarray}
for the ``discrete" monomer labelings in the ``continuous" space.
}
\begin{eqnarray}
\varphi(t,{\bm x})
&\equiv&
\sum_{i=1}^M
\int_0^Ndn\,
\delta({\bm x}-\hat{\bm x}_{i,n}),
\label{rho_poly_continuous_def}
\end{eqnarray}
whose Dirac's delta functions may be smeared out by modeling in parallel with practical use.
$\varphi(t,{\bm x})$ is quite distinct from $\rho(t,{\bm x}^*)$ by $\prod_n (\cdot)$ (see Eqs.~(\ref{NumDen_polymer_2}) and (\ref{rho_xn_Xq})).
\if0
To fill the gap from $\rho(t,{\bm x})$ to $\varphi(t,{\bm x})$, we may introduce the linear operator compatible with the number density $\rho(t,{\bm x})$:
\begin{eqnarray}
{\sf L}
\equiv
\sum_{i=1}^M \prod_{j\neq i}\int d{\bm x}_j,
\qquad
{\sf L}
\equiv
\sum_{i=1}^M \prod_{j\neq i}\sum_{{\bm x}_j},
\end{eqnarray}
such that $\varphi(t,{\bm x})={\sf L}\rho(t,{\bm x}^*)$.
Note that the left or the right equation is for the continuous or the discrete descriptions, respectively.
\fi
To pave a road toward the Flory-Huggins framework and also the modern polymer framework,}
combining two distinct projections (Eqs.~(\ref{NumDen_polymer_2}) and (\ref{Projection_tag_monomer})) would help us understand a structure of a next necessary step of projection or coarse-graining more clearly.
%% Note that just combining these projections (eqs.~(\ref{NumDen_polymer_2}),\,(\ref{Projection_tag_monomer})) is not approximation without changing the spatiotemporal resolutions.
However, the developments of the mathematical techniques to deal with the colored noises appearing after eliminating degrees of freedom in the projections are not sufficient.
This issue is interesting, but left for the future work.

\subsection{Applicability of heat field maps}

In intracellular imaging~\cite{PRL_Weber_2010,PhysToday_Barkai_2012,NuclAcidsRes_Kuwada_2013,ChemRev_Norregaard_2017,ColdSpringHarbPerspectBiol_Cremer_Cremer_2010}, both \red{the particle-level and the density field  observations} have been used.
Our framework implies that the heat in stochastic thermodynamics is measured without tracing the particles; that is, we first obtain $d'Q^*/dt$ and $dS^{(\rho)}/dt$ in Eq.~(\ref{heat_heat_relation}) and then determine $d'Q/dt$.
Applying the conversion from the \red{density field} to the experimental data would be interesting.

\section{Concluding remarks}
\label{Conclusion}

This article has investigated \red{the heat differences naturally defined in the particle  and the density field observed on the spatiotemporally discrete or continuous descriptions.
The two heat forms in the particle level and the density field are not generally identical by the entropy of the number density (Eq.~(\ref{heat_heat_relation})),} which represents the indistinguishability of the particles if they take the same point up.

We discussed Eq.~(\ref{heat_heat_relation}) in Langevin and Dean--Kawasaki equations falling into the spatiotemporally continuous \red{descriptions.}
As the heat was defined by stochastic trajectories of the particles proposed by Sekimoto, our framework employed the analogous form constructed with the stochastic number density in the Dean--Kawasaki equations.
The entropic term of the instantaneous number density as the difference in heat between the Langevin and Dean--Kawasaki equations was discovered, but the entropic term exbibits little temporal variations in plausible situations because of the sparse distributions of the point particles almost everywhere.
On the other hand, the differences in the heat between the \red{particle level} and the \red{density field} in the spatiotemporally discrete \red{descriptions} presents the entropic term Eq.~(\ref{heat_heat_relation}) exhibiting the recognizable temporal variations.

The arguments have been also developed from a viewpoint of the fluctuation theorem. 
If the particles occupy the same place, the indistinguishability of the particles at the temporal-axis ends creates the combinational numbers of the particles' placements or the exponential of the entropic term of the number density, whose ratio turns out to relate the ratios between the forward and the reverse path probabilities for the \red{particle level} to those for the \red{density field} (multiple-step version of Eq.~(\ref{Discrete_FT_x_rho}) or Eq.~(\ref{FT_x_to_n_exact})).

We have also discussed the relevant topics in the energy balance about, e.g., the heat differences on the ensemble average, and the projection of the polymer systems.

The \red{density field observations} might occasionally be more easily accessible than the particle's tracking in experiments.
We hope that the heat difference (Eq.~(\ref{heat_heat_relation})) would help us estimate the heat defined on the \red{particle level} from the observation of those on the \red{density field.}

\section*{Acknowledgement}
The authors would like to thank Prof. Uneyama for giving critical comments.
T.S. thanks Dr. Y. Norizoe for giving advice.
\red{The authors appreciate the anonymous reviewers' comments and suggestions.}
The authors thank FORTE Science Communications (https://www.forte-science.co.jp/) for English language editing.
%% The authors thank T. Sakaue for a fruitful discussion.

\section*{Appendix}

The appendix presents guides for technical calculi.

\subsection{\red{Derivation of Eq.~(\ref{Shannon_entropy_diff_discrete})}}

\red{
This section relates the Shannon entropy at the particle level to that in the density field for the discrete description.
}

\red{
The indistinguishability of the particles in the discrete models at the time $t$ quantifies the differences in the probability:
\begin{eqnarray}
{\cal P}[\{ \rho_{t,{\bm x}} \}]
=
\frac{M!}{\prod_{{\bm x}} n_{t,{\bm x}}!}
{\cal P}[\{ {\bm x}_{t,i} \}].
\label{Pro_ini_x_rho}
\end{eqnarray}
Bear in mind that ${\cal P}[\{ {\bm x}_{t,i} \}]$ is a representative one of the probabilities of the labeled particles' positions with the identical density field.
We here attempt to relate ${\cal P}[\{ {\bm x}_{t+1,i} \}]$ to ${\cal P}[\{ \rho_{t+1,{\bm x}} \}]$ at the subsequent time $t+1$ under an initial condition with Eq.~(\ref{Pro_ini_x_rho}) at the time $t$.
}

\red{
Let ${\cal P}[\{ \rho_{t+1,{\bm x}} \},\alpha]$ be the probability of the number density at the time $t+1$, which is produced by the flow map $\alpha$ from the time $t$ to $t+1$.
The density-field probability ${\cal P}[\{ \rho_{t+1,{\bm x}} \},\alpha]$ is associated with the probability ${\cal P}[\{ {\bm x}_{t+1,i} \},\{ {\bm x}_{t,i} \}]$ at the particle level through
\begin{eqnarray}
{\cal P}[\{ \rho_{t+1,{\bm x}} \},\alpha]
&=&
{\cal P}[\{ \rho_{t+1,{\bm x}} \},\alpha|\{ \rho_{t,{\bm x}} \}]
{\cal P}[\{ \rho_{t,{\bm x}} \}]
\nonumber \\
&=&
\left(
{\cal P}[\{ {\bm x}_{t+1,i} \} | \{ {\bm x}_{t,i} \}]
\prod_{{\bm x}}
\frac{n_{t,{\bm x}}!}{\prod_{{\bm x}'} n_{t,{\bm x}\rightarrow{\bm x}'}^{(\alpha)} !}
\right)
\nonumber \\
&&
\times
\frac{M!}{\prod_{{\bm x}} n_{t,{\bm x}}!}
{\cal P}[\{ {\bm x}_{t,i} \}] 
\nonumber \\
&=&
\frac{M!}{\prod_{{\bm x}} \prod_{{\bm x}'} n_{t,{\bm x}\rightarrow{\bm x}'}^{(\alpha)} !}
{\cal P}[\{ {\bm x}_{t+1,i} \},\{ {\bm x}_{t,i} \}],
\nonumber \\
\end{eqnarray}
where the second line is obtained from the combination of the particle flows together with Eq.~(\ref{Pro_ini_x_rho}).
Note that the joint probability of the labeled particles at the time $t$ and $t+1$ is denoted by ${\cal P}[\{ {\bm x}_{t+1,i} \},\{ {\bm x}_{t,i} \}]$.
To avoid the complicated notation, the arguments $\{ {\bm x}_{t+1,i} \},\{ {\bm x}_{t,i} \}$ of ${\cal P}[\{ {\bm x}_{t+1,i} \},\{ {\bm x}_{t,i} \}]$ is abused to represent one of the temporally sequential maps of the labeled particles with the identical transition rate.
For example, the distinct maps of differently labeled particles $\{ {\bm x}_{t,i} \}$ may reach the identical map at the time $t+1$ that has the same particle labeling $\{ {\bm x}_{t+1,i} \}$ with the same flow $\alpha$.
}

\red{
Keeping this in mind, we take a summation over $\alpha$ and have
\begin{eqnarray}
&&
{\cal P}[\{ \rho_{t+1,{\bm x}} \}]
\nonumber \\
&=&
\sum_\alpha
{\cal P}[\{ \rho_{t+1,{\bm x}} \},\alpha]
\nonumber \\
&=&
\sum_\alpha 
\frac{M!}{\prod_{{\bm x}} \prod_{{\bm x}'} n_{t,{\bm x}\rightarrow{\bm x}'}^{(\alpha)} !}
{\cal P}[\{ {\bm x}_{t+1,i} \},\{ {\bm x}_{t,i} \}] 
\nonumber \\
&=&
\frac{M!}{\prod_{{\bm x}'} n_{t+1,{\bm x}'}!}
\sum_\alpha 
\prod_{{\bm x}'}
\frac{n_{t+1,{\bm x}'}!}{\prod_{{\bm x}} n_{t,{\bm x}\rightarrow{\bm x}'}^{(\alpha)} !}
{\cal P}[\{ {\bm x}_{t+1,i} \},\{ {\bm x}_{t,i} \}] 
\nonumber \\
&=&
\frac{M!}{\prod_{{\bm x}} n_{t+1,{\bm x}}!}
{\cal P}[\{ {\bm x}_{t+1,i} \}].
\label{P_rho_x_comb_cal}
\end{eqnarray}
In the last line, recalling that ${\cal P}[\{ {\bm x}_{t+1,i} \},\{ {\bm x}_{t,i} \}]$ expresses a representative one of the joint probabilities with the particles labeled at the time $t$ and $t+1$, we use a fact found from the combination arguments:
\begin{eqnarray}
{\cal P}[\{ {\bm x}_{t+1,i} \}]
=
\sum_\alpha 
\prod_{{\bm x}'}
\frac{n_{t+1,{\bm x}'}!}{\prod_{{\bm x}} n_{t,{\bm x}\rightarrow{\bm x}'}^{(\alpha)} !}
{\cal P}[\{ {\bm x}_{t+1,i} \},\{ {\bm x}_{t,i} \}].
\nonumber \\
\end{eqnarray}
The consequence ${\cal P}[\{ \rho_{t+1,{\bm x}} \}]=M!/\prod_{{\bm x}} n_{t+1,{\bm x}}!{\cal P}[\{ {\bm x}_{t+1,i} \}]$ of Eq.~(\ref{P_rho_x_comb_cal}) with Eq.~(\ref{Pro_ini_x_rho}) leads to Eq.~(\ref{Shannon_entropy_diff_discrete})
}

\subsection{FDR of the second kind for the \red{density} field (Eq.~(\ref{FDR2nd_field}))}

Let us derive Eq.~(\ref{FDR2nd_field}).
The covariance of the fluctuating component in the current $\delta{\bm j}(t,{\bm x})$ is given by
\begin{eqnarray}
\left< \delta{\bm j}_\alpha(t,{\bm x}) \delta{\bm j}_{\beta}(t',{\bm x}') \right>
=
\frac{2k_BT}{\gamma}
{\rho}(t,\red{{\bm x}})
\delta(t-t')\delta({\bm x}-{\bm x}')
\delta_{\alpha \beta},
\nonumber \\
\end{eqnarray}
where the subscript Greek $\alpha,\beta=x,y,z$ denotes the Cartesian components of the vectors.
Applying the dot products ${\bm \nabla}_{{\bm x}} \cdot = (\partial/\partial {\bm x}) \cdot$ and ${\bm \nabla}_{{\bm x}'} \cdot = (\partial/\partial {\bm x}') \cdot$, we obtain
\begin{eqnarray}
&&
\sum_{\alpha,\beta}
\frac{\partial}{\partial x_\alpha}
\frac{\partial}{\partial x'_\beta}
\left< \delta{\bm j}_\alpha(t,{\bm x}) \delta{\bm j}_\beta(t',{\bm x}') \right>
\nonumber \\
&=&
\sum_{\alpha,\beta}
\frac{\partial}{\partial x_\alpha}
\frac{\partial}{\partial x'_\beta}
\frac{2k_BT}{\gamma}
{\rho}(t,{\bm x})
\delta(t-t')\delta({\bm x}-{\bm x}')
\delta_{\alpha \beta}
\nonumber \\
&=&
\sum_{\alpha}
\frac{\partial}{\partial x_\alpha}
\frac{\partial}{\partial x'_\alpha}
\frac{2k_BT}{\gamma}
{\rho}(t,{\bm x})
\delta(t-t')\delta({\bm x}-{\bm x}').
\label{FDR_j_cal1}
\end{eqnarray}
Equation~(\ref{FDR_j_cal1}) turns out to be Eq.~(\ref{FDR2nd_field}) expressed with the vector notation ${\bm \nabla}_{{\bm x}} \cdot {\bm \nabla}_{{\bm x}'}=\sum_{\alpha}(\partial/\partial x_\alpha)(\partial/\partial x'_\alpha)$.

\subsection{Response function (Eq.~(\ref{M_def}))}

An explicit expression of Eq.~(\ref{M_def}) is obtained with the functional derivative as follows:
\begin{eqnarray}
&&
\frac{\delta}{\delta (-\Delta {\mu}(s,{\bm x}',\boldsymbol{\lambda}))}
\left(
\frac{\partial \left< {\rho}(t,{\bm x}) \right>_\xi }{\partial t}
%% \red{\frac{\partial \left< \hat{\rho}(t,{\bm x}) \right> }{\partial t}}
\right)
\nonumber \\
&=&
\frac{\delta}{\delta (-{\mu}(s,{\bm x}',,\boldsymbol{\lambda}))}
\left(
\frac{1}{\gamma}
{\bm \nabla}_{{\bm x}} \cdot {\rho}(t,{\bm x}) {\bm \nabla}_{{\bm x}}{\mu}(t,{\bm x},\boldsymbol{\lambda})
\right)
\nonumber \\
&=&
\frac{1}{\gamma}
{\bm \nabla}_{{\bm x}} \cdot {\rho}(t,{\bm x}) \delta(t-s){\bm \nabla}_{{\bm x}'}\delta({\bm x}-{\bm x}')
\nonumber \\
&=&
\frac{1}{\gamma}
\delta(t-s)
{\bm \nabla}_{{\bm x}}\cdot {\bm \nabla}_{{\bm x}'} {\rho}(t,{\bm x}) \delta({\bm x}-{\bm x}').
\end{eqnarray}
Notably, the bracket $\left< (\cdot) \right>_\xi$ does not surround ${\rho}(t,{\bm x})$ in the second line because of its definition.

Assuming boundary terms drop upon integration by parts, we can ensure that Eq.~(\ref{EOM_field_rho}) is reduced to Eq.~(\ref{BasicEq_n}) because the convolution part is transformed as \begin{eqnarray}
&&\int_{-\infty}^t ds\int_\Omega d{\bm x'}\,R(t-s,{\bm x},{\bm x'}) 
(-\Delta {\mu} (s,{\bm x'},\boldsymbol{\lambda}))
\nonumber \\
&=&
\frac{2}{\gamma}
\int_{-\infty}^t ds\int_\Omega d{\bm x'}\,
\nonumber \\
&&
\times \delta(t-s) {\bm \nabla}_{{\bm x}} \cdot \left[ {\bm \nabla}_{{\bm x}'} \rho(s,{\bm x}) \delta({\bm x}-{\bm x}') \right] (-\Delta {\mu} (s,{\bm x'},\boldsymbol{\lambda}))
\nonumber \\
&=&
\frac{1}{\gamma}
{\bm \nabla}_{{\bm x}} \cdot [\rho(t,{\bm x}) {\bm \nabla}_{{\bm x}} \Delta {\mu} (t,{\bm x},\boldsymbol{\lambda})]
\nonumber \\
&=&
\frac{1}{\gamma}
{\bm \nabla}_{{\bm x}} \cdot [\rho(t,{\bm x}) {\bm \nabla}_{{\bm x}} {\mu} (t,{\bm x},\boldsymbol{\lambda})].
\label{R_mu_IntegralByParts}
\end{eqnarray}

Note that $R(t-s,{\bm x},{\bm x'})$ in the formalism contains stochastic variable $\rho(t,{\bm x})$.
Hence, the ensemble average of Eq.~(\ref{EOM_field_rho}) is
\begin{eqnarray}
\frac{\partial \left< {\rho}(t,{\bm x}) \right>}{\partial t}
&=&
\int_{-\infty}^t ds\int_\Omega d{\bm x'}\,\left< R(t-s,{\bm x},{\bm x'}) (-\Delta {\mu} (s,{\bm x'},\boldsymbol{\lambda})) \right>
\label{EOM_field_rho_average}
\nonumber \\
\end{eqnarray}
As mentioned with respect to Eq.~(\ref{BasicEq_n}), the equilibrium state requires no average current $\left< {\bm j}(t,{\bm x}) \right>=-(1/\gamma)\left<  \rho(t,{\bm x}) {\bm \nabla}_{{\bm x}} {\mu} (t,{\bm x}) \right>={\bm 0}$, which means $\partial \left< {\rho}(t,{\bm x}) \right>/\partial t=0$.
\red{Indeed}, Eq.~(\ref{EOM_field_rho_average}) is found to satisfy $\partial \left< {\rho}(t,{\bm x}) \right>/\partial t=0$ for $\left< {\bm j}(t,{\bm x}) \right>={\bm 0}$ via Eq.~(\ref{R_mu_IntegralByParts}).

\subsection{Equation~(\ref{heat_heat_relation}) for Langevin \& Dean--Kawsaki equations}
\label{appendix_dif_heat_particle}

Let us first focus on Eqs.~(\ref{heat_def_n}) and (\ref{heat_def_n_local}) to derive Eq.~(\ref{heat_heat_relation}) for a many-particle system.
Substituting Eq.~(\ref{EOM_GLE_mu}) into Eq.~(\ref{heat_def_n_local}) and integrating it with respect to ${\bm x}$, we obtain a total heat rate at time $t$:
\begin{eqnarray} 
\int_{\Omega} d{\bm x}\,
\frac{d'q^*(t,{\bm x})}{dt}
&=&
\int_{\Omega} d{\bm x}\,
\Delta \mu (t,{\bm x},\boldsymbol{\lambda})
\circ
\frac{\partial {\rho}(t,{\bm x})}{\partial t}.
\end{eqnarray}
This equation is transformed into
\begin{eqnarray} 
\int_{\Omega} d{\bm x}\,
\frac{d'q^*(t,{\bm x})}{dt}
&=&
-\int_{\Omega} d{\bm x}\,
\Delta \mu (t,{\bm x},\boldsymbol{\lambda})
\nonumber \\
&&
\times \sum_{i=1}^M
\frac{d\hat{\bm x}_i}{dt}
\odot
{\bm \nabla}_{\bm x}\delta({\bm x}-\hat{\bm x}_i),
\end{eqnarray}
by noting that
\begin{eqnarray} 
\frac{\partial {\rho}(t,{\bm x})}{\partial t}
&=&
\frac{\partial}{\partial t}
\sum_i 
\delta({\bm x}-\hat{\bm x}_i) 
\nonumber \\
&=&
\sum_{i=1}^M
\frac{d\hat{\bm x}_i}{dt}
\odot
{\bm \nabla}_{\hat{\bm x}_i}\delta({\bm x}-\hat{\bm x}_i)
\nonumber \\
&=&
-\sum_{i=1}^M
\frac{d\hat{\bm x}_i}{dt}
\odot
{\bm \nabla}_{{\bm x}}\delta({\bm x}-\hat{\bm x}_i)
\end{eqnarray}
with ${\bm \nabla}_{\hat{\bm x}_i}\delta({\bm x}-\hat{\bm x}_i) =-{\bm \nabla}_{{\bm x}}\delta({\bm x}-\hat{\bm x}_i)$.
Integration by parts then leads to
\begin{eqnarray} 
&&
\int_{\Omega} d{\bm x}\,
\frac{d'q^*(t,{\bm x})}{dt}
\nonumber \\
&=&
\int_{\Omega} d{\bm x}\,
{\bm \nabla}_{{\bm x}}\mu (t,{\bm x},\boldsymbol{\lambda})
\odot
\sum_{i=1}^M
\frac{d\hat{\bm x}_i}{dt}
\delta({\bm x}-\hat{\bm x}_i)
\nonumber \\
&=&
\sum_{i=1}^M
\left[
\int_{\Omega} d{\bm x}\,
{\bm \nabla}_{{\bm x}}\mu (t,{\bm x},\boldsymbol{\lambda})
\delta({\bm x}-\hat{\bm x}_i)
\right]
\odot
\frac{d\hat{\bm x}_i}{dt}.
\label{cal_q_mu_v}
\end{eqnarray}

For later use, the mechanical potential parts included in ${\bm \nabla}_{\hat{\bm x}_i}\mu (t,\hat{\bm x}_i)$ are transformed as
\begin{eqnarray}
&&
{\bm \nabla}_{\hat{\bm x}_i}
\left[
\int_\Omega d{\bm x}'\, {\rho}(t,{\bm x}') u(\hat{\bm x}_i-{\bm x}',\lambda_2)+u_e(\hat{\bm x}_i,\lambda_1)
\right]
\nonumber \\
&=&
{\bm \nabla}_{\hat{\bm x}_i} \left( \frac{1}{2} \sum_{i',j'} u(\hat{\bm x}_{i'}-\hat{\bm x}_{j'},\lambda_2) \right)
+
{\bm \nabla}_{\hat{\bm x}_i} \sum_{i'}u_e(\hat{\bm x}_{i'},\lambda_1)
\nonumber \\
&=&
{\bm \nabla}_{\hat{\bm x}_i}U_T(\{ \hat{\bm x}_{i'} \},\boldsymbol{\lambda}).
\end{eqnarray}
Using this equation, we rewrite the gradients of the chemical potentials as ${\bm \nabla}_{{\bm x}}\mu (t,{\bm x},\boldsymbol{\lambda})={\bm \nabla}_{\hat{\bm x}_i}U_T(\{ \hat{\bm x}_{i'} \},\boldsymbol{\lambda})+{\bm \nabla}_{{\bm x}}k_BT\ln{\hat{\rho}(t,{\bm x})}$.
Then, substituting the rewritten gradients into Eq.~(\ref{cal_q_mu_v}) and using Eq.~(\ref{Lang_colloid}) leads to 
\begin{eqnarray} 
&&
\int_{\Omega} d{\bm x}\,
\frac{d'q^*(t,{\bm x})}{dt}
\nonumber \\
&=&
\sum_{i=1}^M
\left( -\gamma \frac{d\hat{\bm x}_i}{dt} +{\bm \zeta}_i(t) \right)
\odot
\frac{d\hat{\bm x}_i}{dt} 
\nonumber \\
&&
+
\sum_{i=1}^M
\left[
\int_{\Omega} d{\bm x}\,
{\bm \nabla}_{{\bm x}}
k_BT\ln{{\rho}(t,{\bm x})}
\delta({\bm x}-\hat{\bm x}_i)
\right]
\odot
\frac{d\hat{\bm x}_i}{dt}
\nonumber \\
\label{heat_conversion_cal_2}
\end{eqnarray}
The last term on the right-hand side is replaced by
\begin{eqnarray} 
&&
\int_{\Omega} d{\bm x}\,
\sum_i \delta ({\bm x}-\hat{\bm x}_i)
\left(
{\bm \nabla}_{{\bm x}} k_BT\ln{{\rho}(t,{\bm x})}
\odot
\frac{d\hat{\bm x}_i}{dt}
\right)
\nonumber \\
&=&
-\int_{\Omega} d{\bm x}\,
k_BT\ln{{\rho}(t,{\bm x})}
\left(
{\bm \nabla}_{{\bm x}} \sum_i \delta ({\bm x}-\hat{\bm x}_i)
\odot
\frac{d\hat{\bm x}_i}{dt}
\right)
\nonumber \\
&=&
\int_{\Omega} d{\bm x}\,
k_BT\ln{{\rho}(t,{\bm x})}
\left(
\sum_i {\bm \nabla}_{\hat{\bm x}_i} \delta ({\bm x}-\hat{\bm x}_i)
\odot
\frac{d\hat{\bm x}_i}{dt}
\right)
\nonumber \\
&=&
\int_{\Omega} d{\bm x}\,
k_BT\ln{{\rho}(t,{\bm x})}
\circ
\frac{\partial {\rho}(t,{\bm x})}{\partial t},
\label{heat_conversion_cal_3}
\end{eqnarray}
where integration by parts is used to obtain the third line and the fourth line follows from ${\bm \nabla}_{\hat{\bm x}_i}\delta({\bm x}-\hat{\bm x}_i) =-{\bm \nabla}_{{\bm x}}\delta({\bm x}-\hat{\bm x}_i)$.
Because the total number of particles in the system is assumed to be conserved, i.e., $(\partial/\partial t) \int_{\Omega} d{\bm x}\,{\rho}(t,{\bm x})=0$, we further transform it as
\begin{eqnarray} 
&&
\int_{\Omega} d{\bm x}\,
k_BT\ln{{\rho}(t,{\bm x})}
\circ
\frac{\partial {\rho}(t,{\bm x})}{\partial t}
\nonumber \\
&=&
\frac{\partial}{\partial t}
\int_{\Omega} d{\bm x}\,
\times k_BT{\rho}(t,{\bm x})\ln{{\rho}(t,{\bm x})}.
\label{Shannon_entropy_term}
\end{eqnarray}
Substituting Eqs.~(\ref{heat_conversion_cal_3}) and (\ref{Shannon_entropy_term}) into Eq.~(\ref{heat_conversion_cal_2}), we finally arrive at 
\begin{eqnarray} 
\int_{\Omega} d{\bm x}\,
\frac{d'q^*(t,{\bm x})}{dt}
&=&
\sum_i 
\left( -\gamma \frac{d\hat{\bm x}_i}{dt} +{\bm \zeta}_i(t) \right)
\odot
\frac{d\hat{\bm x}_i}{dt} 
\nonumber \\
&&
+
\frac{\partial}{\partial t}
\int_{\Omega} d{\bm x}\,
k_BT{\rho}(t,{\bm x})\ln{{\rho}(t,{\bm x})}.
\nonumber \\
\label{Lang_heat_particle_dff}
\end{eqnarray}
The first term on the right-hand side corresponds to the heat along the particles' trajectories, as in Eq.~(\ref{heat_particle}).

\subsection{Fluctuation theorem for Eqs.~(\ref{FT_field}) and (\ref{FT_particle})}

\subsubsection{Equation~(\ref{FT_particle}) \red{at the particle level}}
\label{sec_FT_particle_der}

We begin with discussing the fluctuation theorem for a many-particle system, obeyed by \red{equation~(\ref{Lang_colloid}) of motion.} To obtain the ratio of the probability of a forward path \red{at the particle level} to that of the reverse, we constructed the path probabilities using the Onsager--Machlup approach~\cite{PhysRev_Onsager_Machlup_1953,JStatMech_Ohkuma_Ohta_2007}.

A set of the Gaussian-distributed noises $\{ {\bm \zeta}_i \}_t=\{ {\bm \zeta}_1(t), {\bm \zeta}_2(t),\cdots, {\bm \zeta}_M(t) \}$ at a time $t$ acts on the particles indexed from $i=1$ to $M$. 
The probability of a temporal sequence of the sets $\{ {\bm \zeta}_i \}_{t_0}^{t_f} =\{ \{ {\bm \zeta}_i(t_0) \} \rightarrow \{ {\bm \zeta}_i(t_1) \} \rightarrow \{ {\bm \zeta}_i(t_2) \} \rightarrow \cdots \rightarrow \{ {\bm \zeta}_i(t_f) \} \}$ is quantified with the mean and the variance (the FDR of the second kind Eq.~(\ref{FDR_element_Lang}))) as 
\begin{eqnarray}
{\cal P}[\{{\bm \zeta}_i(\cdot) \}]
&\sim&
\exp{}\Biggl[-\int_{t_0}^{t_f} dt \sum_{i=1}^M\,\frac{{\bm \zeta}_i(t)^2}{4\gamma k_BT}
\Biggr].
\end{eqnarray}
Changing the variables from $\{ {\bm \zeta}_i \}_{t_0}^{t_f}$ to $\{ \hat{\bm x}_i \}_{t_0}^{t_f}$ reveals that the path probability ${\cal P}[ \{ \hat{\bm x}_i \}_{t_0}^{t_f} | \{ \hat{\bm x}_i \}_{t_0} ]$ along a trajectory $\{ \hat{\bm x}_i \}_{t_0}^{t_f}=\{ \{ \hat{\bm x}_i(t_0) \} \rightarrow \{ \hat{\bm x}_i(t_1) \} \rightarrow \{ \hat{\bm x}_i(t_2) \} \rightarrow \cdots \rightarrow \{ \hat{\bm x}_i(t_f) \} \}$ given initial positions $\{ \hat{\bm x}_i \}_{t_0}$ satisfies ${\cal P}[ \{ {\bm \zeta}_i \}_{t_0}^{t_f}]=J {\cal P}[\{ \hat{\bm x}_i \}_{t_0}^{t_f}|\{ {\bm x}_i \}_{t_0} ]$, where $J$ denotes the Jacobian from $\{ {\bm \zeta}_i \}_{t_0}^{t_f}$ to $\{ \hat{\bm x}_i \}_{t_0}^{t_f}$~\cite{SekimotoBook,JStatMech_Ohkuma_Ohta_2007,PRE_Hochberg_1999}.

We now assume that a reverse trajectory $\{ \hat{\bm x}_i^\dagger \}_{t_0}^{t_f}=\{ \{ \hat{\bm x}_i^\dagger(t_0) \} \rightarrow \{ \hat{\bm x}_i^\dagger(t_1) \} \rightarrow \{ \hat{\bm x}_i^\dagger(t_2) \} \rightarrow \cdots \rightarrow \{ \hat{\bm x}_i^\dagger(t_f) \} \}=\{ \{ \hat{\bm x}_i(t_f) \} \rightarrow \cdots \rightarrow \{ \hat{\bm x}_i(t_2) \} \rightarrow \{ \hat{\bm x}_i(t_1) \} \rightarrow \{ \hat{\bm x}_i(t_0) \} \}$ is generated by a sequence of sets of noises $\{ {\bm \zeta}_i^{(R)} \}_{t_0}^{t_f}$ satisfying
\begin{eqnarray}
\gamma \frac{d\hat{\bm x}_i^\dagger(t)}{dt}
=
-{\bm \nabla}_i U_T(\{ \hat{\bm x}_{i'}^\dagger(t) \},\boldsymbol{\lambda}^\dagger)
+
{\boldsymbol \zeta}_i^{(R)}(t^\dagger)
\label{Lang_colloid_reverse_0}
\end{eqnarray}
with $t^\dagger\equiv t_f-(t-t_0)$ and $\hat{\bm x}_i^\dagger(t) \equiv \hat{\bm x}_i(t^\dagger)$.
A form of the probability of a sequence of the noises in the reverse trajectory defined with $t^\dagger=t_f-(t-t_0)\rightarrow t$ does not change under the time reflection, which is also written as
\begin{eqnarray}
{\cal P}[\{{\bm \zeta}_i^{(R)} \}_{t_0}^{t_f}]
&\sim&
\exp{}\Biggl[-\int_{t_0}^{t_f} dt^\dagger \sum_{i=1}^M\,\frac{{\bm \zeta}_i^{(R)}(t^\dagger)^2}{4\gamma k_BT}
\Biggr]
\nonumber \\
&=&
\exp{}\Biggl[-\int_{t_0}^{t_f} dt \sum_{i=1}^M\,\frac{{\bm \zeta}_i^{(R)}(t)^2}{4\gamma k_BT}
\Biggr].
\end{eqnarray}

Keeping this result in mind, we focus on the ratio of the probability of the forward path to that of the reverse path, which is found to be
\begin{eqnarray}
&&
\frac{{\cal P}[\{ \hat{\bm x}_i \}_{t_0}^{t_f}|\{ {\bm x}_i \}_{t_0} ]}{{\cal P}[\{ \hat{\bm x}_i^\dagger \}_{t_0}^{t_f}|\{ {\bm x}_i^\dagger \}_{t_0} ]}
\nonumber \\
&=&
\frac{{\cal P}[\{{\bm \zeta}_i \}_{t_0}^{t_f}]}{{\cal P}[\{{\bm \zeta}_i^{(R)} \}_{t_0}^{t_f}]}
\nonumber \\
&=&
\exp{}\Biggl[-\int_{t_0}^{t_f} dt \sum_{i=1}^M\,\frac{1}{4\gamma k_BT}
\left[ {\bm \zeta}_i(t)^2-{\bm \zeta}_i^{(R)}(t)^2  \right]
\Biggr],
\nonumber \\
\label{ratio_tjrajectory_cal1}
\end{eqnarray}
where a similar procedure to that in the forward path reveals ${\cal P}[ \{ {\bm \zeta}_i^{(R)} \}_{t_0}^{t_f}]=J^\dagger {\cal P}[\{ \hat{\bm x}_i^\dagger \}_{t_0}^{t_f}|\{ {\bm x}_i^\dagger \}_{t_0} ]$ in the reverse; the conventional calculation of the Jacobian~\cite{SekimotoBook,JStatMech_Ohkuma_Ohta_2007} indicates $J=J^\dagger$.

We here refer back to the equation~(\ref{Lang_colloid}) of motion for the forward path generated by $\{ {\bm \zeta}_i \}_{t_0}^{t_f}$ because Eq.~(\ref{Lang_colloid}) should include Eq.~(\ref{Lang_colloid_reverse_0}).
By redefining the variable $t^\dagger=t_f-(t-t_0)\rightarrow t$, we transform Eq.~(\ref{Lang_colloid_reverse_0}) into
\begin{eqnarray}
-\gamma \frac{d\hat{\bm x}_i(t)}{dt}
=
-{\bm \nabla}_i U_T(\{ \hat{\bm x}_{i'}(t) \},\boldsymbol{\lambda})
+
{\boldsymbol \zeta}_i^{(R)}(t).
\label{Lang_colloid_reverse}
\end{eqnarray}
Notably, the time reversal inverts the sign of the velocity as $d\hat{\bm x}_i^\dagger(t)/dt \rightarrow -d\hat{\bm x}_i(t)/dt$.
In addition, algebraic manipulation of Eq.~(\ref{Lang_colloid}) reveals that
\begin{eqnarray}
-\gamma \frac{d\hat{\bm x}_i(t)}{dt}
&=&
-{\bm \nabla}_i U_T(\{ \hat{\bm x}_{i'}(t) \},\boldsymbol{\lambda})
\nonumber \\
&&
+
\left[ 2{\bm \nabla}_i U_T(\{ \hat{\bm x}_{i'}(t) \},\boldsymbol{\lambda}) -{\boldsymbol \zeta}_i(t) \right].
\label{Lang_colloid_FtoR}
\end{eqnarray}
Both Eq.~(\ref{Lang_colloid_reverse}) and Eq.~(\ref{Lang_colloid_FtoR}) trace back the identical forward path.
Hence, comparing Eq.~(\ref{Lang_colloid_reverse}) with Eq.~(\ref{Lang_colloid_FtoR}), we discover that the noises in the reverse trajectory are represented by those in the forward trajectory as
\begin{eqnarray}
{\boldsymbol \zeta}_i^{(R)}(t)
=
2{\bm \nabla}_i U_T(\{ \hat{\bm x}_{i'}(t) \},\boldsymbol{\lambda}) -{\boldsymbol \zeta}_i(t).
\label{reverse_noise_trajectory}
\end{eqnarray}
Substituting Eq.~(\ref{reverse_noise_trajectory}) into Eq.~(\ref{ratio_tjrajectory_cal1}) and then organizing it with Eq.~(\ref{Lang_colloid}), we arrive at
\begin{eqnarray}
&&
\frac{{\cal P}[\{ \hat{\bm x}_i \}_{t_0}^{t_f}|\{ {\bm x}_i \}_{t_0} ]}{{\cal P}[\{ \hat{\bm x}_i^\dagger \}_{t_0}^{t_f}|\{ {\bm x}_i^\dagger \}_{t_0} ]}
\nonumber \\
&=&
\exp{}\Biggl[
-\frac{1}{k_BT}
\int_{t_0}^{t_f} dt \sum_{i=1}^M\,
\left( -\gamma \frac{d\hat{\bm x}_i}{dt} 
+
{\bm \zeta}_i(t)
\right) \odot \frac{d\hat{\bm x}_i}{dt}
\Biggr].
\nonumber \\
\label{FT_particle_app}
\end{eqnarray}
The heat defined by Sekimoto (Eq.~(\ref{heat_particle})) appears as a factor in the argument of the exponent, and Eq.~(\ref{FT_particle_app}) is equivalent to Eq.~(\ref{FT_particle}).

\subsubsection{Equation~(\ref{FT_field}) for the \red{density field}}

In finding the fluctuation theorem for the \red{density field,} we employ a strategy analogous to that in the above argument for tracing the particle trajectory.
A forward sequence of the noises that determines the stochastic forward trajectory given the initial \red{density field} is denoted by
\begin{eqnarray}
{\cal P}[ \{ \xi^{(\mu)} \}_{t_0}^{t_f} ]
&\sim&
\exp{}\Biggl[-\int_{t_0}^{t_f} dt\int_\Omega d{\bm x}\int_\Omega d{\bm x}'\,
\nonumber \\
&&\times \frac{\sigma({\bm x},{\bm x}')}{4k_BT} \xi^{(\mu)}(t,{\bm x})\xi^{(\mu)}(t,{\bm x}')
\Biggr].
\end{eqnarray}
A variable change then constructs the forward path probability of the number density ${\cal P}[ \{ {\rho} \}_{t_0}^{t_f}| \{ \rho \}_{t_0} ]$ from ${\cal P}[ \{ \xi^{(\mu)} \}_{t_0}^{t_f} ]=J {\cal P}[ \{ \rho \}_{t_0}^{t_f} | \{ \rho \}_{t_0} ]$, where $J$ is the Jacobian from $\{ \xi^{(\mu)} \}_{t_0}^{t_f}$ to $\{ \rho \}_{t_0}^{t_f}$.

Let us here write the reverse trajectory with Eq.~(\ref{EOM_GLE_mu}) under a sequence of noise $\{ \xi^{(\mu)(R)}\}_{t_0}^{t_f}$:
\begin{eqnarray}
\int_\Omega d{\bm x'}\,\sigma^{-1}({\bm x},{\bm x'}) \frac{\partial {\rho}^\dagger(t,{\bm x}')}{\partial t}
&=&
-\Delta \mu^\dagger (t,{\bm x},\boldsymbol{\lambda})
+
\xi^{(\mu)(R)}(t^\dagger,{\bm x}).
\nonumber \\
\label{EOM_mu_Reverse}
\end{eqnarray}
For convenience of mathematical expression, by changing the notation of the time as $t^\dagger=t_f-(t-t_0)\rightarrow t$ with $dt^\dagger=-dt$, we rewrite Eq.~(\ref{EOM_mu_Reverse}) as
\begin{eqnarray}
-\int_\Omega d{\bm x'}\,\sigma^{-1}({\bm x},{\bm x'}) \frac{\partial {\rho}(t,{\bm x}')}{\partial t}
&=&
-\Delta \mu (t,{\bm x},\boldsymbol{\lambda})
+
\xi^{(\mu)(R)}(t,{\bm x}),
\nonumber \\
\label{EOM_mu_Reverse_2}
\end{eqnarray}
We are aware that the left-hand side has a minus sign.
At the same time, Eq.~(\ref{EOM_GLE_mu}) can be reduced to the form of the reverse Eq.~(\ref{EOM_mu_Reverse_2}) written with the forward noise sequence $\{ \xi^{(\mu)} \}_{t_0}^{t_f}$.
Indeed, algebraic manipulation of Eq.~(\ref{EOM_GLE_mu}) reveals that
\begin{eqnarray}
-\int_\Omega d{\bm x'}\, \sigma^{-1}({\bm x},{\bm x'}) \frac{\partial {\rho}(t,{\bm x}')}{\partial t}
&=&
-\Delta \mu (t,{\bm x},\boldsymbol{\lambda})
\nonumber \\
&&
+
\left[
2\Delta \mu (t,{\bm x},\boldsymbol{\lambda})
-\xi^{(\mu)}(t,{\bm x})
\right].
\nonumber \\
\label{EOM_mu_Reverse_forward}
\end{eqnarray}
Both Eqs.~(\ref{EOM_mu_Reverse_2}) and (\ref{EOM_mu_Reverse_forward}) trace the identical trajectory; thus, we obtain
\begin{eqnarray}
\xi^{(\mu)(R)}(t,{\bm x})
=
2\Delta \mu (t,{\bm x},\boldsymbol{\lambda})
-\xi^{(\mu)}(t,{\bm x}).
\label{noise_forward_reverse}
\end{eqnarray}

Keeping Eq.~(\ref{noise_forward_reverse}) in mind, we construct ${\cal P}[ \{ {\rho}^\dagger \}_{t_0}^{t_f} | \{ {\rho}^\dagger \}_{t_0} ]$.
Because the temporal reverse does not alter the kernel as $R(t-s,{\bm x},{\bm x}')=R^{\dagger}(t-s,{\bm x},{\bm x}')=(2/\gamma) \delta(t-s) {\bm \nabla}_{{\bm x}} \cdot {\bm \nabla}_{{\bm x}'}{\rho}({\bm x},t)\delta({\bm x}-{\bm x}')$ and the FDR of the second kind $\left< \xi^{(\mu)(R)}(t^\dagger,{\bm x})\xi^{(\mu)(R)}(s^\dagger,{\bm x}') \right>=k_BTR^{-1\,\dagger}(t-s,{\bm x},{\bm x'})=k_BTR^{-1}(t^\dagger-s^\dagger,{\bm x},{\bm x'})$, the probability of the noise sequence in the reverse process is given by
\begin{eqnarray}
{\cal P}[ \{ \xi^{(\mu)(R)} \}_{t_0}^{t_f} ]
&\sim&
\exp{}\Biggl[-\int_{t_0}^{t_f} dt\int_\Omega d{\bm x}\int_\Omega d{\bm x}'\,
\nonumber \\
&&\times \frac{\sigma({\bm x},{\bm x}')}{4k_BT} \xi^{(\mu)(R)}(t,{\bm x})\xi^{(\mu)(R)}(t,{\bm x}')
\Biggr].
\nonumber \\
\end{eqnarray}
Similarly, the path probability of the reverse process is obtained by the variable transformation as ${\cal P}[ \{ \xi^{(\mu)(R)} \}_{t_0}^{t_f} ]=J^\dagger {\cal P}[ \{ \rho^\dagger \}_{t_0}^{t_f} | \{ \rho^\dagger \}_{t_0} ]$.
Recalling Eq.~(\ref{noise_forward_reverse}), we find that
\begin{eqnarray}
&&(J^\dagger)
{\cal P}[ {\rho}^\dagger(\cdot,\cdot)|{\rho}^\dagger(t_0,\cdot) ]
\nonumber \\
&\sim&
\exp{}\Biggl[-\int_{t_0}^{t_f} dt\int_\Omega d{\bm x}\int_\Omega d{\bm x}'\,
\nonumber \\
&&\times \frac{\sigma({\bm x},{\bm x}')}{4k_BT}
\left( 2\Delta \mu (t,{\bm x},\boldsymbol{\lambda}) -\xi^{(\mu)}(t,{\bm x})
\right)
\nonumber \\
&&\times
\left( 2\Delta \mu (t,{\bm x}',\boldsymbol{\lambda}) -\xi^{(\mu)}(t,{\bm x}') \right).
\end{eqnarray}
Noting $J=J^\dagger$, we obtain ${\cal P}[ \{ \xi^{(\mu)} \}_{t_0}^{t_f} ]/{\cal P}[ \{ \xi^{(\mu)(R)} \}_{t_0}^{t_f} ]={\cal P}[ \{ {\rho} \}_{t_0}^{t_f}| \{ {\rho} \}_{t_0} ]/{\cal P}[ \{ \rho^\dagger \}_{t_0}^{t_f} | \{\rho^\dagger \}_{t_0}]$; the ratio of the forward path probability to that of the reverse path probability is
\begin{eqnarray}
&&
\frac{{\cal P}[ \{ {\rho}_{t_0}^{t_f} | \{ \rho \}_{t_0} ]}{{\cal P}[ \{ \rho^\dagger \}_{t_0}^{t_f} | \{\rho^\dagger \}_{t_0}] }
\nonumber \\
&=&
\exp{}
\Bigg[
-\frac{1}{k_BT}
\int_{t_0}^{t_f} dt\int_\Omega d{\bm x}
\nonumber \\
&&
\Bigg(-\int_\Omega d{\bm x'}\,\sigma^{-1}({\bm x},{\bm x}') \frac{\partial {\rho}(t,{\bm x}')}{\partial t}  
+\xi^{(\mu)}(t,{\bm x})
\Biggr)
\red{\circ}
\frac{\partial {\rho}(t,{\bm x})}{\partial t}
\Biggr].
\nonumber \\
\end{eqnarray}
Thus, the heat in the \red{density field} defined by Eqs.~(\ref{heat_def_n}) and (\ref{heat_def_n_local}) is found to satisfy Eq.~(\ref{FT_field}).

\subsection{Guides for mode analyses}

This section summarizes the technical guides for the mode analyses.
Let us focus only on a single polymer and disregard the indices $i$ that distinguish polymers (e.g., ${\bm X}_{i,q} \rightarrow {\bm X}_{q}$).

\subsubsection{Orthogonality}

Let us see orthogonality spanning over the continuous $n$- or discrete $q$-index.

The $q$ is an integer, resulting in an integer $qn$ at lower or upper bounds ($n=0,\,N$) in integration of Eqs.~(\ref{int_n_h_h_dagger})--(\ref{int_n_h_dagger_h_dagger}) over $n$.
Hence, the integer $qn$ at lower or upper bounds simplifies the consequences as the orthogonality in $n$-space: 
\begin{eqnarray}
\int_0^Ndn\,
h_{q,n}h_{q',n}^\dagger
&=&
\delta_{qq'},
\label{int_n_h_h_dagger}
\\
\int_0^Ndn\,
h_{q,n}h_{q',n}
&=&
\frac{c_q}{N}
\delta_{qq'},
\label{int_n_h_h}
\\
\int_0^Ndn\,
h_{q,n}^\dagger h_{q',n}^\dagger
&=&
\frac{N}{c_q}
\delta_{qq'}.
\label{int_n_h_dagger_h_dagger}
\end{eqnarray}
For instance, Eq.~(\ref{int_n_h_h_dagger}) is verified as
\begin{eqnarray}
&&
\int_0^Ndn\,
h_{q,n}h_{q',n}^\dagger
\nonumber \\
&=&
\int_0^Ndn\,
\frac{1}{c_qN}
\cos{\left( \frac{\pi qn }{N} \right)}
\cos{\left( \frac{\pi q'n}{N} \right)}
\nonumber \\
&=&
\frac{1}{2c_qN}
\int_0^Ndn\,
\Bigg[
\cos{\left( \frac{\pi n(q-q') }{N} \right)}
+
\cos{\left( \frac{\pi n(q+q')}{N} \right)}
\Biggr]
\nonumber \\
&=&
\frac{1}{2c_qN}
N
\Bigg[
\delta_{qq'}
+
\delta_{q,-q'}
\Biggr]
\nonumber \\
&=&
\delta_{qq'}.
\end{eqnarray}
To obtain the last line, we note that $\delta_{q,-q'}=0$ for $q\geq 1$ and that $\delta_{qq'}=\delta_{q,-q'}=1$ for $q=q'=0$.
Similarly, we find the orthogonality for Eqs.~(\ref{int_n_h_h}) and (\ref{int_n_h_dagger_h_dagger}) but need the prefactor in front of the Kronecker delta.

Subsequently, we consider the summation over $q$.
If the upper bound of $q$ is infinite, a useful formula is
\begin{eqnarray}
\sum_{q=0}^\infty
h_{q,n}h_{q,n'}^\dagger
&=&
\delta (n-n')+\delta (n+n').
\label{h_h_dagger_q}
\end{eqnarray}
Let us briefly show a derivation.
Because cosine is an even function, by noting $1/c_q=2$ for $q\geq 1$ and $1/c_q=1$ for $q=0$, we transform the left-hand side of Eq.~(\ref{h_h_dagger_q}) into
\begin{eqnarray}
\sum_{q=0}^\infty
h_{q,n}h_{q,n'}^\dagger
&=&
\sum_{q=-\infty}^\infty
\frac{1}{2N}
\Bigg[
\cos{\left( \frac{\pi q(n-n') }{N} \right)}
\nonumber \\
&&
+
\cos{\left( \frac{\pi q(n+n')}{N} \right)}
\Biggr].
\label{h_h_dagger_q_cos}
\end{eqnarray}

We wish to specify the above infinite series; however, its answer from a direct calculation appears to be difficult.
Instead, we attempt to consider the Dirac's delta function in the cosine series.
The Dirac's delta function $\delta (n-n')$ is assumed to be expressed with the series of $(d_q(n')/2)\exp{\left( i\pi qn/N \right)}$:
\begin{eqnarray}
\delta (n-n')=\frac{1}{2}\sum_{q=-\infty}^{+\infty}d_q(n') \exp{\left( +\frac{i\pi qn}{N} \right)},
\label{delta_pre}
\end{eqnarray}
where $i$ is an imaginary number and orthogonality in this expansion is satisfied by
\begin{eqnarray}
\frac{1}{2N}\int_{-N}^{+N}dn\,
\exp{\left( +\frac{i\pi q'n}{N} \right)}
\exp{\left( -\frac{i\pi qn}{N} \right)}
&=&
\delta_{qq'}.
\nonumber \\
\label{Dirac_delta_pre_1}
\end{eqnarray}
Then, $d_q(n')$ is determined as
\begin{eqnarray}
d_q(n')
&=&
\sum_{q'=-\infty}^{+\infty} d_{q'}(n')\delta_{qq'}
\nonumber \\
&=&
\sum_{q'=-\infty}^{+\infty}d_{q'}(n')
\int_{-N}^Ndn\,
\frac{1}{2N}
\exp{\left( +\frac{i\pi q'n}{N} \right)} 
\nonumber \\
&& \times
\exp{\left( -\frac{i\pi qn}{N} \right)}
\nonumber \\
&=&
\int_{-N}^{+N}dn\,
\frac{1}{N}\delta(n-n')
\exp{\left( -\frac{i\pi qn}{N} \right)}
\nonumber \\
&=&
\frac{1}{N}\exp{\left( -\frac{i\pi qn'}{N} \right)},
\end{eqnarray}
where the second line follows from Eq.~(\ref{Dirac_delta_pre_1}) and the assumption of Eq.~(\ref{delta_pre}) is included in the third line.
Hence, we arrive at
\begin{eqnarray}
\delta (n-n')
&=&
\sum_{q=-\infty}^{+\infty}\frac{1}{2N}\exp{\left( \frac{i\pi q(n-n')}{N} \right)}
\nonumber \\
&=&
\sum_{q=0}^{+\infty}\frac{1}{2c_qN}\cos{\left( \frac{\pi q(n-n')}{N} \right)},
\end{eqnarray}
where $q$ runs from $0$ to $+\infty$ in the second line by folding the summation with $c_q$.

\subsubsection{Derivatives}

Derivatives in the mode space are converted into those in the real space as follows:
\begin{eqnarray}
\frac{\partial}{\partial {\bm X}_{q}}
&=&
\int_0^Ndn\,
\frac{\partial {\bm x}_{n}}{\partial {\bm X}_{q}}
\frac{\partial}{\partial {\bm x}_n}
=
\int_0^Ndn\,
h_{q,n}^\dagger
\frac{\partial}{\partial {\bm x}_n},
\label{covariant_dev}
\end{eqnarray}
where ${\bm x}_{n}=\sum_{q=0}^\infty{\bm X}_{q}h_{q,n}^\dagger$ is used.

\subsubsection{Coefficients for a polymer}

We here mention the transformation pertinent to coefficients such as $\gamma_q$.

Let us consider the frictional coefficients.
The frictional kernel and its inverse are respectively defined as
\begin{eqnarray}
\gamma_{nm}
&\equiv&
\sum_{q=0}^\infty
\frac{N\gamma_{q}}{c_q}
h_{q,n}
h_{q,m}
=
\sum_{q=0}^\infty
\gamma_{q}
h_{q,n}
h_{q,m}^\dagger
\label{Gamma_q_series}
\end{eqnarray}
\begin{eqnarray}
\gamma_{nn'}^{-1}
\equiv
\sum_{q=0}^\infty
\frac{c_q}{N\gamma_q}
h_{q,n}^\dagger h_{q,n'}^\dagger
=
\sum_{q=0}^\infty
\frac{1}{\gamma_q}
h_{q,n} h_{q,n'}^\dagger
\label{inv_Gamma_q_series}
\end{eqnarray}
where $h_{q,n}=(c_q/N)h_{q,n}^\dagger$.
Note that these kernels are symmetric from the definition; i.e., $\gamma_{nm}=\gamma_{mn}$ and $\gamma_{nn'}^{-1}=\gamma_{n'n}^{-1}$.

Although a long-range interaction with $\gamma_{nm}\sim |n-m|^{-2+\nu}$ is roughly found for non-draining $z=3$, the free-draining $z=2+(1/\nu)$ provides the local friction.
Indeed, the above definition easily recovers that of the Rouse polymer belonging to the local friction model as $\gamma_{nm}=\sum_{q=0}^\infty\gamma_{q}h_{q,n}h_{q,m}^\dagger=\gamma \delta (n-m)$.
In addition, $\gamma_{nm}$ and $\gamma_{nn'}^{-1}$ successfully satisfy an invertibility: \begin{eqnarray}
\int_0^Ndm\,
\gamma_{nm}
\gamma_{mn'}^{-1}
=
\delta (n-n'),
\label{Gamma_invertibility}
\end{eqnarray}
which is verified by substituting Eqs.~(\ref{Gamma_q_series}) and (\ref{inv_Gamma_q_series}) into the left-hand side of Eq.~(\ref{Gamma_invertibility}) as
\begin{eqnarray}
&&
\int_0^Ndm\,
\sum_{q'\geq 0}
\gamma_{q'}
h_{q',n}
h_{q',m}^\dagger
\sum_{q\geq 0} 
\gamma_q^{-1}
h_{q,m} h_{q,n'}^\dagger 
\nonumber \\
&=&
\sum_{q'\geq 0}
\sum_{q\geq 0} 
\gamma_{q'}\gamma_q^{-1}
\delta_{qq'}
h_{q',n}
h_{q,n'}^\dagger 
\nonumber \\
&=&
\sum_{q\geq 0}
h_{q,n}
h_{q,n'}^\dagger 
=\delta (n-n').
\end{eqnarray}

We here demonstrate an example with $\gamma_{nn'}$.
The heat rate in the real space for a particle's trajectory is also expressed as
\begin{eqnarray}
\frac{d'Q}{dt}
&=&
\sum_{i=1}^M
\int_0^N dn\,
\Biggl( 
-
\int_0^Ndn'\,
\gamma_{nn'}\frac{d\hat{\bm x}_{i,n'}(t)}{dt}
\nonumber \\
&&
+
{\bm \zeta}_{i,n}(t)
\Biggr)
\odot
\frac{d\hat{\bm x}_{i,n}(t)}{dt}.
\label{heat_rate_polymer_real_space}
\end{eqnarray}

Although the calculi are not listed here, we can make similar arguments about the spring constant $k_q$.

\subsection{FDR of the second kind for the polymer field}

In this section, we verify the consistency of the FDR of the second kind (Eq.~(\ref{FDR_field_mode})) with the \red{basic field equation~(\ref{BE_rho_q})} for the many-polymer system.
The derivation proceeds through the steps detailed below.

To specify the noise term, as with the derivation of Eq.~(\ref{BasicEq_n}), we start with performing a temporal derivative of ${\rho}(t,{\bm X}^*)=\sum_{i=1}^M \prod_{q=0}^\infty {\rho}_{i,q}(t,{\bm X}_q)$ under $u_{T,q}({\bm X}_q,\boldsymbol{\lambda})\equiv u_{q}({\bm X}_q,\lambda_2)+u_{e,q}({\bm X}_q,\lambda_1)$:
\begin{eqnarray}
\frac{\partial {\rho}(t,{\bm X}^*)}{\partial t}
&=&
\sum_q {\bm \nabla}_{{\bm X}_{q}} \cdot \left({\rho} \frac{1}{\gamma_q} {\bm \nabla}_{{\bm X}_{q}} u_{T,q} \right)
\nonumber \\
&&
+
\frac{k_BT}{N} \sum_q \frac{c_q}{\gamma_q} {\bm \nabla}_{{\bm X}_{q}}^2 {\rho}
\nonumber \\
&&
-\sum_i\sum_q {\bm \nabla}_{{\bm X}_{q}} \cdot \left({\rho}_{i} \frac{1}{\gamma_q} {\bm Z}_{i,q}\right).
\label{BE_field_poly_pre1}
\end{eqnarray}
The first or second term on the right-hand side is rephrased with a derivative of ${\bm X}^*$, respectively:
\begin{eqnarray}
\sum_q {\bm \nabla}_{{\bm X}_{q}} \cdot \left({\rho} \frac{1}{\gamma_q} {\bm \nabla}_{{\bm X}_{q}} u_{T,q} \right)
=
{\bm \nabla}_{{\bm X}^*} \cdot \left({\rho} {\sf c} {\bm \gamma}^{*\,-1} {\bm \nabla}_{{\bm X}^*} u_T \right),
\nonumber \\
\label{BE_field_poly_pre2}
\\
\frac{k_BT}{N} \sum_q \frac{c_q}{\gamma_q} {\bm \nabla}_{{\bm X}_{q}}^2 {\rho}
=
\frac{k_BT}{N}
{\bm \nabla}_{{\bm X}^*} \cdot
\left(
 {\sf c} {\bm \gamma}^{*\,-1} {\bm \nabla}_{{\bm X}^*} {\rho}
 \right),
 \nonumber \\
\label{BE_field_poly_pre3}
\end{eqnarray}
where $u_{T}({\bm X}^*)\equiv \sum_q c_q^{-1} u_{T,q}({\bm X}_q)$, as in Eqs.~(\ref{harmonic_chain_q}) and (\ref{external_q}).
The last term of Eq.~(\ref{BE_field_poly_pre1}) turns out to be the noise ${\Xi}^{*(\rho)}(t,{\bm X}^*)$ for the \red{density field equation}~(\ref{BE_rho_q}).
Algebraic manipulation transforms the noise ${\Xi}^{*(\rho)}(t,{\bm X}^*)$ of Eq.~(\ref{BE_Xi_q}) into
\begin{eqnarray}
{\Xi}^{*(\rho)}(t,{\bm X}^*)
=
\sum_{i,q}
\prod_{q'\neq q} \delta ({\bm X}_{q'}-\hat{\bm X}_{i,q'})
{\Xi}_{i,q}^{(\rho)}(t)
\label{BE_field_poly_pre4}
\end{eqnarray}
with a linear combination of
\begin{eqnarray}
{\Xi}_{i,q}^{(\rho)}(t,{\bm X}_q)
&=&
-{\bm \nabla}_{{\bm X}_{q}} \cdot \left({\rho}_{i,q} (t,{\bm X}_q) \frac{1}{\gamma_q} {\bm Z}_{i,q}(t)\right).
\label{elementary_noise_i_q}
\end{eqnarray}
Note that ${\rho}_{i,q} (t,{\bm X}_q)\equiv \delta ({\bm X}_q-\hat{\bm X}_{i,q})$.
Combining Eqs.~(\ref{BE_field_poly_pre1})--(\ref{elementary_noise_i_q}), we encounter Eqs.~(\ref{BE_rho_q})--(\ref{BE_Xi_q}).

A remaining issue is to check the FDR of the second kind (Eq.~(\ref{FDR_field_mode})).
To access Eq.~(\ref{FDR_field_mode}), we first construct the correlation of the elementary noise ${\Xi}_{i,q}^{(\rho)}(t,{\bm X}_q)$:
\begin{eqnarray}
&&
\left<
{\Xi}_{i,q}^{(\rho)}(t,{\bm X}_q)
{\Xi}_{i',q'}^{(\rho)}(t',{\bm X}'_{q'})
\right>
\nonumber \\
&=&
\Biggl<
\left( {\bm \nabla}_{{\bm X}_{q} }\cdot ({\rho}_{i,q} \frac{1}{\gamma_q} {\bm Z}_{i,q}(t)) \right)
\nonumber \\
&&
\times 
\left( {\bm \nabla}_{{\bm X}_{q'}'}\cdot ({\rho}_{i',q'} \frac{1}{\gamma_{q'}} {\bm Z}_{i',q'}(t')) \right)
\Biggr>
\nonumber \\
&=&
\frac{1}{\gamma_q\gamma_{q'}}
\left( {\bm \nabla}_{{\bm X}_{q}} \cdot \right)\left( {\bm \nabla}_{{\bm X}_{q'}'} \cdot \right)
\left<
{\rho}_{i,q}{\rho}_{i',q'} {\bm Z}_{i,q}(t) \otimes {\bm Z}_{i',q'}(t')
\right>
\nonumber \\
&=&
\frac{1}{\gamma_q\gamma_{q'}}
\left( {\bm \nabla}_{{\bm X}_{q}} \cdot \right)\left( {\bm \nabla}_{{\bm X}_{q'}'} \cdot \right)
\Biggl<
\delta ({\bm X}_q-\hat{\bm X}_{i,q}) \delta ({\bm X}'_{q'}-\hat{\bm X}_{i',q'})
\nonumber \\
&&
\times
{\bm Z}_{i,q}(t) \otimes {\bm Z}_{i',q'}(t')
\Biggr>
\nonumber \\
&=&
\frac{2c_qk_BT}{N\gamma_q}\delta (t-t')\delta_{qq'}\delta_{ii'}
\nonumber \\
&&
\times \left( {\bm \nabla}_{{\bm X}_{q}} \cdot \right)\left( {\bm \nabla}_{{\bm X}_{q'}'} \cdot \right)
{\rho}_{i,q}
\delta ({\bm X}_q-{\bm X}_{q'}'){\sf I}.
\end{eqnarray}
where the last equality follows from $\delta ({\bm X}_q-\hat{\bm X}_{i,q}) \delta ({\bm X}'_{q}-\hat{\bm X}_{i,q})=\delta ({\bm X}_q-\hat{\bm X}_{i,q}) \delta ({\bm X}_q-{\bm X}'_{q})$ and Eq.~(\ref{FDR_polymer_mode_1}).

A next step is to construct $\left<{\Xi}_i^{(\rho)}(t,{\bm X}^*){\Xi}_{i'}^{(\rho)}(t',{\bm X}'^*)\right>$.
Recalling Eq.~(\ref{BE_field_poly_pre4}), we superimpose the elementary noises ${\Xi}_{i,q}^{(\rho)}(t,{\bm X}_q)$, ${\Xi}_{i',q'}^{(\rho)}(t,{\bm X}'_{q'})$ over $q$, $q'$ (not yet over $i$, $i'$) as
\begin{eqnarray}
&&
\left<
{\Xi}_i^{(\rho)}(t,{\bm X}^*)
{\Xi}_{i'}^{(\rho)}(t',{\bm X}'^*)
\right>
\nonumber \\
&=&
\sum_{q}\sum_{q'}
\Biggl<
\prod_{q_1\neq q} \delta ({\bm X}_{q_1}-\hat{\bm X}_{i,q_1})
{\Xi}_{i,q}(t,,{\bm X}_q)
\nonumber \\
&&
\times
\prod_{q_2\neq q'} \delta ({\bm X}'_{q_2}-\hat{\bm X}_{i',q_2})
{\Xi}_{i',q'}(t',,{\bm X}'_{q'})
\Biggr>
\nonumber \\
&=&
\frac{2k_BT}{N}\delta (t-t')\delta_{ii'}
{\bm \nabla}_{{\bm X}^*} \cdot {\bm \nabla}_{{\bm X}^{*'}} \cdot 
\nonumber \\
&& 
\times
{\sf c}
{\bm \gamma}^{*\,-1}
{\rho}_i({\bm X}^*)
\delta ({\bm X}^*-{\bm X}'^*),
\label{FDR_single_poly_mode}
\end{eqnarray}
where the last line is obtained from
\begin{eqnarray}
&& 
\delta ({\bm X}_q-\hat{\bm X}_{i,q})
\delta ({\bm X}_q-{\bm X}'_{q})
\nonumber \\
&&
\times
\prod_{q_1\neq q} \delta ({\bm X}_{q_1}-\hat{\bm X}_{i,q_1})
\prod_{q_2\neq q} \delta ({\bm X}'_{q_2}-\hat{\bm X}_{i,q_2})
\nonumber \\
&=&
\delta ({\bm X}^*-\hat{\bm X}_i^*)
\delta ({\bm X}^*-{\bm X}'^*).
\end{eqnarray}
Summing over $i$ and $i'$ in Eq.~(\ref{FDR_single_poly_mode}), we arrive at Eq.~(\ref{FDR_field_mode}).

\subsection{Differences in the heat for the many-polymer system}

From the derivations of the heat difference for the many-particle system around Eq.~(\ref{heat_conversion_cal_2}), the analogous formalism between the many-particle and the many-polymer systems reveals that Eq.~(\ref{heat_polymer_field_def}) is transformed like Eq.~(\ref{heat_conversion_cal_2}) into
\begin{eqnarray}
\frac{d'Q^*}{dt}
&=&
\sum_{i=1}^M
N{\sf c}^{-1}
\left( 
-{\bm \gamma}^* \frac{d{\bm X}_{i}^*(t)}{dt}
+{\bm Z}_{i}^*(t)
\right)
\odot
\frac{d{\bm X}_i^*(t)}{dt}
\nonumber \\
&&
+
\sum_{i=1}^M
\Biggl[
\int d{\bm X}^*\,
N{\sf c}^{-1}
\nonumber \\
&& 
\times
{\bm \nabla}_{{\bm X}^*} \frac{{\sf c}k_BT}{N}\ln{{\rho}(t,{\bm X}^*)}
\delta({\bm X}^*-{\bm X}^*_i)
\Biggr]
\odot
\frac{d\hat{\bm X}_i^*(t)}{dt}.
\nonumber \\
\label{heat_field_polymer_cal_1}
\end{eqnarray}
Note that the first term on the right-hand side corresponds to the heat along particles' trajectories and that $N{\sf c}^{-1}$ is present, unlike the particles' system.
The last term is subsequently transformed in a manner similar to Eq.~(\ref{heat_conversion_cal_3}) as
\begin{eqnarray}
&&
\sum_{i=1}^M
\Biggl[
\int d{\bm X}^*\,
N{\sf c}^{-1}
\nonumber \\
&& 
\times
{\bm \nabla}_{{\bm X}^*} \frac{{\sf c}k_BT}{N}\ln{{\rho}(t,{\bm X}^*)}
\delta({\bm X}^*-{\bm X}^*_i)
\Biggr]
\odot
\frac{d\hat{\bm X}_i^*(t)}{dt}
\nonumber 
\end{eqnarray}
\begin{eqnarray}
&=&
\int d{\bm X}^*\,k_BT\ln{{\rho}(t,{\bm X}^*)} 
\circ
\frac{\partial {\rho}(t.{\bm X}^*)}{\partial t}
\nonumber \\
&=&
\frac{\partial }{\partial t}
\int d{\bm X}^*\,k_BT {\rho}(t.{\bm X}^*) \ln{{\rho}(t,{\bm X}^*)}.
\label{polymer_log_cal1}
\end{eqnarray}
Combining Eq.~(\ref{polymer_log_cal1}) with Eq.~(\ref{heat_field_polymer_cal_1}), we eventually arrive at the same formalism (Eq.~(\ref{heat_heat_relation})) even in the many-polymer system.

\subsection{\red{Discrete many-polymer models}}

\red{
This section summarizes attempts to develop the discrete descriptions for many particles modeled in Sec.~\ref{sec_Discrete} into those for many polymers.
There are two principal types of the constructions.
First constructions are intended for closed descriptions that retain the Markov-processes by dealing with each polymer as a single particle on the hyperdimensional descriptions as in Sec.~\ref{ManyPolymers}.
The number density defined in hyperdimensions, however, needs further projections or coarse-grainings to render it practical.
The other one is based on descriptions with the number density built by the projections into the (ordinary) spatial dimensions, where the time evolutions are generally given in the nonMarkov processes, but the fundamental equations or the transition rules closed within a local moment may not be known.
}

\red{
The time and the space are discretized as in Sec.~\ref{sec_Discrete}.
The basic setup for the polymers is the same as that in Sec.~\ref{ManyPolymers}.
Let a system consist of $M$ polymers, each of which forms a liner chain with $N$ monomers.
In this section, we are restricted ourselves into three dimensions as the physical real space, where the position of $n$-th monomer of the $i$-th polymer is denoted by ${\bm x}_{i,n}=(x_{i,n},y_{i,n},z_{i,n})$.
A configuration of the $i$-th polymer is identified by ${\bm x}_i^*\equiv {\bm x}_{i,1} \otimes {\bm x}_{i,2} \otimes \cdots \otimes {\bm x}_{i,N}$ in $3N$ dimensions.
}

\subsubsection{\red{Markov processes on hyperdimensions}}

\red{
We begin with the first formulation intended for the Markov processes.
The density field is defined in the $3N$ dimensions:
\begin{eqnarray}
\rho_{t,{\bm x}^*}
\equiv
\sum_{i=1}^M \delta_{{\bm x}^*,{\bm x}_{t,i}^*}.
\label{rho_discrete_polymer}
\end{eqnarray}
Note that internal symmetricity of the polymer chain is not taken into account now for simplicity whereas a polymer chain may be arbitrarily labeled from either end.
The number density defined by Eq.~(\ref{rho_discrete_polymer}) in $3N$ dimensions does not look practical, but easy to handle with for some purposes.
A single polymer with $N$ monomers in 3 dimensions is viewed as a single particle in $3N$ dimensions.
Thus, many of the consequences discovered for the $M$ particles in 3 dimensions are applicable to $M$ polymers as $M$ particles in $3N$ dimensions.
}

\red{
Through just a replacement of the variables, i.e., from ${\bm x}$, $\rho_{t,{\bm x}}$ in Sec.~\ref{sec_Discrete} to ${\bm x}^*$, $\rho_{t,{\bm x}^*}$,
we have the fluctuation theorem at the particle level for a single step $t\rightarrow t+1$:
\begin{eqnarray}
\frac{{\cal P}[\{ {\bm x}_{t+1,i}^* \}|\{ {\bm x}_{t,i}^* \}]}{{\cal P}[\{ {\bm x}_{t+1,i}^{*\dagger} \}|\{ {\bm x}_{t,i}^{*\dagger} \}]}
&=&
\exp{\left(-\frac{
d' Q_{t}
}{k_BT}\right)},
\label{Q_x_polymer_discrete}
\end{eqnarray}
and also that in the density field is
\begin{eqnarray}
\frac{{\cal P}[\{ \rho_{t+1,{\bm x}^*} \}|\{ \rho_{t,{\bm x}^*} \}]}{{\cal P}[\{ \rho_{t+1,{\bm x}^*}^\dagger \}|\{ \rho_{t,{\bm x}^*}^\dagger \}]}
&=&
\exp{\left(-\frac{
d' Q^*_{t}
}{k_BT}\right)}.
\label{Q_rho_polymer_discrete}
\end{eqnarray}
The entropy of the number density in the polymer configuration space is defined as
\begin{eqnarray}
S^{(\rho)}
=
-k_BT \sum_{{\bm x}^*} \ln{\rho(t,{\bm x}^*)!}.
\label{S_polymer_discrete}
\end{eqnarray}
The heat and the entropy of the number density defined with Eqs.~(\ref{Q_x_polymer_discrete}),\,(\ref{Q_rho_polymer_discrete}), and (\ref{S_polymer_discrete}) are found to be associated through Eq.~(\ref{heat_heat_relation}).
}

\red{
Equations~(\ref{Q_x_polymer_discrete}),\,(\ref{Q_rho_polymer_discrete}), and (\ref{S_polymer_discrete}) are considered as the formal extension to satisfy Eq.~(\ref{heat_heat_relation}), but the number density $\rho_{t,{\bm x}^*}$ in $3N$ dimensions is not suitable for practical use.
Hence, additional projections or coarse-grainings over the framework maintaining $d'Q^*=d'Q-TdS^{(\rho)}$ should be asked for.
}

\subsubsection{\red{nonMarkov processes on ordinary spatial dimensions}}

\red{
Let us next consider the other formulation with the number density defined in $3$ dimensions, but intrinsically built on the nonMarkov processes.
Recall that the fundamental equations like the memoryless Langevin or the Dean-Kawasaki equations are written as the closed forms on the Markov processes, which means that the fundamental equations are given at a local moment.
Eliminating some degrees of freedom, therein, generally makes the description on the Markov to that on the nonMarkov processes, which may not be closed at a local moment with variables of interests.
This generally holds true for the discrete descriptions.}

\red{
Here, we observe only tagged monomers, and have the projected (ordinary) number density $\overline{\rho}_{t,{\bm x}}$ of the tagged monomers in three-dimensional space.
Incidentally, the projections should transfer the change in the elastic potential into the heat on the projected descriptions as in the continuous-space descriptions~\cite{PRE_Saito_2022_1}.
After eliminating degrees of freedom, instead of Eqs.~(\ref{Q_x_polymer_discrete}) and (\ref{Q_rho_polymer_discrete}), we find
\begin{eqnarray}
\frac{{\cal P}[\{ {\bm x}_{i} \}_{t_0}^{t_f}|\{ {\bm x}_{i} \}_{t_0}]}{{\cal P}[\{ {\bm x}_{i}^{\dagger} \}_{t_0}^{t_f}|\{ {\bm x}_{i}^{\dagger} \}_{t_0}]}
&=&
\exp{\left(-\frac{
\Delta' \overline{Q}_{t}
}{k_BT}\right)},
\label{Q_x_polymer_discrete_projection}
\\
\frac{{\cal P}[\{ \overline{\rho} \}_{t_0}^{t_f} |\{ \overline{\rho} \}_{t_0}]}{{\cal P}[\{ \overline{\rho}^\dagger \}_{t_0}^{t_f}|\{ \overline{\rho }^\dagger \}_{t_0}]}
&=&
\exp{\left(-\frac{
\Delta' \overline{Q}^*_{t}
}{k_BT}\right)}
\label{Q_rho_polymer_discrete_projection}
\end{eqnarray}
such that the difference in the heat $\Delta'\overline{Q}^*=\Delta'\overline{Q}-T\Delta S^{(\overline{\rho})}$ is maintained with $S^{(\overline{\rho})}=-k_B\sum_{{\bm x}}\ln{\overline{\rho}(t,{\bm x})!}$.
The derivation of Eq.~(\ref{Q_rho_polymer_discrete_projection}) from Eq.~(\ref{Q_x_polymer_discrete_projection}) can be made by replacing $(t,t+1)$ in Sec.~\ref{sec_Discrete} with $(t_0,t_f)$ since the arguments in Sec.~\ref{sec_Discrete} do not rely on the details of the transient processes and the transition rules.
In addition, the analogous relation concerning the differences in the Shannon entropy $\Delta\overline{S}^*=\Delta S-T\Delta S^{(\overline{\rho})}$ holds, and also the relation pertinent to first and second laws of thermodynamics are formulated consistently. 
However, we need to be aware that generally
\begin{eqnarray}
\frac{{\cal P}[\{ \overline{\rho} \}_{t_0}^{t_f} |\{ \overline{\rho} \}_{t_0}]}{{\cal P}[\{ \overline{\rho}^\dagger \}_{t_0}^{t_f}|\{ \overline{\rho}^\dagger \}_{t_0}]}
&\neq&
\prod_{t=t_0}^{t_f-1}
\frac{{\cal P}[\{ \overline{\rho} \}_{t+1}|\{ \overline{\rho} \}_{t}]}{{\cal P}[\{ \overline{\rho}^\dagger \}_{t+1}|\{ \overline{\rho}^\dagger \}_{t}]},
\end{eqnarray}
which arises from the nonMarkov processes due to the eliminations of degrees of freedom.
Thus, the transition rules are quite different from those in the Markov processes.
}

\red{
The relations on the energy or entropy balances are analogously maintained in the above projected descriptions.
To the best of our knowledge, however, the closed simple fundamental equations or formulations related to the transition rules is not clear as an open issue.
This is left for the future studies.
}

\subsection{no-stationary memoryless processes}

Equation~(\ref{nsLangevin_FE}) is rewritten with $d{\bm W}_i(t)$ denoting an infinitesimal in Wiener process for $i$-th particle as
\begin{eqnarray}
\frac{d\hat{\bm x}_i(t)}{dt}
=
-
\frac{1}{\gamma(t,t_0)}
{\bm \nabla}_i U_T(\{ \hat{\bm x}_{i'} \},\boldsymbol{\lambda})
+
\sqrt{\frac{2k_BT}{\gamma(t,t_0)}}
d{\bm W}_i(t).
\nonumber \\
\label{nsLang_dW}
\end{eqnarray}
Applying the Ito formula together with Eq.~(\ref{nsLang_dW}) into ${\rho}(t,{\bm x})=\sum_{i=1}^M \delta({\bm x}-\hat{\bm x}_i)$, we have 
\begin{eqnarray}
\frac{\partial}{\partial t}\sum_{i=1}^M \delta({\bm x}-\hat{\bm x}_i) dt
&=&
-\sum_{i=1}^M \frac{1}{\gamma(t,t_0)}{\bm \nabla}_i U_T(\{ \hat{\bm x}_{i'} \},\boldsymbol{\lambda}) {\bm \nabla}_{\hat{\bm x}_i}  \delta({\bm x}-\hat{\bm x}_i) dt
\nonumber \\
&& + \frac{1}{2}\sum_{i=1}^M\left( \sqrt{\frac{2k_BT}{\gamma(t,t_0)}} \right)^2 {\bm \nabla}_{\hat{\bm x}_i}^2  \delta({\bm x}-\hat{\bm x}_i) dt
\nonumber \\
&&
+\sum_{i=1}^M \sqrt{\frac{2k_BT}{\gamma(t,t_0)}} {\bm \nabla}_{\hat{\bm x}_i}\delta({\bm x}-\hat{\bm x}_i)\cdot d{\bm W}_i(t).
\end{eqnarray}
The similar calculations to that in Sec.~\ref{stochatic_field} yield
\begin{eqnarray}
\frac{\partial {\rho}(t,{\bm x})}{\partial t}
&=&
{\bm \nabla}_{{\bm x}}\cdot \frac{1}{\gamma(t,t_0)} {\rho}(t,{\bm x}) {\bm \nabla}_{{\bm x}}{\mu}(t,{\bm x},\boldsymbol{\lambda})
+
\xi^{(\rho)}(t,{\bm x}),
\nonumber \\
\end{eqnarray}
where the noise is explicitly expressed as
\begin{eqnarray}
\xi^{(\rho)}(t,{\bm x})dt
=
-\sum_{i=1}^N \sqrt{\frac{2k_BT}{\gamma(t,t_0)}} {\bm \nabla}_{{\bm x}}  \delta({\bm x}-\hat{\bm x}_i)\cdot d{\bm W}_i(t).
\nonumber \\
\end{eqnarray}
Accordingly, the FDR of second kind is satisfied as
\begin{eqnarray}
\left< \xi^{(\rho)}(t,t_0,{\bm x}) \xi^{(\rho)}(t',t_0,{\bm x}') \right>
&=&
\frac{2k_BT}{\gamma (t,t_0)}\delta(t-t')
\nonumber \\
&& 
\times {\bm \nabla}_{{\bm x}} \cdot {\bm \nabla}_{{\bm x}'}
{\rho}({\bm x},t)
\delta({\bm x}-{\bm x}').
\nonumber \\
\end{eqnarray}

\end{document}